# Nanometer resolution imaging and tracking of fluorescent molecules with minimal photon fluxes

Francisco Balzarotti[1]†, Yvan Eilers[1]†, Klaus C. Gwosch[1]†, Arvid H. Gynnå[2],Volker Westphal[1], Fernando D. Stefani[3,4], Johan Elf[2], Stefan W. Hell[1]*

[1]Department of NanoBiophotonics, Max Planck Institute for Biophysical Chemistry, Göttingen, Germany.

[2]Department of Cell and Molecular Biology, Science for Life Laboratory, Uppsala University, Uppsala, Sweden.

[3]Centro de Investigaciones en Bionanociencias (CIBION), Consejo Nacional de Investigaciones Científicas y Técnicas (CONICET), Buenos Aires, Argentina.

[4]Departamento de Física, Facultad de Ciencias Exactas y Naturales, Univeristy of Buenos Aires, Buenos Aires, Argentina

*Correspondence to: shell@gwdg.de; Phone +495512012500.

†Equal contributors.

**Abstract**: We introduce MINFLUX, a concept for localizing photon emitters in space. By probing the emitter with a local intensity minimum of excitation light, MINFLUX minimizes the fluorescence photons needed for high localization precision. A 22-fold reduction of photon detections over that required in popular centroid-localization is demonstrated. In superresolution microscopy, MINFLUX attained ~1 nm precision, resolving molecules only 6 nm apart. Tracking single fluorescent proteins by MINFLUX increased the temporal resolution and the localizations per trace by 100-fold, as demonstrated with diffusing 30S ribosomal subunits in living E. coli. Since conceptual limits have not been reached, we expect this localization modality to break new ground for observing the dynamics, distribution, and structure of macromolecules in living cells and beyond.

**One Sentence Summary:** Probing with local excitation intensity zeros minimizes the emissions needed for localizing emitters with nanometer precision.

Superresolution fluorescence microscopy or nanoscopy methods, such as those called STED (*1, 2*) and PALM/STORM (*3-5*), are strongly impacting modern biology because they can discern fluorescent molecules or features that are closer together than half the wavelength of light. Despite their different acronyms, all these methods ultimately discern densely packed features or molecules in the same way: only one of them is allowed to emit, while the neighbors are prepared to remain silent (*6*). Although this sequential on- and off-switching is highly effective at making neighboring molecules discernible, it does not provide their location in space, which is the second requirement for obtaining a superresolution image. Regarding this point, these methods strongly depart from each other, broadly falling into two categories.

In the so-called coordinate-targeted versions (*6*), which most prominently include STED microscopy, the position of the emitting molecules is established by illuminating the sample with a pattern of light featuring points of ideally zero intensity, such as a doughnut-shaped spot or a standing wave. The intensity and the wavelength of the pattern are adjusted such that the fluorescence ability of the molecules is switched off (or on) everywhere - except at the minima where this process cannot happen. As it is injected by the pattern, the emitter position is always known through the device controlling the position of the minima. In contrast, the coordinate-stochastic superresolution modalities PALM/STORM switch on the molecules individually and randomly in space, implying that the molecular position is established subsequently, using emitted rather than injected photons. Concretely, the emitter position is estimated from the centroid of the fluorescence diffraction pattern produced by the emitter on a camera (*7*). Called 'localization', this process can reach a precision given by the standard deviation of the diffraction fluorescence pattern ($\sigma_{PSF} \backsim 100$ nm) divided by $\sqrt{N}$, with $N$ being the number of photon detections (*8-11*). While $N = 400$ should yield precisions of $\sigma \backsim 5$ nm, obtaining these limits is commonly challenged by other factors such as the typically unknown orientation of the fluorophore emission dipole (*12, 13*).

Camera based localization is also the method of choice for tracking individual molecules (*14-16*). Here, the sum of molecular emissions determines the track length, whereas the emission rate determines the spatio-temporal resolution. Unfortunately, large emission rates reduce the track length by exacerbating bleaching. Alternatively, the molecule can be localized with scanning confocal arrangements (*17*), but also in this case the need for large photon numbers $N$ remains. Therefore, improving localization has so far concentrated on increasing molecular emission, particularly through anti-bleaching agents (*18*), special fluorophores (*19*), cryogenic conditions (*20*), transient (fluorogenic) labels (*21, 22*), and fluorophore-metal interactions (*23*). However, apart from the fact that all these remedies entail restrictions when applied to (living) cells, none of them have addressed the problem of limited emission budget fundamentally.

Here we introduce MINFLUX, a concept for establishing the coordinates of a molecule with (*min*imal) emission *flux*es, originating from a local excitation *min*imum. Compared to centroid-based localization, MINFLUX attains nanoscale precision with a much smaller number of detected photons $N$ and records molecular trajectories with $> 100$-fold higher temporal resolution (*24*). Moreover, our concept is surprisingly simple and can be realized in both scanning beam and standing wave microscopy arrangements.

## Basic concept

To start off, let us recall that in a background-free STED fluorescence microscope with true molecular (1 nm) resolution, detecting a single photon from the position of the doughnut zero is enough to identify a molecule at that coordinate (*25*). Detecting more than one photon is utterly redundant. Let us now perform a *gedanken* experiment in which we seek to establish the trajectory of a molecule diffusing in space. Instead of using uniform widefield excitation and a camera, we now excite with a reasonably bright focal doughnut that can be moved rapidly throughout the focal plane. If we, or a demon, now managed to target the zero of the doughnut-shaped excitation beam exactly at the molecule, steering it so that it is constantly overlapped with the molecule in space, the doughnut-targeting device would map the molecule in perfection without eliciting a single emission. On the other hand, a single emission (e.g. due to a minimal misplacement) would be enough to know that the molecule is not at the location of the doughnut zero.

Unfortunately, we cannot know the position of the molecule in advance and place the doughnut to that coordinate in a single shot, which is why perfect localization without emissions will remain the privilege of the demon. Yet, this *gedanken* experiment suggests that multiple-shot probing of the position of a molecule with an intensity zero should reduce the emissions required for localization. This is because, in our picture, the fluorescence emissions are the price to be paid for not knowing the position, and the closer the zero gets in the course of probing, the lower will be the price. As a matter of fact, the emissions are highly valuable because, apart from confirming the presence of the molecule, they convey information about its distance to the probing zero. Therefore, in a typical realization of MINFLUX (*26, 27*), the location of the molecule is probed with an intensity minimum while the 'residual' emissions reveal the molecular position. Clearly, this strategy entails a favorable photon economy: the approximate position is injected by the many photons from the light source (*25*), whereas the precious fluorescence emissions are just for fine-tuning.

MINFLUX can be implemented with many types of patterns, including standing waves which, after localizing in one dimension (1D), can be rotated to localize in other directions, too. Nonetheless, some key characteristics of MINFLUX hold for any pattern. To derive them, we now assume an arbitrary 1D intensity pattern $I(x)$ with $I(x = 0) = 0$. One can picture a standing wave (fig. 1A) of wavelength $\lambda$, but we explicitly make no restrictions as to the pattern shape. Let us now probe the location $x_m$ of a molecule, ignoring photon statistics for a while. If the pattern is moved, such that the zero sweeps over the probing range $-L/2 < x < L/2$, the molecular fluorescence $f(x) = C\,I(x_m - x)$ vanishes at $x_m$. $C$ is a prefactor that is proportional to the molecular brightness and the detection sensitivity, as well as to a parameter describing the molecular orientation in space. The solution $x_m$ is now easily obtained by solving $f(x_m) = 0$.

Since $C$ is just a prefactor, the molecular orientation has no influence on the solution. This should be contrasted with camera-based localization, where unidentified molecular orientations can induce systematic errors in the tens of nanometer range (*12, 13*). Moreover, because $I(x)$ is known or can be determined experimentally, two probing measurements with the zeros of $I(x)$ placed around the molecule are sufficient for establishing $x_m$ (fig. 1B). Clearly, this also holds for the two 'endpoints' of the $L$-sized probing range, where the signal is given by $f_0 = C\,I(x_m + L/2) = C\,I_0(x_m)$ and $f_1 = C\,I(x_m - L/2) = C\,I_1(x_m)$; note that we have redefined the two displaced intensity functions with the subscripts 0 and 1. If $L$ is so small that

$f(x)$ can be approximated quadratically around $x_m$, any dependence on $\lambda$ disappears. $f(x_m) = C(x_m - x)^2 = 0$ then yields the solution $x_m = L[1/(1 + \sqrt{f_1/f_0}) - 1/2]$ (see supplementary note 3). In other words, for small distances between the zero and the molecular position ($L \ll \lambda/\pi$), MINFLUX localization does not depend on the wavelength creating the light pattern and, since the fluorescence is just collected, the quality of the localization estimate does not depend on any wavelength.

In practice, $f_0$ and $f_1$ are the averages of the acquired photon counts $n_0$ and $n_1$ obeying Poissonian statistics. Therefore, $x_m$ is actually the expected value of the localization with the individual measurements fluctuating around it. The conditional probability distribution of photons $P(n_0, n_1|N)$ follows a binomial distribution $P(n_0, n_1|N) \sim \text{Binomial}(p_0, N)$, where $p_0$ is the probability of assigning a photon to the first probing measurement $I_0$. This success probability is given by $p_0(x) = f_0(x)/[f_0(x) + f_1(x)] = I_0(x)/[I_0(x) + I_1(x)]$ considering the dependence on both $I(x)$ and $L$. We calculated $p_0(x)$ for three distances $L = 50, 100, 150$ nm of a standing wave of $\lambda = 640$ nm showing that between $x = -L/2$ and $x = L/2$, it steeply spans the whole range between zero and unity (fig. 1C). With decreasing $L$, the steepness increases and, in the quadratic approximation, we have $p_0(x) = \frac{1}{2}(2x/L + 1)^2/[(2x/L)^2 + 1]$.

The position of the emitter $x_m$ can be estimated using a maximum likelihood approach. The maximum likelihood estimator (MLE) of $\hat{x}_m$ is such that $p_0(\hat{x}_m) = n_0/(n_0 + n_1) = \hat{p}_0$, where $\hat{p}_0$ is the MLE of the success probability $p_0(x_m)$. We can say that $p_0(x)$ maps the statistics of $n_0$ and $n_1$ into the position estimation, giving the distribution of the position estimator $P(\hat{x}_m|N, L)$. The smaller $L$ is, the more sharply distributed is $\hat{x}_m$ (fig. 1C). Statistical modeling of MINFLUX allows us to calculate the Fisher information of the emitter position and its Cramér-Rao bound (CRB, see supplementary note 1), which determines the best localization precision attainable with any unbiased estimator (fig. 1D). For the quadratic approximation, the CRB is given by $\sigma_{CRB}(x) = L/(4\sqrt{N})[(2x/L)^2 + 1]$ (eq. (S19h)). Unlike in camera-based localization, where the precision is homogeneous throughout the field of view, here it reaches a minimal value $\sigma_{CRB}(0) = L/(4\sqrt{N})$ (fig. 1D) at the center of the probing range. Note that, for example, two measurements with the zero targeted to coordinates within a distance $L = 50$ nm localize a molecule with $\leq 2.5$ nm precision using just 100 detected photons.

Analytical expressions of $p_0(x)$ and $\sigma_{CRB}(x)$ are equally well derived for doughnut beams and other types of patterns, as well as extended in 2D (fig. S1 and supplementary note 2). In fact, a doughnut excitation beam displays similar mathematical behavior around its minimum as a standing wave, but provides 2D information. Moreover, it can be advantageously combined with confocal detection for background suppression. Hence, we decided to explore the MINFLUX concept in a scanning confocal arrangement featuring a doughnut-shaped excitation beam, similarly to our *gedanken* experiment (fig. 2A). Moving the doughnut across a large sample area ($\sim 20 \times 20\ \mu m^2$) was realized by piezoelectric beam deflection, whereas fine positioning was performed electro-optically (see fig. S13 and Materials and Methods). The latter allowed us to set the doughnut zero within $< 5\ \mu s$ and with $\ll 1$ nm precision to arbitrary coordinates $\bar{r}_i$, concomitantly defining the distance $L$ (fig. 2B).

2D MINFLUX localization requires at least three coordinates $\bar{r}_1$, $\bar{r}_2$, and $\bar{r}_3$ of the doughnut zero, preferably arranged as an equilateral triangle (fig. 2B). Considerations and simulations show that adding a fourth doughnut right at the triangle center $\bar{r}_0$ helps removing ambiguities in the position estimation (see fig. S2 and supplementary note 2). Thus, a set of four emitted photon counts $n_0$, $n_1$, $n_2$, and $n_3$ corresponding to the four positions of the doughnut yields the molecular location $(x_m, y_m)$ within an approximate range of diameter $L$, referred to as the field of view (fig. 2B). As we can move and zoom the field of view quickly, our setup entails three basic modes of operation: i) fluorescence nanoscopy (fig. 2C); ii) short range tracking of individual emitters that move within the field of view (fig. 2D); and iii) long range tracking and nanoscopy in microns sized areas, where the field of view is shifted in space in order to cover the large areas (fig. 2E).

The success probability $\bar{p}(\bar{r})$, which maps the statistics of $n_0$, $n_1$, $n_2$, and $n_3$ into the position estimation, is now a multivariate function as is the CRB of the estimator (see fig. S3 and supplementary note 2). Like in the one-dimensional case, the CRB scales linearly with $L$ at the origin and the dependence on $\lambda$ vanishes with increasing validity of the quadratic approximation. We employed two types of position estimators in our experiments. The MLE is used for imaging because its precision was found to converge to the CRB for $N = \sum n_i \gtrsim 100$ photons. If $N < 100$, as is the case for quick position estimation in tracking, a modified least mean square estimator (mLMSE) is more suitable and can be implemented directly in the electronics hardware. Since the mLMSE is biased, the recorded trajectories are corrected afterwards using a numerically unbiased mLMSE (numLMSE) (see fig. S9 and supplementary note 3).

**Localization precision, nanoscopy, and molecular tracking**

To investigate the localization precision of MINFLUX, we repeatedly localized a single fluorescent emitter at different positions throughout the field of view. We used an ATTO 647N molecule in ROXS buffer (*18*) and divided the field of view into an array of $35 \times 35$ pixels separated by 3 nm in both directions. The excitation intensity and pixel dwell time were chosen such that each pixel contained $\lesssim 2$ counts on average. A stack of $\sim 6000$ arrays allowed us to perform an MLE- and numLMSE-based MINFLUX localization on each pixel using varying subsets of $N$ photons. Repeating this procedure with different $N$-sized subsets and comparing each result with the pixel coordinate provided the localization precision at that pixel as a function of $N$ (fig. 3A, fig. S8). At the center of an $L = 100$ nm field of view, $N = 500$ photons were sufficient for obtaining 2 nm precision (fig. 3A-D). Note that localization precision and localization error can be considered equivalent as the bias (accuracy) of the position estimations is negligible. Generally, the precision obtained with MINFLUX is higher than that achievable by a camera (fig. 3D-E). The measurements also confirm the inverse square-root dependence on $N$ (fig. 3E). Throughout the field of view, the precision obtained with MINFLUX agrees very well with the CRB (fig. 3B and 3D), indicating that photon information has indeed been used optimally.

To investigate the resolution obtainable with MINFLUX nanoscopy, we set out to discern fluorophores on immobilized labeled DNA origamis (*28*) featuring distances of 11 nm and 6 nm from each other (fig. 4). After identifying an origami by widefield microscopy, we moved it as close as possible to the center of the field of view ($\bar{r}_0$). As fluorophores we used the popular Alexa Fluor 647 which, in conjunction with suitable chemical environment (*29*), $\lambda = 405$ nm illumination for on-switching, and $\lambda = 642$ nm excitation light, provided the on-off switching rates needed for keeping all but one molecule non-fluorescent. Imaging was performed by identifying the position of each emitting molecule as it emerged stochastically within the field of view. We used $L = 70$ nm and $L = 50$ nm for the 11 nm and the 6 nm origami, respectively. By applying a hidden Markov model (HMM, see Materials and Methods) to the fluorescence emission trace, we discriminated the recurrent single molecule emissions from multiple molecule events and from the background. Recording $n_0$, $n_1$, $n_2$, and $n_3$ for each burst and applying MINFLUX on those with $N = \sum n_i \geq 500$ and $\geq 1000$ for the 11 nm and the 6 nm origami, respectively, allowed us to assemble a map of localizations yielding nanoscale resolution images (fig. 4). Although the individual molecules emerged very clearly, we further applied a $k$-means cluster analysis to classify the localization events into nano-domains representing fully discerned molecules at 11 nm and 6 nm distance. MINFLUX clearly resolves the molecules at 6 nm distance with 100 % modulation (fig. 4N) proving that true molecular scale resolution has been reached at room temperature.

We made a rigorous comparison of MINFLUX nanoscopy with PALM/STORM. For the latter, we considered a noise-free ideal camera, so as to obtain an optimal performance independent of camera characteristics such as dark and gain-dependent noise. To this end, we redistributed the photon counts of each emission event of our MINFLUX images, so that each one comprised exactly $N = 500$ or 1000 counts for the 11 nm and 6 nm origami, respectively. For each nano-domain, the spread (covariance) of the localizations was calculated and displayed as a bivariate Gaussian distribution centered on each nano-domain (fig. 4G,L). For PALM/STORM we also considered $N = 500$ and 1000 photons per measured localization point for the larger and smaller origami, respectively. We then calculated an ideal PALM/STORM image using the CRB of camera-based localization under the conditions that the camera has no read-out noise and the signal to background ratio ($SBR_c$) is 500. For the 11 nm origami, we obtained a localization precision of $\sigma = 5.4$ nm by PALM/STORM and an average $\sigma$ of 2.1 nm for MINFLUX (see supplementary note 4). For the 6 nm origami, the corresponding values were $\sigma = 3.8$ nm for PALM/STORM and just $\sigma = 1.2$ nm for the average MINFLUX precision. While the CRB-based PALM/STORM images represent ideal recordings, the MINFLUX data may still contain influences by sample drift and other experimental imperfections, implying that further improvements are possible.

Next we tracked single 30S ribosomal protein subunits fused to the photoconvertible fluorescent protein mEos2 (*30*) in living Escherichia coli (fig. 5A). MINFLUX tracking became possible after ensuring that i) the switched-on molecules were in the field of view, ii) the four-doughnut measurement was carried out so fast, that it was hardly blurred by motion, and iii) the molecular position was estimated so quickly that repositioning the field of view kept the molecule largely centered. Additionally, the tracking algorithm had to be robust against losing the molecule by blinking, i.e. by the irregular mEos2 on and off intermittencies of 2.2 ms and 0.6 ms average duration, respectively (see fig. S12E and Materials and Methods). These hurdles were overcome by implementing position estimation and decision-making routines in hardware (fig. 2A, see Material and Methods) which, together with our electro-optical and piezoelectric beam steering devices, provided a ~µs response time across a micrometer in an overall observation area of several tens of microns (fig. 2E). The localization frequency of MINFLUX was set to 8 kHz and the mLMS and numLMS position estimators were used in the live and post recording stages, respectively.

A collection of 1535 single molecule tracks was recorded from 27 living E. coli cells. Typical measured trajectories (fig. 5B-E) show that the central ($\bar{r}_0$) doughnut produces a lower count rate, indicating that a single molecule is well centered while tracking. The reconstructed trajectories are naturally constituted of short ∼ms traces (fig. 5C), as the localization procedure is repeatedly interrupted by blinking of the fluorescent probe. The on and off states were identified by applying an HMM to the total collected photons per time interval (see Materials and Methods), thus discriminating the valid localizations.

For each trajectory, the apparent diffusion coefficient $D$ and the localization precision $\sigma$ were estimated for sliding windows of 35 ms. Both parameters were obtained from optimal least square fits (OLSF) of the mean square displacement (MSD, see supplementary note 5). The time dependence of $D$ (fig. 5D) reveals transient behavioral changes with unprecedented temporal resolution for these kinds of probes. It is worth noting that *each point* of this curve utilizes more than 100 valid localizations, which greatly surpasses the typical trajectory length ($\lesssim 15$ samples, see table S2) of classical camera tracking with single fluorescent proteins.

Plotting the mean localization precision $\sigma$ against the mean number of photons per localization $N$ (fig. 5H) proves that the photon efficiency of MINFLUX tracking is 5-10 fold higher than that of its camera-based counterpart (even for an ideal detector, see fig. S6 and supplementary note 4). A mean localization precision of $< 48$ nm was obtained by detecting, on average, just 9 photons per localization with a time resolution $\Delta t$ of 125 µ$s$. MINFLUX tracking was primarily limited by the blinking of mEos2, as it prevents the molecule from being tightly followed by the center of the beam pattern, where photon efficiency is highest. A non-blinking probe would then be tracked more closely to the center, allowing for a smaller pattern size $L$ and further reducing the average tracking error.

Any method that tracks a finite photon-budget probe will suffer from a tradeoff between the trajectory length $S$ and the spatial resolution $\sigma$. Our MINFLUX tracking experiments have been tuned in favor of the tracking length, as it has been shown to be the best strategy regarding the estimation of $D$ (*31*). This can be appreciated in the contour levels of the relative CRB of $D$, $\sigma_D^{CRB}/D$ (fig. 5I), as a function of the tracking length $S$ and the so-called reduced square localization precision $X = \sigma^2/D\Delta t - 2R$ ($R$: blurring coefficient, (*32*)). The latter can be thought of as the squared localization precision in units of the diffusion length within the integration time. In this $X$-$S$ plane, a scatter plot represents each measured trajectory (red), using average values per track. The average trajectory length was 157 ms with 742 valid localizations (which represents a ∼100-fold improvement, see table S2), with a photon budget of ∼5800 collected photons. This shows that MINFLUX tracking can measure apparent diffusion coefficients with precisions $< 20\%$ on average, while camera-base implementations (gray ellipse) center around 70 %.

## Discussion and outlook

Among the reasons why MINFLUX excels over centroid-based localization is that, in the latter, the origin of any detected photon has a spatial uncertainty given by the diffraction limit; in MINFLUX each detected photon is associated with an uncertainty given by the size $L$. Hence, adjusting $L$ below the diffraction limit renders the emitted photons more informative. A perfect example is the origami imaging (fig. 4) where adjusting $L = 50, 70$ nm improves the localization precision substantially. However, making $L$ smaller must not be confused with exploiting external a priori information about molecular positions; no Bayesian estimation approach is needed. MINFLUX typically starts at the diffraction limit, but as soon as some position information is gained, $L$ can be reduced and the uncertainty range 'zoomed in', making the detected photons continually more informative. Therefore, we can also regard MINFLUX as an acronym for *m*aximally *inf*ormative *lu*minescence *ex*citation probing. While we have not really exploited the zooming-in option here, decreasing $L$ repeatedly during the localization procedure will further augment the power of MINFLUX, also eliminating the anisotropies prevalent at large $L$ (fig. 3 C, 4G, 5L). Iterative MINFLUX variants bear enormous potential for investigating macromolecules or interacting macromolecular complexes, potentially rivaling current Förster resonance energy transfer (*33*) and camera-localization-based approaches to structural biology (*34*).

Like in any other concept operating with intensity minima, the practical limits of MINFLUX will be set by background and aberrations blurring the intensity zero of $I(x)$. In our experiments, the doughnut minimum amounted to $< 0.2\%$ of the doughnut crest (fig. S7). Regarding aberration corrections, in camera-based localization one has to correct a faint single molecule emission wavefront containing a few tens or hundreds photons of broad spectral range (100-200 nm), whereas in MINFLUX the corrections are applied to the bright and highly monochromatic (laser) wavefront producing $I(x)$, making the application of spatial light modulators straightforward. Moreover, the correction has to be optimized for the $L$-defined range only. This brings about the important advantage that in iterative MINFLUX implementations it is sufficient to compensate aberrations in the last (smallest $L$) iteration step, where their effect is minimal.

Spatial wavefront modulators can also be used to target the coordinates $\bar{r}_i$ with patterns $I_i(x)$ of varying shape and intensity, which is another degree of freedom for engineering the field of view towards uniform localization precision and for adapting the field of view towards the molecular motion. Since we have achieved molecular scale resolution with the standard fluorophores already, the new frontiers of MINFLUX will not be given by the resolution values but by the number of photons needed to attain that (single nanometer digit) resolution. Conversely, we can expect MINFLUX to enable tracking and nanoscopy of fluorophores that provide much fewer emissions, including auto- and other types of luminophores.

A fundamental difference between MINFLUX and STED nanoscopy is that in the latter, the doughnut pattern simultaneously performs both the localization and the on-off transition. Creating on-off state disparities between two neighboring points requires intensity differences that are large enough to create the off (or on) state with certainty. Since in MINFLUX nanoscopy the doughnut is used just for localization, such certain ('saturated') transitions are obsolete.

Given that probing with an intensity maximum and solving for $\max(f = C\,I(x))$ is equally possible, it is now interesting to ask whether the same localization precision can also be achieved by probing with an intensity maximum. The answer is no, because at a local emission maximum, small displacements of the emitter will not induce detection changes of similar significance for a small distance $L$ (fig. S4). Yet, it will be possible to accommodate multiple fields of view in parallel using arrays of minima provided by many doughnuts or standing waves. Further obvious expansions of our work include multicolor, 3D localization (e.g. by using a z-doughnut (*2*)), discerning emission spectra, polarization, or lifetime. Besides providing isotropic molecular resolution, such expansions should enable observation of inter- and intraprotein dynamics at their characteristic time scales. MINFLUX can also be implemented in setups featuring light sheet illumination (*35*), optical tweezers (*36*) and anti-Brownian electrokinetic trapping (*37*). In fact, MINFLUX should become the method of choice in virtually all experiments that localize single molecules and are limited by photon budgets or slow recording, such as the method called PAINT (*22*). Since it keeps or even relaxes the requirements for sample mounting, our concept should be widely applicable not only in the life sciences but also in other areas where superresolution and molecular tracking bear strong potential.

Finally, it is worthwhile reflecting over the fact that MINFLUX nanoscopy has attained the resolution scale ($\lesssim 6$ nm) where fluorescence molecules start to interact with each other, i.e. the ultimate limit attainable with fluorophores. While fluorescence on-off switching remains the cornerstone for breaking the diffraction barrier, in MINFLUX this breaking is augmented by the fact that, for small distances between a molecule and the intensity zero, the emitter localization does not depend on any wavelength. A consequence of this arguably staggering finding is that superresolution microscopy should also be expandable to low numerical aperture lenses, wavelengths outside the visible spectrum as well as to hitherto inapplicable luminophores. More staggering, however, is the implication that focusing by itself is becoming obsolete, meaning that it should be possible to design microscopy modalities with molecular (1 nm) resolution without employing a single lens.

**Acknowledgments:** We thank Fredrik Persson (Uppsala) for helping with the E coli sample preparation and Elisa D'Este and Steffen Sahl for critical reading. One of us (KCG) thanks the Cusanuswerk for a stipend.

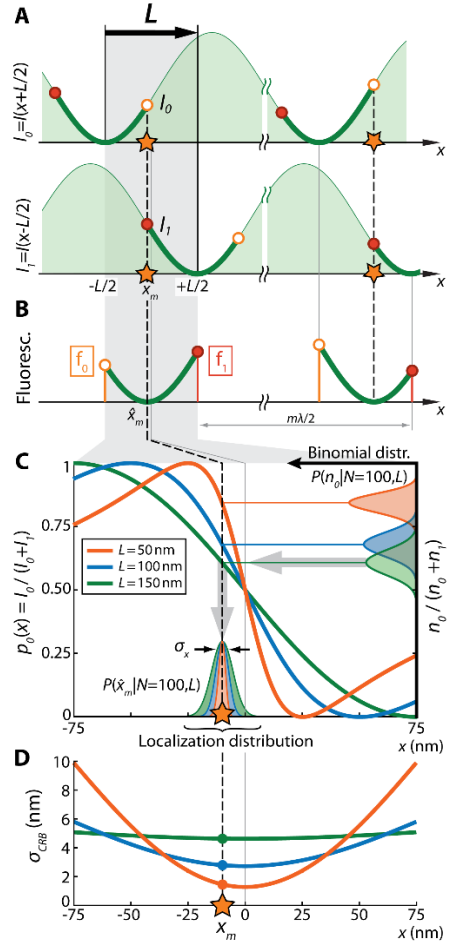

**Fig. 1. Principles of MINFLUX illustrated in a single dimension ($x$) using a standing light wave of wavelength $\lambda$.** (**A**) The unknown position $x_m$ of a fluorescent molecule is determined by translating the standing wave, such that one of its intensity zeros travels from $x = -L/2$ to $L/2$, with $x_m$ being somewhere in between. (**B**) As the molecular fluorescence $f(x)$ becomes zero at $x_m$, solving $f(x_m) = 0$ readily yields a positon estimation $\hat{x}_m$. Equivalently, the emitter can also be located by the two exposures $I_0(x)$ and $I_1(x)$, with their zero at $x = -L/2$ and $L/2$, respectively. The same localization procedure can be performed in parallel with another zero, targeting molecules further away than $\lambda/2$ from the first one. (**C**) Success probability $p_0(x)$ for beam separations $L$ of 50 nm (orange), 100 nm (blue) and 150 nm (green) for $\lambda = 640$ nm. The photon count distribution $P(n_0|N = n_0 + n_1 = 100)$ conditioned to a total 100 photons is plotted along the right vertical axis of normalized counts $n_0/N$ for each $L$. The distribution of counts is mapped into the position axis $x$ through the corresponding $p_0(x, L)$ function (gray arrows), delivering the localization distribution $P(\hat{x}_m|N = 100)$. The position estimator distribution contracts as the distance $L$ is reduced. (**D**) Cramér Rao bound for each $L$. Precision is maximal halfway between the two points where the zeros are placed. For $L = 50$ nm detecting just 100 photons yields a precision of 1.3 nm.

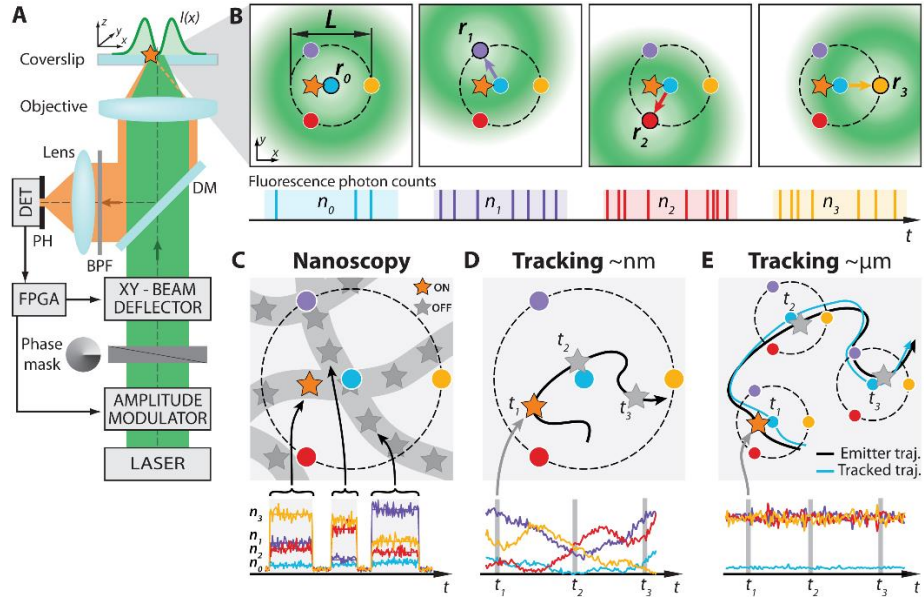

**Fig. 2. Setup, measurement strategy, and various application fields of the two-dimensional MINFLUX implementation.** (**A**, **B**) Simplified setup (details in Materials and Methods). An excitation laser beam (green) is shaped by a vortex phase mask forming a doughnut intensity spot in the focal plane of the objective lens. The intensity of the beam is modulated and deflected such that its central zero is sequentially placed at the four focal plane positions $\bar{r}_{0,1,2,3}$, indicated by blue, violet, red, and yellow dots, respectively. Photons emitted by the fluorescent molecule (star) are collected by the objective lens and directed towards a fluorescence bandpass filter (BPF) and a confocal pinhole (PH), using a dichroic mirror (DM). The fluorescence photons $n_{0,1,2,3}$ counted for each doughnut position $\bar{r}_{0,1,2,3}$ by the detector (DET) are used to extract the molecular location. Intensity modulation and deflection, as well as the photon counting are controlled by a field-programmable-gate-array (FPGA). (**C**-**E**) Basic application modalities of MINFLUX. (C) Nanoscopy: a nanoscale object features molecules whose fluorescence can be switched on and off, such that only one of the molecules is on within the detection range. They are distinguished by abrupt changes in the ratios between the different $n_{0,1,2,3}$ or by intermissions in emission. (D) Short (nanometer) range tracking: the same procedure can be applied to a single emitter that moves within the localization region of size $L$. As the emitter moves, different ratios are observed that allow the localization. (E) Long (microns) range tracking: if the emitter leaves the initial $L$-sized field of view, the triangular set of positions of the doughnut zeros is (iteratively) displaced to the last estimated position of the molecule. By keeping it around $\bar{r}_0$ by means of a feedback loop, photon emission is expected to be minimal for $n_0$ and balanced between $n_1$, $n_2$ and $n_3$, as shown.

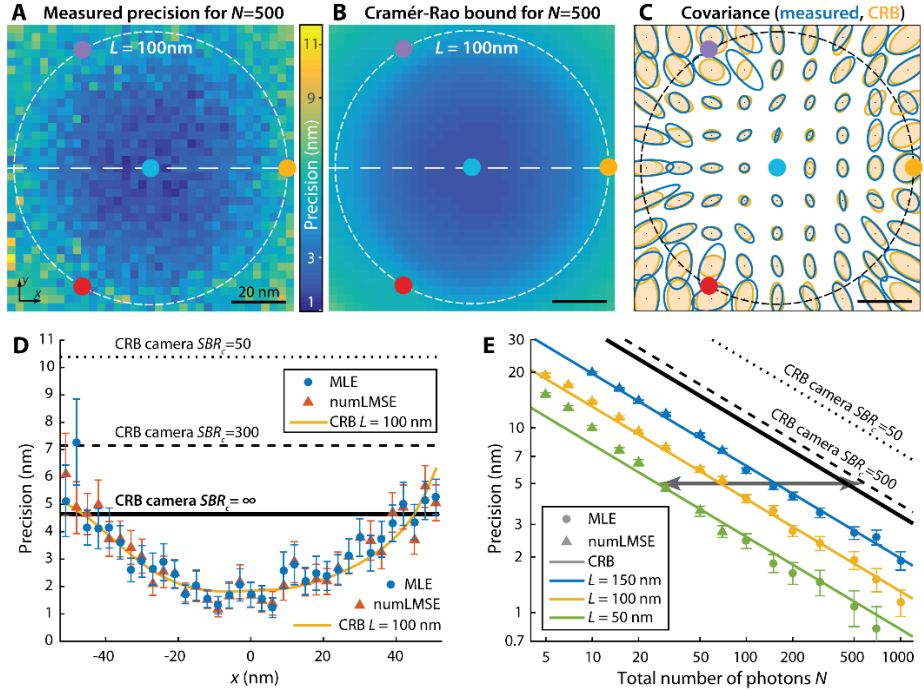

**Fig. 3. Localization precision of MINFLUX measured on a single ATTO 647N molecule.**
(**A**) Measured localization precision obtained by maximum-likelihood estimation (MLE) based MINFLUX. The localization error reaches down to 2 nm for $N = 500$ detected photons using a beam separation of $L = 100$ nm. The signal-to-background ratio ($SBR$) was 13.6 at the central pixel. (**B**) The best precision possible (Cramér-Rao bound, CRB) under the same conditions as in (A). (**C**) Representation of the measured and theoretical localization uncertainty covariances (as ellipses of contour level $e^{-1/2}$), same conditions as in (A, B). (**D**) Measured localization precision along the $x$-axis in (A, B) for MINFLUX localization performed with the MLE (circles) and with the numerically unbiased position estimator (numLMSE, triangles), and the corresponding CRB of MINFLUX (yellow line). The CRB on the localization precision of an ideal camera with realistic signal-to-background ratio (dashed lines) is worse than that provided by MINFLUX (see supplementary note 4). The ultimate limit for the ideal camera (infinite $SBR_c$) is shown by the solid black line. (**E**) Localization precision at the center of the excitation pattern as a function of total number of detected photons $N$: decreasing the beam separation $L$ improves the localization precision more effectively than increasing the number of detected photons; note the logarithmic scales. For the low photon regime ($N < 100$) the numerically unbiased position estimator (numLMSE, see supporting note 3.2.3) was employed, while the MLE was used for $N \geq 100$ detected photons. For most regimes, the measured MINFLUX localization precision reaches the theoretical limit under the measurement conditions (CRB, solid lines). For comparison, the CRB of an ideal camera localization is shown (dashed lines). The camera case of infinite $SBR_c$ is shown by the solid black line. Measurement and theory show that obtaining a localization precision of 5 nm requires ~600 photon counts with an ideal camera ($SBR_c = 500$), while MINFLUX with $L = 50$ nm requires only ~27 photon counts (gray arrow in (E)).

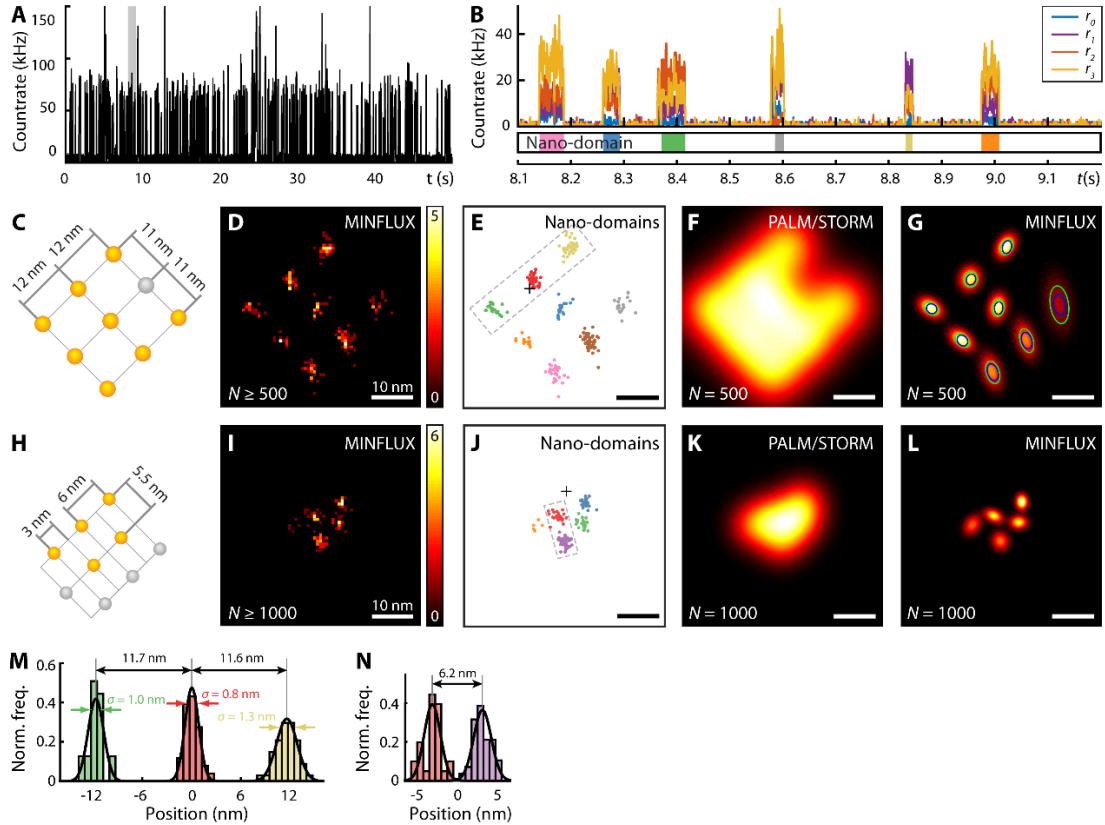

**Fig. 4. MINFLUX nanoscopy of labeled DNA origamis. (A)** Time trace of total photon count rate from a single DNA origami. Time bins: 1 ms. **(B)** Zoomed-in trace interval showing count rates for the four doughnut positions and the resulting classification of each localization in nano-domains (lower panel); color corresponds to the cluster assignments shown in (E). Time bins: 1 ms. **(C)** Arrangement of up to nine on-off switchable fluorophores on the origami (those remaining off throughout the measurement shown in gray). **(D)** Nanoscopy image rendered by spatial binning of direct MINFLUX localizations. Events yielding $N < 500$ detected photons were discarded. Bin size: 0.75 nm. **(E)** Scatter plot of MINFLUX localizations. The coloring shows the classification into nano-domains as described in (B). The dashed gray line indicates the profile for the profile displayed in (M). The position of the central doughnut zero $\bar{r}_0$ is marked by a black cross. **(F, G)** Comparison between practical MINFLUX nanoscopy and ideal PALM/STORM imaging (simulated, see supporting note 4) of the origami using $N = 500$ photons. The rendering shows bivariate normal distributions with the experimental or theoretical covariance, respectively. The green and blue ellipses in (G) illustrate the $e^{-1/2}$ level (diameter $2\sigma$) of the experimental covariance and the Cramér-Rao bound (CRB), respectively. **(H-L)** Analogous to (C-G) for the smaller DNA origami sketched in (H). Events yielding $N < 1000$ detected photons were discarded. **(M)** Projected line profile of the larger origami (C) as indicated by the dashed gray rectangle in (E). Bin size: 0.75 nm. **(N)** Projected line profile of the smaller DNA origami as indicated by the dashed gray rectangle in (J). Bin size: 0.75 nm. Owing to its higher localization precision, MINFLUX nanoscopy displays fundamentally improved resolution over PALM/STORM, reaching single nanometer resolution with $N = 500$ photons at room temperature.

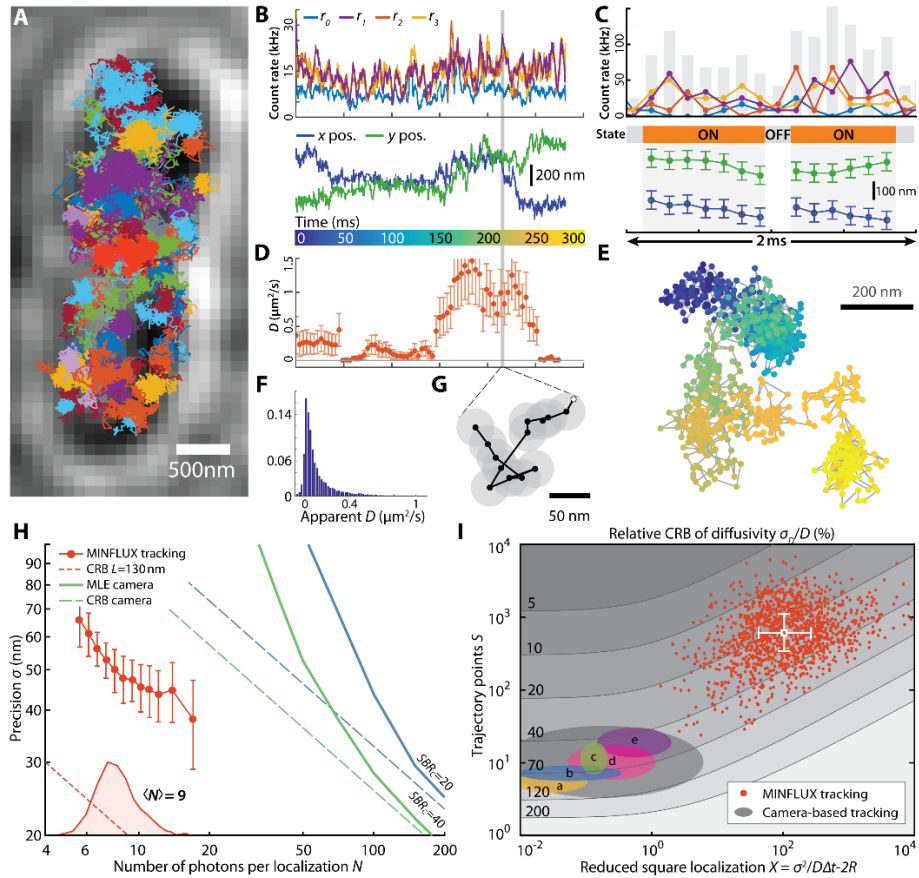

**Fig. 5. Single-molecule MINFLUX tracking in living E. coli bacteria.** Single 30S ribosomal protein subunits fused to the switchable fluorescent protein mEos2 are tracked. (**A**) Transmission image of a bacterium overlaid with 77 independent tracks. (**B**) Details of a track. Upper panel: low pass filtered count rate of the four exposures (blue: $\bar{r}_0$, violet: $\bar{r}_1$, red: $\bar{r}_2$, yellow: $\bar{r}_3$), average total count rate 52 kHz. Lower panel: extracted $x$ and $y$ coordinates of the trajectory. (**C**) 2 ms excerpt of the trace in (B) (marked in gray at time point 210 ms). Upper panel: counts per exposure are shown together with their sum (gray bars) used for on/off classification. Middle panel: on/off classification. Lower panel: Extracted $x$ and $y$ coordinates (cf. (G)); the average tracking error is 48 nm. (**D**) Apparent diffusion constants $D$ for a sliding window of 35 ms with their approximated error bars. (**E**, **G**) Trajectories shown in (B) and (C), respectively. The diameter of the shaded circles in (G) visualize the average tracking error. (**F**) Normalized occurrences of apparent diffusion constants $D$ for all measured tracks. (**H**) Mean localization precision $\sigma$ vs. average counts per MINFLUX localization $N$ for all measured tracks (red circles, marginal distribution of $N$ plotted along the horizontal axis), CRB of MINFLUX static localization for the measuring condition $L = 130$ nm (red dashed), idealized static camera localization performance (CRB: dashed, MLE: line) for two relevant signal to background ratios $SBR_c$ of 20 and 40. (**I**) Contour lines of the relative CRB of the diffusivity $\sigma_D^{CRB}/D$, as a function of the trajectory length $S$ and the reduced localization precision $X$. The white cross shows the quartiles (25 %, 50 % and 75 %) of the marginal distributions of $S$ and $X$ for the experimental data. Simulations showed that the diffusion estimator, though not optimal, provides acceptable results (see fig. S10D and supplementary note 5). The gray ellipse represents how

well the diffusion coefficient has been identified with state-of-the-art camera tracking of fluorescent proteins (colored ellipses refer to table S2).

**Supplementary Materials:**

# Supplementary Materials for

## Nanometer resolution imaging and tracking of fluorescent molecules with minimal photon fluxes

**Authors:** Francisco Balzarotti[1]†, Yvan Eilers[1]†, Klaus C. Gwosch[1]†, Arvid H. Gynnå[2],Volker Westphal[1], Fernando D. Stefani[3,4], Johan Elf[2], Stefan W. Hell[1]*

correspondence to: shell@gwdg.de

**This PDF file includes:**

# Contents of Supplementary Materials

# 1   Fisher information and Cramér-Rao bounds for MINFLUX

In order to quantify how the proposed measurement scheme indeed increases the photon efficiency regarding the estimation of an emitter position, we calculate its Fisher information and its Cramér-Rao Bound (CRB).

We begin with a description of the statistics of photon collection, followed by the construction of a likelihood function. This is used for calculating the Fisher information that a given photon count measurement holds on the parameters of its statistics. Upon a multivariate space transformation, we obtain the Fisher information that said photons hold of the position of the emitter.

An emitter at a position $\bar{r}_m \in \mathbb{R}^d$ is exposed to a number of $K$ different light intensities $\{I_0(\bar{r}), \ldots, I_{K-1}(\bar{r})\}$ yielding a collection $\bar{n} = \{n_0, n_1, \ldots, n_{K-1}\}$ of acquired photons per exposure (see fig. 1). Each number of photon $n_i$ follows Poissonian statistics with a mean $\lambda_i$ that depends on the corresponding light intensity $I_i(\bar{r}_m)$, where background and detector dark counts contributions are assumed to be negligible (background contributions are takin into account in eq. (S27)). When the emitter saturation is avoided, the Poissonian mean $\lambda_i$ can be approximated by

$$\lambda_i = c_e q_e \sigma_a I_i(\bar{r}_m) \tag{S1}$$

where $c_e$ is the collection efficiency of the system, $q_e$ is the quantum yield and $\sigma_a$ is the absorption cross-section of the emitter at the wavelength of the illumination $I_i$.

To facilitate the study in terms of the total acquired photons

$$N = n_0 + \cdots + n_{K-1} \tag{S2}$$

and to make the localization scheme independent of the intrinsic brightness of the molecules, the probabilities for measuring the collection of photons $P(\bar{n})$ is conditioned to $N$, yielding multinomial statistics:

$$P(n_i) \sim \text{Poisson}(\lambda_i) \text{ with } i \in [0, K-1] \Longrightarrow P(\bar{n}|N) \sim \text{Multinomial}(\bar{p}, N)$$

$$P(\bar{n}|N) = \frac{N!}{n_0! \cdots n_{K-1}!} \prod_{i=0}^{K-1} p_i^{n_i} \tag{S3}$$

with the components of the parameter vector $\bar{p}$ being

$$p_i^{(0)}(\bar{r}_m) = \frac{\lambda_i}{\sum_{j=0}^{K-1} \lambda_j} \approx \frac{I_i(\bar{r}_m)}{\sum_{j=0}^{K-1} I_j(\bar{r}_m)} \text{ with } i \in [0, \ldots, K-1] \tag{S4}$$

where, for the case of negligible dark counts (denoted by the (0) superscript), the brightness of the molecule is canceled out.

It should be stressed that, for the multinomial distribution, the number of *independent* photon acquisitions $n_i$ (and parameters $p_i$) is $K-1$, as stated in eq. (S2) and eq. (S4). Therefore, the number of independent elements $p_i$ span a $(K-1)$-dimensional space, termed *reduced $\bar{p}$-space* henceforth. $P(\bar{n}|N)$ can thus be written as:

$$P(\bar{n}|N) = \frac{N!}{n_0! \cdots n_{K-1}!} \left( \prod_{i=0}^{K-2} p_i^{n_i} \right) \left( 1 - \sum_{j=0}^{K-2} p_j \right)^{n_{K-1}} \tag{S5}$$

To quantify how much information the measured photon collection $\bar{n}$ holds on the $d$-dimensional position of the molecule $\bar{r}_m = [r_{m1} \dots r_{md}]^T$, the Fisher information matrix $F_{\bar{r}_m}$ can be calculated as:

$$F_{\bar{r}_m} = \mathcal{J}^{*T} F_{\bar{p}} \mathcal{J}^* \tag{S6}$$

where $\mathcal{J}^* \in \mathbb{R}^{(K-1) \times d}$ is the Jacobian matrix of the transformation from the $\bar{r}$-space to the *reduced $\bar{p}$-space* and $F_{\bar{p}}$ is the Fisher information on the $(K-1)$-dimensional reduced parameter vector $\bar{p}$, which follows the definition

$$F_{\bar{p}_{ij}} = E\left( -\frac{\partial^2}{\partial p_i \partial p_j} \ln \mathcal{L}(\bar{p}|\bar{n}) \middle| \bar{p} \right) \text{ with } i,j \in [0, \dots, K-2] \tag{S7}$$

Here $\mathcal{L}(\bar{p}|\bar{n}) = P(\bar{n}|\bar{p})$ is the likelihood of measuring a collection of photons $\bar{n}$ when the molecule is at the position $\bar{r}_m$, such that $\bar{p} = \bar{p}(\bar{r}_m)$. Operating with the definitions of equations (S5) and (S7) yields an expression for the Fisher information matrix for $\bar{p}$:

$$F_{\bar{p}_{ij}} = N\left( \frac{1}{p_{K-1}} + \delta_{ij} \frac{1}{p_i} \right) \text{ with } i,j \in [0, \dots, K-2] \tag{S8}$$

where $\delta_{ij}$ is the Kronecker delta function. Note that failing to take into account the interdependence of the parameters $p_i$ in (S5) (i.e., that ), results in an incorrect Fisher information matrix $F_{\bar{p}}$.

Combining eq. (S6) and eq. (S8) finally yields an expression for the Fisher information on the molecule position $\bar{r}_m$:

$$F_{\bar{r}_m} = \mathcal{J}^{*T} N \begin{bmatrix} \frac{1}{p_{K-1}} + \frac{1}{p_0} & \cdots & \frac{1}{p_{K-1}} \\ \vdots & \ddots & \vdots \\ \frac{1}{p_{K-1}} & \cdots & \frac{1}{p_{K-1}} + \frac{1}{p_{K-2}} \end{bmatrix} \mathcal{J}^*, \text{ with } \mathcal{J}^* = \begin{bmatrix} \frac{\partial p_0}{\partial r_{m1}} & \cdots & \frac{\partial p_0}{\partial r_{md}} \\ \vdots & \ddots & \vdots \\ \frac{\partial p_{K-2}}{\partial r_{m1}} & \cdots & \frac{\partial p_{K-2}}{\partial r_{md}} \end{bmatrix} \tag{S9}$$

This expression can be simplified to the form:

$$F_{\bar{r}_m} = N \sum_{i=0}^{K-1} \frac{1}{p_i} \begin{bmatrix} \left( \frac{\partial p_i}{\partial r_{m1}} \right)^2 & \cdots & \frac{\partial p_i}{\partial r_{m1}} \frac{\partial p_i}{\partial r_{md}} \\ \vdots & \ddots & \vdots \\ \frac{\partial p_i}{\partial r_{md}} \frac{\partial p_i}{\partial r_{m1}} & \cdots & \left( \frac{\partial p_i}{\partial r_{md}} \right)^2 \end{bmatrix} \tag{S10}$$

Once the Fisher information matrix is obtained, a lower bound for the covariance matrix of the molecule position $\Sigma(\bar{r}_m)$ can be derived from the Cramér-Rao inequality:

$$\Sigma(\bar{r}_m) \geq \Sigma_{CRB}(\bar{r}_m) = F_{\bar{r}_m}^{-1} \tag{S11}$$

Thus, a CRB $\Sigma_{CRB}(\bar{r}_m)$ is obtained for each possible molecule position $\bar{r}_m$. We choose to analyze the arithmetic mean $\tilde{\sigma}_{CRB}$ of the eigenvalues $\sigma_i^2$ of this matrix and their isotropy $\mathbb{I}$ as performance metrics.

$$\tilde{\sigma}_{CRB} = \sqrt{\frac{1}{d} tr(\Sigma_{CRB})} \tag{S12}$$

$$\mathbb{I} = \frac{\min_{i \in [1,d]} \sigma_i}{\max_{i \in [1,d]} \sigma_i} \tag{S13}$$

The use of $\tilde{\sigma}_{CRB}$ is equivalent to the commonly used norm (*38*) for the Fisher information matrix $\phi_q[F] = (\text{tr}(F^q)/d)^{1/q}$ with $q = -1$.

Table S1 displays explicit expressions for eq. (S4), (S8), (S12) and the *reduced* Jacobian for a set of dimensionalities $d$ and exposures $K$.

## 2 Cramér-Rao bounds for geometries of interest

The general results of the previous section will now be evaluated for a number of specific conditions of interest, namely, dimensionality, number of exposures, beam shapes and arrangement.

For the sake of simplicity in the experimental design, in this work we focus on localization schemes where the respective beam intensities $I_i(\bar{r})$ are all equal except for a displacement by a distance $\bar{r}_{b_i}$, such that

$$I_i(\bar{r}) = I(\bar{r} - \bar{r}_{b_i}) \tag{S14}$$

The beams of interest are the quadratic beam, standing wave, doughnut beam and Gaussian beam. The first one, though not physically possible, is used as a first relevant order approximation of the doughnut beam and the standing wave around its minima. The following definitions will be used:

Quadratic $\qquad I_{quad}(\bar{r}) = A_{quad} r^2 \tag{S15}$

Doughnut $\qquad I_{doughnut}(\bar{r}) = A_0 4 e \ln 2 \dfrac{r^2}{fwhm^2} e^{-4 \ln 2 \frac{r^2}{fwhm^2}} \tag{S16}$

Standing wave $\qquad I_{sw}(\bar{r}) = A_i \sin^2(\bar{k} \cdot \bar{r}) \tag{S17}$

Gaussian $\qquad I_{gauss}(\bar{r}) = A_0 \, e^{-4 \ln 2 \frac{r^2}{fwhm^2}} \tag{S18}$

where $A_{quad}$ is the concavity of the parabolic beam, $A_0$ is the peak intensity for the doughnut and Gaussian cases, $fwhm$ is a size-related parameter for the doughnut case (peak diameter occurs at $\ln(2)^{-1/2} fwhm \approx 1.2 fwhm$) and the full-width-half-maximum for the Gaussian beam, $A_i$ are the respective peak intensities of the standing waves and $\bar{k}$ determines the spacing between successive intensity minima and its direction. The quadratic approximation of the doughnut beam ($A_{quad} = A_0 4 e \ln 2 / fwhm^2$) holds as long as

$$4 \ln 2 \, r^2 \ll fwhm^2 \tag{S19}$$

### 2.1 One-dimensional localization with two exposures

In this case, the CRB of eq. (S12) is reduced to the simple expression

$$\tilde{\sigma}_{CRB}\big|_{K=2}^{d=1} = \frac{1}{\sqrt{N}} \frac{\sqrt{p_0(1-p_0)}}{\left|\frac{dp_0}{dr_0}\right|} \tag{S20}$$

where $p_0$ is the parameter of a binomial distribution. By having the beams at positions $r_{b_0} = -L/2$ and $r_{b_1} = L/2$ (separated by a distance $L$) and using the beam definitions of eq. (S15)-(S16), with $\bar{r} = x$, the following expressions are obtained for the corresponding binomial parameters and CRB:

$$p_0^{quad}(x) = \frac{1}{2}\frac{\left(1 + \frac{x}{L/2}\right)^2}{1 + \left(\frac{x}{L/2}\right)^2} \tag{S21a}$$

$$\tilde{\sigma}_{CRB}^{quad}(x) = \frac{1}{\sqrt{N}}\frac{L}{4}\left[1 + \left(\frac{x}{L/2}\right)^2\right] \tag{S21b}$$

$$\tilde{\sigma}_{CRB}^{quad}(0) = \frac{1}{\sqrt{N}}\frac{L}{4} \tag{S21c}$$

$$p_0^{doughnut}(x) = \frac{\frac{1}{2}\left(1 + \frac{x}{L/2}\right)^2 e^{-\frac{4\ln 2\, xL}{fwhm^2}}}{\left[1 + \left(\frac{x}{L/2}\right)^2\right]\cosh\left(\frac{4\ln 2\, xL}{fwhm^2}\right) - 2\frac{x}{L/2}\sinh\left(\frac{4\ln 2\, xL}{fwhm^2}\right)} \tag{S21d}$$

$$\sigma_{CRB}^{doughnut}(x) = \frac{1}{\sqrt{N}}\frac{L}{4}\frac{\left|\left[1 + \left(\frac{x}{L/2}\right)^2\right]\cosh\left(\frac{4\ln 2\, xL}{fwhm^2}\right) - 2\frac{x}{L/2}\sinh\left(\frac{4\ln 2\, xL}{fwhm^2}\right)\right|}{\left|1 + \ln 2\frac{L^2}{fwhm^2}\left[\left(\frac{x}{L/2}\right)^2 - 1\right]\right|} \tag{S21e}$$

$$\tilde{\sigma}_{CRB}^{doughnut}(0) = \frac{1}{\sqrt{N}}\frac{L}{4}\frac{1}{1 - \ln 2\frac{L^2}{fwhm^2}} \tag{S21f}$$

$$p_0^{sw}(x) = \frac{\sin^2\left[k\left(x + \frac{L}{2}\right)\right]}{\sin^2\left[k\left(x + \frac{L}{2}\right)\right] + \sin^2\left[k\left(x - \frac{L}{2}\right)\right]} \tag{S21g}$$

$$\tilde{\sigma}_{CRB}^{sw}(x) = \frac{\left|\cos\left[2k\left(x + \frac{L}{2}\right)\right] + \cos\left[2k\left(x - \frac{L}{2}\right)\right] - 2\right|}{4k\sqrt{N}\sin kL} \tag{S21h}$$

$$\tilde{\sigma}_{CRB}^{sw}(0) = \frac{1}{2k\sqrt{N}}\tan\frac{kL}{2} \tag{S21i}$$

Figure S1 shows, for each beam shape, the intensities $I_i(x)$ (A-C), the binomial success probability parameter $p(x)$ (D-F) and the CRB (G-I) for beam separations $L$ of 25 nm, 50 nm and 150 nm, a beam size parameter ($fwhm$) of 300 nm, a $k$-value of $k = 2\pi/\lambda$, with $\lambda = 600$ nm and a total number of photons $N = 100$.

The case of doughnut beams –displayed in fig. S1A,D,G– produced the intricate expressions eq. (S21d)-(S21f), all of which can be approximated by their quadratic beam counterparts eq. (S21a)-(S21c) as long as (i) the emitter lays in the central region (described by eq. (S19)) and (ii) the beam separation is smaller than the overall beam size ($L \ll fwhm/\sqrt{\ln 2}$). These conditions encompass the regime of interest, therefore the quadratic beam case displayed in fig. S1B,E,H is the one to be analyzed further.

In the quadratic approximation, the CRB scales linearly with the beam separation $L$, which makes the localization more precise the closer the beams are. There is, though, a tradeoff between precision at the origin and the field of view. Remarkable low CRB values below 5 nm are achieved with only 100 measured photons within a region of 100 nm centered around the origin.

The case of standing waves is displayed in fig. S1C,F,I. The depicted functions are based on the eq. (S21g)-(S21i), which can again be approximated by their quadratic beam counterparts (S21a)-(S21c) as long as (i) the emitter is close to an intensity zero ($\Delta r \ll \pi/k$) and (ii) the beam separation is smaller than half the sine period ($L \ll \pi/k$). In the case of $L = 150$ nm and $k = 2\pi/600$ nm$^{-1}$ these conditions are not fulfilled anymore. Note that the CRB is flat in this case.

The case of two Gaussian beams can be treated in the same way as the other cases, despite the absence of local minima. Using the same definitions as above and eq. (S18) yields

$$p_0^{gauss}(x) = \frac{e^{\frac{4\ln 2\, xL}{fwhm^2}}}{2\cosh\frac{4\ln 2\, xL}{fwhm^2}} \tag{S22a}$$

$$\tilde{\sigma}_{CRB}^{gauss}(x) = \frac{1}{\sqrt{N}}\frac{fwhm^2}{4\ln 2\, L}\cosh\frac{4\ln 2\, xL}{fwhm^2} \tag{S22b}$$

$$\tilde{\sigma}_{CRB}^{gauss}(0) = \frac{1}{\sqrt{N}}\frac{fwhm^2}{4\ln 2\, L} \tag{S22c}$$

Figure S4 shows the intensities $I_i(x)$ (A), the binomial parameter $p(x)$ (C) and the CRB (E) for beam separations $L$ of 25 nm, 50 nm, 150 nm, 300 nm, 600 nm, and 1200 nm. The performance of the Gaussian beam case is poor for small beam separations; this is visible in the slope of the binomial parameter $p(x)$ in fig. S4C and leads to the high values of the CRBs in fig. S4E, in contrast to the cases of beams with minima shown in fig. S1.

It should be noted from eq. (S22c) that the localization CRB for Gaussian beams scales as $\propto L^{-1}$ at the origin. This implies that the further apart the beams are, the more precise the emitter localization will be (as shown in fig. S4). This makes sense, as the series expansion of largely separated Gaussian beams around the $x = 0$ has significant contributions only from order 2 (as long as the field of view $\ll 3/[2\ln(2)\,L]$). Therefore, they are not distinguishable from the parabolic beams discussed previously, mimicking intensity minima.

Increasing the beam separation also produces a faster growth of the CRB away from the origin (given the $\propto xL/fwhm^2$ dependence in eq. (S22b), reducing the method's effective *field of view*. Practical applications with these beams may be highly affected by background collection and aberrations.

It is straightforward to state that a proper manipulation of the beams $I_i(\bar{r})$ enables great flexibility of the CRB which can be adapted for different applications. We have illustrated this point by using the doughnut and Gaussian beam shapes and the separation $L$ as a parameter, but the possibilities are not at all limited to these cases. Arbitrary beam shapes generated, for example, by a spatial light modulator (SLM) could be used to optimize for specific properties, e.g. a spatially flatter CRB. Optimizations of the beams according to the application specifications (dimensionality, speed, field of view, precision, etc.) and to the implementation technology will follow on further studies. In this work we focus on the many benefits on speed, precision and observation time that a collection of displaced doughnut beams can bring for single molecule localization applied for tracking and imaging.

## 2.2  Two-dimensional localization with four doughnut exposures

Localizing an emitter in a plane requires, through dimensional arguments, at least three exposures due to the constraining eq. (S2). In this work, the pattern shown in fig. 2 is extensively utilized; it is composed of four exposures with doughnut shaped excitation profiles eq. (S16): three equidistant beams with their zero on a circle of diameter $L$ and a fourth one placed at the origin. The respective displacements are chosen to be:

$$\bar{r}_{b_0} = [0,0]^T$$

$$\bar{r}_{b_i} = \frac{L}{2} \cdot [\cos(\alpha_i), \sin(\alpha_i)]^T, \quad \text{for } \alpha_i = i \cdot \frac{2\pi}{3} \text{ and } i \in [1,3]$$

(S23)

This particular choice of an excitation beam pattern (EBP) has some advantages over other more straightforward possible choices, e.g., a square four-beam pattern or a triangular three-beam pattern. Doughnut beams conveniently increase the Fisher information at the central region of the EBP, as discussed previously for the one-dimensional case. However, successive maxima and minima can lead to indeterminations in the position. This means the same parameter vector $\bar{p}$ is obtained at multiple positions of the emitter, and therefore similar measured photon collections $\bar{n}$. The central doughnut beam is included to counteract such effect.

As an example, fig. S2A shows the likelihood function $\mathcal{L}(\bar{r}_m|\bar{n}) = P(\bar{n}|N, \bar{r}_m)$ as a function of the molecule position $\bar{r}_m$ for an arbitrary combination of collected photons $\bar{n} = \{n_1, n_2, n_3\}$ from <u>only three</u> peripherical doughnut exposures. Though multiple positions exhibit local maxima for the likelihood function $\mathcal{L}$, a unique ML estimator is defined as long as one of them is absolute. However, fluctuation-driven hopping between two local maxima with similar likelihood values make such an estimator badly behaved.

The inclusion of a fourth doughnut exposure at the origin aids the localization procedure by including radial information. This is readily visible from the $\bar{p}(\bar{r})$ function visualization in fig. S7D-G (note that the shown function results from an experimentally measured doughnut). The radial information breaks apart most indeterminations, especially in the region surrounding the beam pattern origin, leading to a better behaved estimator. This is illustrated in fig. S2B where the likelihood function $\mathcal{L}$ displays a single maximum. Interestingly, the photon distribution is *exactly the same* as in the previous example, except for the collection of *zero photons* during the central doughnut exposure. This illustrates how a zero of intensity allows for the extraction of position information even without emission from the sample.

It should be stressed that the inclusion of additional exposures (or structured detection) is not the only way to avoid multiple local maxima in the likelihood function. It is to be expected that, by freeing the constraint of using identical beams, it may be possible to neutralize this inconvenience without the use of further exposures; this, however, is left for future explorations.

Using eq. (S10) and eq. (S11), an explicit expression for the two-dimensional ($d = 2$) localization CRB $\Sigma_{CRB}$ and the arithmetic mean of its eigenvalues $\tilde{\sigma}_{CRB}$ can be derived:

$$\Sigma_{CRB}(\bar{r}_m) = \frac{1}{N}\left(\left[\sum_{i=0}^{K-1}\frac{1}{p_i}\left(\frac{\partial p_i}{\partial x}\right)^2\right]\left[\sum_{i=0}^{K-1}\frac{1}{p_i}\left(\frac{\partial p_i}{\partial y}\right)^2\right] - \left[\sum_{i=0}^{K-1}\frac{1}{p_i}\frac{\partial p_i}{\partial x}\frac{\partial p_i}{\partial y}\right]^2\right)^{-1}$$

$$\times \sum_{i=0}^{K-1}\frac{1}{p_i}\begin{bmatrix}\left(\frac{\partial p_i}{\partial y}\right)^2 & -\frac{\partial p_i}{\partial x}\frac{\partial p_i}{\partial y}\\ -\frac{\partial p_i}{\partial x}\frac{\partial p_i}{\partial y} & \left(\frac{\partial p_i}{\partial x}\right)^2\end{bmatrix} \tag{S24}$$

$$\tilde{\sigma}_{CRB} = \sqrt{\frac{1}{2N}\frac{\sum_{i=0}^{K-1}\frac{1}{p_i}\left[\left(\frac{\partial p_i}{\partial y}\right)^2+\left(\frac{\partial p_i}{\partial x}\right)^2\right]}{\left[\sum_{i=0}^{K-1}\frac{1}{p_i}\left(\frac{\partial p_i}{\partial x}\right)^2\right]\left[\sum_{i=0}^{K-1}\frac{1}{p_i}\left(\frac{\partial p_i}{\partial y}\right)^2\right]-\left[\sum_{i=0}^{K-1}\frac{1}{p_i}\frac{\partial p_i}{\partial x}\frac{\partial p_i}{\partial y}\right]^2}} \tag{S25}$$

Combining these expressions with the doughnut beam definition from eq. (S16) and the beam displacements of eq. (S23), it is possible to calculate analytically a performance metric of the proposed localization scheme. The functional dependence of $\tilde{\sigma}_{CRB}$, obtained by solving eq. (S25) at the origin, is given by:

$$\tilde{\sigma}_{CRB}(\bar{r} = \bar{0}) = \frac{L}{2\sqrt{2N}}\left(1 - \frac{L^2\ln(2)}{fwhm^2}\right)^{-1} \tag{S26}$$

For a beam separation that is smaller than the overall beam size ($L \ll fwhm/\sqrt{\ln 2}$) the CRB scales linearly with the beam separation $L$. Therefore, as in the one dimensional case, we have the possibility of increasing the photon efficiency of the localization process by means of geometrical degrees of freedom in the excitation pattern.

Figure S3A visualizes this through the spatial dependence of $\tilde{\sigma}_{CRB}$ for different pattern diameters $L$ and a total number of $N = 100$ photons. Analogously to the one dimensional case (see figure S1H), the CRB is lowest around the origin and retains a low value within a diameter $L$. Figure S3B shows the covariance matrix $\Sigma_{CRB}$ as a quadratic form for a discrete grid of positions separated $9\,\text{nm}$ for $N = 1000$. Note, that the localization precision is not isotropic. Figure S3C illustrates the dependence of $\tilde{\sigma}_{CRB}$ as a function of the total number of photons $N$ for different pattern sizes $L$ at the origin.

The result in eq. (S26) is obtained in the limit of an infinitesimal small background contribution. In order to evaluate the expected performance of our method in a more realistic manner, the influence of the background – which is unavoidable in every realistic implementation of the method – on the CRB will be discussed. Most importantly, the definition of the parameter vector $\bar{p}^{(0)}$ stated in eq. (S4) has to be adapted. We assume that all relevant background contributions follow Poissonian statistics such that they can be described by a Poissonian distribution with mean $\lambda_{bi}$. The modified parameter vector $\bar{p}$ is then given by:

$$p_i(\bar{r}_m) = \frac{\lambda_i + \lambda_{bi}}{\sum_{j=0}^{K-1}(\lambda_j + \lambda_{bj})} \approx \frac{c_e q_e \sigma_a I_i(\bar{r}_m) + \lambda_{bi}}{\sum_{j=0}^{K-1}(c_e q_e \sigma_a I_j(\bar{r}_m) + \lambda_{bj})} \text{ with } i \in [0, K-1] \quad \text{(S27)}$$

The convenient cancellation of the intrinsic brightness of the molecule, as in eq. (S4), is not applicable in this situation anymore. We can define a signal-to-background ratio $SBR$ as

$$SBR(\bar{r}_m) = \frac{\sum_{j=0}^{K-1} \lambda_j}{\sum_{j=0}^{K-1} \lambda_{bj}} \approx \frac{c_e q_e \sigma_a \sum_{j=0}^{K-1} I_j(\bar{r}_m)}{\sum_{j=0}^{K-1} \lambda_{bj}} \quad \text{(S28)}$$

with the definitions as above. It should be noted that the signal-to-background ratio $SBR$ is a function of the set of exposures $\{I_i(\bar{r})\}$ and the position of the emitter $\bar{r}_m$. With this definition, $p_i(\bar{r}_m)$ can be written as

$$\begin{aligned}
p_i(\bar{r}_m) &= \frac{SBR(\bar{r}_m)}{SBR(\bar{r}_m)+1} \frac{\lambda_i}{\sum_{j=0}^{K-1} \lambda_j} + \frac{1}{SBR(\bar{r}_m)+1} \frac{\lambda_{bi}}{\sum_{j=0}^{K-1} \lambda_{bj}} \\
&\approx \frac{SBR(\bar{r}_m)}{SBR(\bar{r}_m)+1} \frac{I_i(\bar{r}_m)}{\sum_{j=0}^{K-1} I_j(\bar{r}_m)} + \frac{1}{SBR(\bar{r}_m)+1} \frac{\lambda_{bi}}{\sum_{j=0}^{K-1} \lambda_{bj}} \\
&\approx \frac{SBR(\bar{r}_m)}{SBR(\bar{r}_m)+1} p_i^{(0)}(\bar{r}_m) + \frac{1}{SBR(\bar{r}_m)+1} \frac{1}{K}
\end{aligned} \quad \text{(S29)}$$

Again, the beam definition and its displacements are taken from eq. (S16) and eq. (S23), respectively. It is assumed that the background contributions depend on the excitation power only, such that these are rendered equal for the respective beams, i.e., $\lambda_{bi} = \lambda_b$ for all $i \in [0, K-1]$. Using eq. (S25) with eq. (S29), the functional dependence of $\tilde{\sigma}_{CRB}$ at the origin is now given by:

$$\tilde{\sigma}_{CRB}(\bar{r} = \bar{0}) = \frac{L}{2\sqrt{2N}}\left(1 - \frac{L^2 \ln(2)}{fwhm^2}\right)^{-1}\sqrt{\left(1 + \frac{1}{SBR(\bar{0})}\right)\left(1 + \frac{3}{4\,SBR(\bar{0})}\right)} \qquad (S30)$$

Note, that the $SBR(\bar{0})$ as defined in (S28) has a dependence on $L$. In an experimental realization, the actual background $\lambda_b$ will normally not depend on the beam displacement. The emitter at $\bar{r} = \bar{0}$ does see a reduced total intensity when $L$ is diminished, though. It follows then that $SBR(\bar{0})$ decreases with a reduction of $L$. Assuming, that $SBR(\bar{0}, L_0)$ (the $SBR$ at the origin for a beam displacement $L_0$) is known, $SBR(\bar{0}, L)$ can be written as:

$$SBR(\bar{0}, L) = \frac{L^2}{L_0^2}\exp\left(\frac{ln(2)}{fwhm^2}(L_0^2 - L^2)\right) \cdot SBR(\bar{0}, L_0) \qquad (S31)$$

Figure S3D visualizes the influence that the $SBR$ has on $\tilde{\sigma}_{CRB}$ at the origin for different $SBR(\bar{0}, L_0 = 100\,\text{nm})$ values. The presence of background increases the value of $\tilde{\sigma}_{CRB}(\bar{0})$ and introduces a limit to the usable beam separation parameter $L$. The localization precision cannot be improved further by reducing the beam separation over that limit. Using different beam shapes, however, might enable flexibility on this aspect.

# 3 Position estimators

Section 2 investigated the Cramér-Rao bounds (CRBs) for different excitation geometries. It was shown that the information content on the position estimation can be increased significantly in a region of interest, and be far higher compared to the information content present in ideal camera localization (no amplification noise, no readout noise, see section 4). This theoretical limit provides the best possible performance that any unbiased estimator can achieve, but not how the actual estimators perform. Section 3.1 investigates the performance of the maximum likelihood estimator (MLE), for different excitation geometries. Section 3.2 investigates further estimators optimized for low photon applications.

## 3.1 Maximum likelihood estimators

In order to estimate the position $\bar{r}_m$ of an emitter, we use the maximum likelihood estimator $\hat{\bar{r}}_m^{MLE}$ defined as:

$$\hat{\bar{r}}_m^{MLE} = \arg\max \mathcal{L}(\bar{r}|\bar{n}) \tag{S32}$$

where $\mathcal{L}(\bar{r}|\bar{n}) = P(\bar{n}|N, \bar{r}_m)$ is the likelihood function dependent on the conditional probability distribution of the measured set of photons $P(\bar{n}|N, \bar{r}_m)$ defined in eq. (S3).

Three instances of MLE calculations follow in this section: (i) the one-dimensional localization with two Parabolas, (ii) the two-dimensional localization with four doughnut exposures including background as described in section 2.2, which is the main method presented in this work, and (iii) the one-dimensional localization with two Gaussian exposures, which is used in the intermediate localization step for tracking described in section 6.4.1.

### 3.1.1 MLE for 1D position with two parabolas

In the vicinity of intensity zeros, a standing wave as well as a doughnut can be approximated in first relevant order by a parabola. It is therefore enlightning to state the MLE for the position in the case of displaced parabolic excitation intensities. Following the definition in eq. (S14), two beams $I_0(x)$ and $I_1(x)$ with parabolic excitation profile (S15) are placed at positions $x_{b_0} = -L/2$ and $x_{b_1} = L/2$ and consequently seperated by a distance $L$. In this excitation geometry, the parameter vector $\bar{p}$ defined in eq. (S4) is given by:

$$p_0 = \frac{\left(1 + \frac{2x}{L}\right)^2}{2\left(1 + \frac{4x^2}{L^2}\right)}, \quad p_1 = \frac{\left(1 - \frac{2x}{L}\right)^2}{2\left(1 + \frac{4x^2}{L^2}\right)} \tag{S33}$$

and leads to the following likelihood function $\mathcal{L}(x|\bar{n})$:

$$\mathcal{L}(x|\bar{n}) = \frac{N!}{n_0! \, n_1!} \prod_{i=0}^{1} p_i^{n_i} = \frac{N!}{n_0! \, n_1!} \frac{\left(1 + \frac{2x}{L}\right)^{2n_0} \left(1 - \frac{2x}{L}\right)^{2n_1}}{\left(2 + \frac{8x^2}{L^2}\right)^N} \tag{S34}$$

The maximum likelihood estimator $\hat{x}_m^{MLE}$ is then easily calculated:

$$\frac{\mathrm{d}}{\mathrm{d}x} \mathcal{L}(\hat{x}_m^{MLE}|\bar{n}) = 0$$

$$\Rightarrow \hat{x}_{m,1}^{MLE}(n_0, N) = -\frac{L}{2} + \frac{L}{1 + \sqrt{\frac{n_1}{n_0}}} \tag{S35}$$

$$\hat{x}_{m,2}^{MLE}(n_0, N) = -\frac{L}{2} + \frac{L}{1 - \sqrt{\frac{n_1}{n_0}}}$$

Keep in mind that the MLE has two solutions, where $\hat{x}_{m,1}^{MLE}$ is the result for the region defined by $-\frac{L}{2} < x < \frac{L}{2}$ and therefore the one of interest. Additional exposures can make the MLE unique.

### 3.1.2 MLE for 2D position with four doughnut-shaped beams

The geometry employed in the two-dimensional localization scheme using four doughnut shaped excitation profiles is depicted in fig. 2 and explained in more detail in section 2.2.

As it is classicly done, we maximize the log-likelihood function and drop the multiplicative factors that only depend on $\bar{n}$. Thus, starting from eq. (S32) we obtain a simpler function $\ell(\bar{r}|\bar{n})$ to maximize

$$\ln \mathcal{L}(\bar{r}|\bar{n}) \propto \ell(\bar{r}|\bar{n}) = \sum_{i=0}^{K-1} n_i \ln p_i \tag{S36}$$

where the multinomial distribution definition from eq. (S3) was used. An intricate expression is obtained for this simplified likelihood function $\ell(\bar{r}|\bar{n})$ when the parameter vector $\bar{p}$ defined in eq. (S27) is evaluated for the four doughnut beam case defined in eq. (S16) and eq. (S23) (see fig. S7D-G for a visualization). Given the complexity of finding and analytical solution to the maximization of $\ell(\bar{r}|\bar{n})$, the problem is solved numerically.

Given a measured count quartet $\bar{n}$, the simplified likelihood function $\ell(\bar{r}|\bar{n})$ is evaluated in successive grid searches that approximate to the maximum by subsequently reducing the grid spacing.

The performance of the MLE compared to the CRB is shown in fig. S9A. Different separations $L$ and multiple molecule positions on the $x$ axis, given a signal-to-background ratio of $SBR = 10$, are depicted. At the origin, the performance of the estimator converges to the CRB for about 100 photons for all shown beam separations $L$. The further away the molecule is, the more photons are needed for the MLE to achieve the CRB. For $x = 50$ nm, $y = 0$ nm and $L = 75$ nm for example, the MLE converges only starting from about $N = 500$ photons.

### 3.1.3   MLE for 1D position with two Gaussian-shaped beams

The maximum likelihood estimator for the case of two Gaussian shaped excitation profiles (S18) can be calculated follwing the same procedure used in section 3.1.1. The parameter vector $\bar{p}$ is:

$$p_0 = \frac{e^{\frac{4\ln 2\, xL}{fwhm^2}}}{2\cosh\dfrac{4\ln 2\, xL}{fwhm^2}}, \quad p_1 = \frac{e^{-\frac{4\ln 2\, xL}{fwhm^2}}}{2\cosh\dfrac{4\ln 2\, xL}{fwhm^2}} \tag{S37}$$

and leads to the following likelihood function $\mathcal{L}(x|\bar{n})$:

$$\mathcal{L}(x|\bar{n}) = \frac{N!}{n_0!\, n_1!}\prod_{i=0}^{1} p_i^{n_i} = \frac{N!}{n_0!\,(N-n_0)!}\frac{e^{4\ln 2\frac{Lx(2n_0-N)}{fwhm^2}}}{\cosh\left(\dfrac{4\ln 2\, xL}{fwhm^2}\right)^N} \tag{S38}$$

Differentiation of this likelihood function yields the maximum likelihood position estimator $\hat{x}_m^{MLE}$:

$$\frac{d}{dx}\mathcal{L}(\hat{x}_m^{MLE}|\bar{n}) = 0$$

$$\Rightarrow \hat{x}_m^{MLE}(n_0, N) = \frac{fwhm^2}{8\ln(2)\, L}\left[\ln(n_0) - \ln(N-n_0)\right] \tag{S39}$$

$$\hat{x}_m^{MLE}(n_0, N) = \frac{fwhm^2}{8\ln(2)\, L}\ln\left(\frac{n_0}{n_1}\right)$$

Thus, the MLE takes a surprisingly simple form, which is easily implemented as a live position estimation, i.e. in the FPGA board. From eq. (S39), it stands out that a position estimation makes sense only if at least one photon is collected from each exposure. In the case of $n_i = 0$, the emitter would be located at infinity. In the presented tracking application, these photon combinations are neglected by defining $\hat{x}_m^{MLE}(n_i = 0|N) \triangleq 0$.

It is of particular interest not only to compute the ML-estimate, but also the localization bias and precision for different emitter positions $x_m$. They can be calculated as:

$$\text{bias}_{MLE}(x_m) = \langle \hat{x}_m^{MLE} \rangle - x_m \tag{S40}$$

$$\tilde{\sigma}_{MLE}^2(x_m) = \sum_{n_0=1}^{N-1} (\hat{x}_m^{MLE} - \langle \hat{x}_m^{MLE} \rangle)^2 \; P(n_0|N) \tag{S41}$$

$$\text{with } \langle \hat{x}_m^{MLE} \rangle = \sum_{n_0=1}^{N-1} \hat{x}_m^{MLE} \, P(n_0|N)$$

Both quantities are visualized in fig. S4B,D for different photon numbers $N$. It is evident that, for a given $L$ and $fwhm$ combination, a minimal $N$ is needed in order to have an acceptable bias in a given field of view around the origin. The standard deviation $\tilde{\sigma}_{MLE}(x_m)$ asymptotically approaches the CRB $\tilde{\sigma}_{CRB}(x_m)$ with growing $N$ in that region. Consequently, the localization precision reproduces (i) the $\propto L^{-1}$ dependence of eq. (S22b) and (ii) the faster divergence of the $\tilde{\sigma}_{CRB}(x_m)$ with increasing beam separation $L$. However, the standard deviation $\tilde{\sigma}_{MLE}(x_m)$ falls below the CRB at positions where the bias is not negligible. This is acceptable, as the CRB is a lower bound for unbiased estimators only.

## 3.2   Other estimators

In addition to the MLE, we investigated further position estimators. The single emitter tracking application, which employs the excitation beam pattern (EBP) introduced in section 2.2, requires an estimator that is suitable for low photon numbers, as well as for live position estimation in the FPGA board. In particular, the position estimation needs to be fast enough such that the molecule does not diffuse out of the high sensitivity region surrounding the EBP origin

We start this section presenting the solution of the least-mean-squared (LMS) estimator, for a linearization of the problem around the origin. We then present a modified LMS estimator (mLMSE) which has a better performance, but keeps the computational simplicity. Finally, we present the position estimator used in post processing, which is a numerically unbiased version of the mLMSE.

| | |
|---|---|
| $\bar{r}$ | Molecule position |
| $\bar{p}(\bar{r})$ | Multinomial parameter |
| $\bar{n}$ | Collected counts |
| $\hat{\bar{p}}(\bar{n})$ | Estimator of the multinomial parameter |
| $\hat{\bar{r}}(\hat{\bar{p}}(\bar{n}))$ | Estimator of the molecule position |
| $\bar{R}(\bar{r}) = E(\hat{\bar{r}})$ | Expectation of the position estimator |

### 3.2.1 Linearized least mean squared (LMS) estimator

The vicinity of the EBP origin is the area with the lowest CRB values and also the place in which the molecule is kept in the tracking application. Therefore, it makes sense to construct an estimator that is especially suited for that region. A first order approximation of the parameter vector $\bar{p}$ (see eq. (S4)) at the beam pattern origin is used:

$$p_i(\bar{r})|_{r=0} \cong p_i(\bar{0}) + \sum_{j=1}^{d} r_j \frac{\partial p_i}{\partial r_j} \qquad (S42)$$

where $d$ depicts the dimensionality of the $\bar{r}$-space and $\bar{0}$ is its null vector.

Given a set of measured counts $\bar{n}$, the position estimate of the molecule $\hat{\bar{r}}$ can be obtained by making use of the invariance property of the MLE:

$$\bar{p}(\hat{\bar{r}}) = \hat{\bar{p}} \qquad (S43)$$

where $\bar{p}$ is the parameter vector defined in eq. (S42) and $\hat{\bar{p}}$ is the MLE, which, in the case of a multinomial distribution, is given by:

$$\hat{p}_i = \frac{n_i}{\sum_j n_j} = \frac{n_i}{N} \qquad (S44)$$

In other words, the estimated $\hat{\bar{p}}$ values are mapped into the position space $\bar{r}$ through the function $\bar{p}(\bar{r})$. A visualization of this concept for a one dimensional example can be found in fig. 1 of the main text.

Inserting eq. (S42) into eq. (S43) yields:

$$\mathcal{J}\hat{\bar{r}} = \hat{\bar{p}} - \bar{p}(\bar{0}) \qquad (S45)$$

with $\mathcal{J} = \bar{\nabla}\bar{p}$ the Jacobian matrix of the transformation from the $\bar{r}$-space to the $\bar{p}$-space (this is not $\mathcal{J}^*$, the Jacobian of the $\bar{r}$-space to *reduced* $\bar{p}$-space transformation, introduced in eq. (S6)).

The solution of the overdetermined linear system in eq. (S45) is obtained by a least mean square projection. Consequently, the following cost function $S$ needs to be minimized:

$$S = \|\hat{\bar{p}} - \bar{p}(\bar{0}) - \mathcal{J}\hat{\bar{r}}\|^2 \qquad (S46)$$

The solution is given by:

$$\hat{\bar{r}}_{LMS}(\hat{\bar{p}}) = (\mathcal{J}^T \mathcal{J})^{-1} \mathcal{J}^T (\hat{\bar{p}} - \bar{p}(\bar{0})) \qquad (S47)$$

For the two dimensional localization with four doughnut shaped excitation profiles introduced in section 2.2, we obtain:

$$\hat{\bar{r}}_{LMS}(\hat{p}) = \frac{L}{2} \cdot \frac{1}{1 - \frac{L^2 \log(2)}{fwhm^2}} \begin{bmatrix} -\hat{p}_3 + \frac{1}{2}(\hat{p}_1 + \hat{p}_2) \\ \frac{\sqrt{3}}{2}(\hat{p}_2 - \hat{p}_1) \end{bmatrix} \tag{S48}$$

Comparing this solution with the beam positions $\bar{r}_{b_i}$ defined in eq. (S23) we can rewrite eq. (S48) into:

$$\hat{\bar{r}}_{LMS}(\hat{p}) = -\frac{1}{1 - \frac{L^2 \log(2)}{fwhm^2}} \sum_{i=1}^{3} \hat{p}_i \, \bar{r}_{b_i} \tag{S49}$$

This is because the gradient of each function $p_i(\bar{r})$ points in the direction of the corresponding beam position $\bar{r}_{b_i}$.

### 3.2.2 Modified least mean squared (mLMS) estimator

The LMS solution in eq. (S49) takes a simple form that could readily be implemented in the FPGA board for live position estimation. Unfortunately, it does not make use of the photons $n_0$ collected from the first doughnut exposure, as the function $p_0(\bar{r})$ has no linear term in its polynomial expansion at the origin. These photons do not hold directional information but they do hold radial information. Especially in the vicinity of the EBP origin, an increase of the $\hat{p}_0$ value indicates an increase of the radial coordinate. To include this property in the position estimator, eq. (S49) is expanded in orders of $\hat{p}_0$ with parameters $\beta_j$:

$$\hat{\bar{r}}_{mLMS}^{(k)}(\hat{p}, \bar{\beta}) = -\frac{1}{1 - \frac{L^2 \log(2)}{fwhm^2}} \left( \sum_{j=0}^{k} \beta_j \hat{p}_0^j \right) \sum_{i=1}^{3} \hat{p}_i \cdot \bar{r}_{b_i} \tag{S50}$$

Note that an expansion in arbitrary orders $\hat{p}_0^k \hat{p}_1^l \hat{p}_2^m \hat{p}_3^n$ with the respective directional vectors could be conducted. This was not studied further in this work, though. In the case of the live position estimator in tracking, the estimator $\hat{\bar{r}}_{mLMS}^{(k=1)}$ was used.

### 3.2.3 Numerically unbiased mLMS (numLMS) estimator

Though the mLMSE is fast to calculate, it has the drawback of being biased. This is especially true for experimental beam shapes, that might deviate slightly from its ideal counterparts for which the mLMSE was calculated. Consequently, the measured trajectories in the tracking application, that rely on the mLMSE, have to be corrected in post processing. Unfortunately, the MLE introduced in section 3.1.2 does not converge to the CRB in the photon range employed in tracking, where in average $\langle N \rangle = 9$ photons were used per localization. Even for the $\langle N \rangle = 150$ photons per localization employed in the short range tracking application, the MLE does not achieve the CRB.

In comparison to the MLE, the mLMSE has the advantage that performance metrics like the covariance and the bias can easily be calculated analytically. The knowledge of the bias enables the possibility to unbias the estimator. We first present the definition and derivation of the analytical bias of the mLMSE, followed by the routine that finds the optimal numerically unbiased mLMSE (numLMSE).

*Bias calculation*

The bias of the mLMSE is given by:

$$\text{bias}_{mLMS}^{(k)}(\bar{r}, \bar{\beta}) = E\left(\hat{\tilde{r}}_{mLMS}^{(k)}(\hat{p}, \bar{\beta})\right)_{(\bar{r})} - \bar{r} \tag{S51}$$

where the explicit dependence on the molecule position $\bar{r}$ is written as a subscript. Though $\hat{\tilde{r}}_{mLMS}^{(k)}$ is a random variable that depends on the *individual realization* of the collected photon counts $\bar{n}$ (eq. (S50)) and the expansion parameter $\bar{\beta}$, its expectation depends on the actual molecule position.

The expectation of the position estimator $\bar{R}_{mLMS}^{(k)} \triangleq E\left(\hat{\tilde{r}}_{mLMS}^{(k)}\right)_{(\bar{r})}$ is

$$\bar{R}_{mLMS}^{(k)}(\bar{r}, \bar{\beta}) = -\frac{1}{1 - \dfrac{L^2 \log(2)}{fwhm^2}} \sum_{j=0}^{k} \sum_{i=1}^{3} \beta_j E(\hat{p}_0^j \hat{p}_i) \cdot \bar{r}_{b_i} \tag{S52}$$

A closed expression can be obtained by making use of the identity $E(\hat{p}_0^j \hat{p}_i) = E(n_0^j n_i)/N^{j+1}$ and the expression taken from (*39*) for the generalized factorial moments of the multinomial distribution. The mean localization $\bar{R}_{mLMS}^{(k=2)}$, for a molecule at position $\bar{r}$ in the case that $N$ photons where collected, is given by:

$$\bar{R}_{mLMS}^{(k=2)}(\bar{r}, \bar{\beta}) = E\left(\hat{\tilde{r}}_{mLMS}^{(k=2)}\right)_{(\bar{r})} = \frac{-1}{1 - \dfrac{L^2 \log(2)}{fwhm^2}}$$

$$\cdot \sum_{i=1}^{3} \left(\beta_0 p_i(\bar{r}) + \frac{(N-1)}{N}\left[\beta_1 + \frac{\beta_2}{N}\right] p_0(\bar{r}) p_i(\bar{r}) + \frac{\beta_2(N-1)(N-2)}{N^2} p_0^2(\bar{r}) p_i(\bar{r})\right) \bar{r}_{b_i} \tag{S53}$$

The order $k$ in eq. (S50) was found to increase the area (around the origin) in which unbiasing is possible. This is related to the function $\bar{R}_{mLMS}^{(k)}(\bar{r}, \bar{\beta})$ being injective up to a maximal radius $|\bar{r}|$ only. The size of this radius depends on the choice of $\bar{\beta}$ and, in particular, of $k$. In this work, we chose $k = 2$ as a compromise, which enables to unbias the mLMSE in a radius of about $L/2$ surrounding the origin.

*Unbiasing*

In order to unbias the estimator $\hat{\bar{r}}_{mLMS}^{(k=2)}$, the following numerical optimization routine is employed:

1. Choose a value $\mathcal{R}_2 \in \mathbb{R}$, a set of positions $\bar{r} \in \mathbb{R}^2$ and a set of query points $\bar{r}_q \in \mathbb{R}^2$ with $|\bar{r}_q| \leq \mathcal{R}_2$. Minimize the loss function $\mathbb{L}(\mathcal{R}_1)$, with $\mathcal{R}_1 \in \mathbb{R}$, $\mathcal{R}_1 \leq \mathcal{R}_2$.

The loss function $\mathbb{L}(\mathcal{R}_1)$ is calculated as follows:

a. <u>Obtain optimal mLMSE in $\mathcal{R}_1$</u>.

Optimize $\bar{\beta}$ such that $\langle \text{bias}_{mLMS}^{(k=2)}(\bar{r},\bar{\beta}) \rangle$ is minimized, where the average is taken over all positions $\bar{r}$, with $|\bar{r}| \leq \mathcal{R}_1$. Let the optimal $\bar{\beta}$-vector be $\bar{\beta}_{opt}$.

b. <u>Unbias the optimal mLMSE</u> by generating an interpolant function $\boldsymbol{\mathcal{F}}_{\mathcal{R}_1}$ such that:

$$\boldsymbol{\mathcal{F}}_{\mathcal{R}_1} \colon \mathbb{R}^2 \to \mathbb{R}^2$$
$$\boldsymbol{\mathcal{F}}_{\mathcal{R}_1}\left(\bar{R}_{mLMS}^{(k=2)}(\bar{r},\bar{\beta}_{opt})\right) = \bar{r}, \quad \forall \bar{r} \colon |\bar{r}| \leq \mathcal{R}_1 \tag{S54}$$

c. <u>Test the generated unbiased estimator</u>.

   i. For each query point $\bar{r}_q$, generate a set $\hat{P}_{\bar{r}_q} = \{\hat{\bar{p}}_0, \ldots, \hat{\bar{p}}_M\}_{\bar{r}_q}$ of $M$ $\bar{p}$-parameter estimates, and calculate the mLMSE $\hat{\bar{r}}_{mLMS}^{(k=2)}(\hat{\bar{p}},\bar{\beta}_{opt})$ for all $\hat{\bar{p}}$-vectors.

   ii. Calculate the mean square error (MSE):

   $$MSE_{\hat{P}_{\bar{r}_q}} = \frac{1}{M}\sum_{i=1}^{M}\left(\boldsymbol{\mathcal{F}}_{\mathcal{R}_1}\left(\hat{\bar{r}}_{mLMS}^{(k=2)}(\hat{\bar{p}}_i,\bar{\beta}_{opt})\right) - \bar{r}_q\right)^2 \tag{S55}$$
   with $\hat{\bar{p}}_i \in \hat{P}_{\bar{r}_q}$

   iii. Calculate the loss function $\mathbb{L}(\mathcal{R}_1)$:

   $$\mathbb{L}(\mathcal{R}_1) = \left\langle \left| \sqrt{MSE_{\hat{P}_{\bar{r}_q}}} - CRB(\bar{r}_q) \right| \right\rangle \tag{S56}$$

   where the average is taken over the set of query points $\bar{r}_q$.

2. Let the region $\mathcal{R}_1$ minimizing the loss function $\mathbb{L}$ be $\mathcal{R}_{opt}$. The numerically unbiased estimator is then given by:

   $$r_{numLMS}^{(k=2)}(\hat{\bar{p}}) = \boldsymbol{\mathcal{F}}_{\mathcal{R}_{opt}}\left(\hat{\bar{r}}_{mLMS}^{(k=2)}(\hat{\bar{p}},\bar{\beta}_{opt})\right) \tag{S57}$$

In this work we chose $\mathcal{R}_2 = \frac{L}{2}$. The set of positions $\bar{r}$ was a rectangular grid with spacing of 1nm. The query points were equally spaced positions on the x and the y axis with $x, y \leq L/2$, respectively. To calculate $\bar{\beta}_{opt}$ the optimization function "fmincon" of Matlab was used, where $\beta_1$ was set to $\beta_1 = 0$ per default. As interpolant $\mathcal{F}$, the "scatteredInterpolant" function from Matlab was employed. The set size $M$ was chosen to be $M = 10^4$. The loss function was optimized using a grid search. Note that the mean localization $\bar{R}_{mLMS}^{(k=2)}(\bar{r}, \bar{\beta})$ depends on $N$ and on the vector parameter $\bar{p}$ (eq. (S52)), and that $\bar{p}$ depends on the SBR (eq. (S29)). To reduce the computation time of the numLMS calculation, the parameters $\mathcal{R}_{opt}$ and $\bar{\beta}_{opt}$ were calculated only for some N and SBR values. Estimates of the optimal parameter for other N and SBR combinations were then inferred by 2D interpolation.

The localization performance of the MLE, compared to the numLMSE, is depicted in fig. S9. It can be seen that the MLE converges to the CRB for $N \gtrsim 100 - 500$ photons only, depending on the emitter position $\bar{r}$ and on $L$. For smaller $N$, the MLE deviates considerably from its information theory limit. In comparison, the divergence of the numLMSE from the CRB is strongly reduced. This is especially true for low photon numbers $N$. Note that the pictured numLMSE was designed to work in a region $|\bar{r}| \lesssim L/2$. The dependence of the localization performance as a function of the SBR is not explicitly shown, but it was found to performs well for the evaluated range of $SBR = 2 - 60$.

# 4 Camera localization

Camera localization is a widely employed method to image or track molecules labeled with a fluorescent emitter. In this work, the localization performance of MINFLUX is frequently compared to the information theory limit and the CRB or MLE performance of camera localization. In this section, both quantities are calculated for an ideal camera, neglecting influences of gain, readout noise and motion blurring.

## 4.1 Static localization precision

When a static single point-like emitter is imaged with a camera, the collected photons generate an intensity distribution on the camera that is described by the point spread function (PSF). The position of the emitter can be estimated with substantially better precision than the spread of the PSF.

For characterization of the achievable localization precision, we calculate the Fisher information and Cramér-Rao bound on the localization precision for a perfect camera. Currently available cameras have additional sources of noise (e.g. read noise, EM excess noise) which make the localization precision worse. Taking all these noise sources into account complicates the model and become less relevant as camera technologies are improved. The CRB for camera localization is derived in the same way as presented in section 2, with the parameter vector $\bar{p}$ being the photon detection probabilities in each pixel of the camera (or rather a small camera region [e.g. $9 \times 9$ pixels] around the emitter, where the fluorescence photons will be concentrated). The entries $p_i$ of the parameter vector $\bar{p}$ are given by the probability of detecting a photon in pixel $i$ of the camera. Approximating the PSF by a symmetric Gaussian function with the width $\sigma_{PSF}$ yields

$$I_{PSFgauss}(x,y) = A_{gauss} e^{-\frac{1}{2}\left(\frac{x-x_m}{\sigma_{PSF}}\right)^2} e^{-\frac{1}{2}\left(\frac{y-y_m}{\sigma_{PSF}}\right)^2} \tag{S58}$$

with the emitter at position $(x_m, y_m)$. Assuming a homogeneous background, $p_i$ can be written as

$$p_i(x_m, y_m) = \frac{1}{\frac{SBR_c}{K}+1} \cdot \frac{1}{K} + \frac{1}{1+\frac{K}{SBR_c}} \cdot \frac{1}{4\sqrt{K}}$$
$$\cdot \left(\text{erf}\left(\frac{x_i + a/2 - x_m}{\sqrt{2}\,\sigma_{PSF}}\right) - \text{erf}\left(\frac{x_i - a/2 - x_m}{\sqrt{2}\,\sigma_{PSF}}\right)\right) \tag{S59}$$
$$\cdot \left(\text{erf}\left(\frac{y_i + a/2 - y_m}{\sqrt{2}\,\sigma_{PSF}}\right) - \text{erf}\left(\frac{y_i - a/2 - y_m}{\sqrt{2}\,\sigma_{PSF}}\right)\right)$$

with the pixel center coordinate $(x_i, y_i)$ of pixel $i$, the emitter position $(x_m, y_m)$, the pixel size $a$, the number of pixels $K$ and the camera signal-to-background ratio $SBR_c$. The $SBR_c$ is defined as

$$SBR_c = \frac{\lambda_{signal}}{\lambda_{background}/K} \tag{S60}$$

where $\lambda_{signal}$ is the average total fluorescence photons, $\lambda_{background}$ is the mean total background counts and $K$ is the number of pixels of the corresponding camera region. The Fisher information and CRB can be calculated using eq. (S10) and eq. (S12), respectively. CRBs with relevant background levels are shown in fig. S6A.

The CRB gives a lower limit on the variance of any unbiased estimator but it does not describe the actual performance of an estimator. Therefore, we investigated the performance of the MLE for single emitter position estimation with a camera. The MLE is known to be asymptotically unbiased and efficient (i.e. it converges to the CRB). The likelihood function is given by

$$\mathcal{L}(x|\bar{n}) = \frac{N!}{\prod_{j=0}^{K-1} n_j!} \prod_{i=0}^{K-1} p_i^{n_i} \tag{S61}$$

where $p_i$ is probability of detecting a photon in pixel $i$ as given by eq. (S59) evaluated at $(x, y)$ and $\{n_i\}$ is the set of collected photons in the pixels of the camera and the total number of detected photons $N = \sum_{i=0}^{K-1} n_i$. The performance of the MLE on simulated images is shown in fig. S6A. Deviations of the MLE from the CRB are visible for low $N$ and $SBR_c$. This is in agreement with the asymptotic convergence of the MLE to the CRB. A representative collection of simulated images for different $SBR_c$ and $N$ is shown in fig. S6B-F. It should be stressed that the MLE requires precise knowledge of the PSF shape and noise of the system (especially the camera) (*40*).

In this work, the following parameters were used for calculation of the localization performance using a camera: width of the detection PSF $\sigma_{PSF}$: ; 100 nm for static localization of a single emitter (fig. 3) and imaging (fig. 4), and 87 nm for tracking (fig. 5); pixel size of the camera $a$: 100 nm; camera ROI around emitter: $9 \times 9$ pixels, i.e. $K = 81$; signal-to-background ratio $SBR_c$: 50 or 500 for static localization of a single emitter (fig. 3), 500 for imaging (fig. 4), and 20 or 40 for tracking (fig. 5).

# 5 Diffusion coefficient estimation

Different algorithms to estimate the apparent diffusion coefficient $D$, as well as the localization precision $\sigma$, from single-particle trajectories can be found in the literature. This section starts by giving a brief overview and motivates the choice of an optimized least-square fit (OLSF) diffusion estimation (*32*) for this work.

In the second part, it is evaluated whether the OLSF is applicable to the trajectories obtained using MINFLUX by using computer simulations. It is found that the OLSF from (*32*) needs to be slightly adapted. Doing so enables $D$ estimation with a performance close to its information-based theoretical limit. Furthermore, the OLSF extracts the tracking error $\epsilon$ from the measured trajectory.

## 5.1 D estimation: algorithm overview

A maximum likelihood estimation (MLE) approach for the diffusion, based on the statistics of the observed molecule displacements, was developed in (*41*). In this approach, the calculation of the exact value of the likelihood function can be very time consuming, especially for large sample sizes $S$. An approximation based on the discrete Fourier transform was employed, rendering the MLE computationally more efficient. In the case that molecular blinking is present, some time-lapse displacements might be missing in the particle trajectory. In that situation, the MLE can still be used, as shown in (*42*). Unfortunately, non-equally spaced time-lapse displacements do not allow the use of the mentioned approximation. This renders the MLE time consuming, especially for long trajectories. In this work, blinking as well as long trajectories are present, such that the MLE was not employed.

Nevertheless, the likelihood function from (*41*) enables the computation of the CRB for the D and $\sigma^2$ parameter estimation. For $S \gg 1$, the theoretical limit on the relative standard deviation $\sigma_D / D$ is given by:

$$\frac{\sigma_D}{D} \geq \sqrt{\frac{2}{d(S-1)}} \cdot \left[1 + 2\sqrt{1+2X}\right]^{\frac{1}{2}} \tag{S62}$$

where $d$ is the dimensionality of the trajectory space. Here $X = \sigma^2 / D\Delta t - 2R$ depicts the *reduced square localization error*, where $R$ is the motion blur coefficient and $\Delta t$ is the time interval between trajectory points.

Vestergaard et. al (*43*) proposed a regression-free covariance-based estimator (CVE) that is computationally very fast. It converges to the CRB of eq. (S62) for small reduced square localization errors $X \lesssim 1$. For $X \gtrsim 1$, however, the CVE departs considerably from the CRB. Given that this work employs time intervals of $\Delta t = 125\mu s$, the reduced square localization error exhibits high values $X \gg 1$ (for typical D and $\sigma^2$ values), such that the CVE is not well-suited.

Michalet and Berglund (*32*) developed an algorithm for an optimized least-square fit (OLSF) to the mean square displacement (MSD) curve. In the case of free isotropic Brownian Motion in $d$ dimensions, the MSD is given by:

$$\rho_s = 2d(\sigma^2 - 2RD\Delta t) + 2dDs\Delta t \qquad (S63)$$

where $R$ is a parameter related to the blurring effect produced by a non-instantaneous illumination. The experimental MSD curve is fitted to this model, using time lags from $s = 0$ to a given value $s$. There is an optimal number points $s = s_{opt}$ to fit, in order to extract the diffusion coefficient estimate $\widehat{D}$ with maximum precision. The value of $s_{opt}$ for $D$ estimation depends solely on the *reduced square localization error $X$* and on the trace length $S$. It was shown that, if $s_{opt}(X, S)$ is employed, the OLSF performance approaches optimality for a wide parameter range. This, in addition to its computational simplicity, makes the OLSF a suitable candidate for extracting diffusion coefficients from the measured trajectories.

## 5.2 Applicability and performance of the OLSF estimator

In the presented single emitter tracking application, the emitter is blinking and large $X$ values are present. Furthermore, the localization precision has a dependence on the relative position of the emitter to the beam pattern origin. This leads to non-uniform localization precisions for different trajectory points (see section 2.2 and fig. S3). Therefore, the prerequisites that allow the utilization of the $s_{opt}(X, S)$ relationship developed in (*32*) are not met. To evaluate whether the OLSF is still applicable, we resorted to an *in silico* evaluation.

### *Simulation*

Two dimensional trajectories with 1200 localizations were generated for free isotropic Brownian motion with diffusion constants in the range of $D \in [0.01,1]\mu m^2/s$ and count rates of $\Gamma \in [20, 200]kHz$. 100 trajectories were generated for each parameter combination. Blinking on,- and off-times were assumed to be exponentially distributed with parameters $t_{on} = 2$ ms and $t_{off} = 0.6$ ms, which are in agreement with what was found experimentally. The time sampling was set to $\Delta t = 125$ µs, where a 12-fold subsampling was used to account for motion blurring during illumination. The excitation geometry was set to that described in section 2.2, with a separation of $L = 130$ nm. The $fwhm$ of the excitation PSF (eq. (S16)) was set to $fwhm = 450$ nm. The respective beam pattern exposures were applied subsequently, with illumination times of 31.25 µs, respectively. Poissonian photon count realizations were drawn in every subsampling interval. Emitter localizations employed the live position estimator $\hat{\overline{r}}_{mLMS}^{(k=1)}$ with $\beta_0 = 1.27$ and $\beta_1 = 38$. The chosen simulation parameters covered the range of the employed values in the single emitter tracking measurement.

In the case of blinking, a MSD curve can still be defined as

$$\bar{\rho}_s = \frac{1}{\sum_{i=1}^{S-s} q_i q_{i+s}} \sum_{i=1}^{S-s} (\bar{r}_{i+s} - \bar{r}_i)^2 \, q_i q_{i+s} \qquad (S64)$$

with

$$q_i = \begin{cases} 1 & \text{emitter on at sample } i \\ 0 & \text{emitter off at sample } i \end{cases} \qquad (S65)$$

The raw trajectory $\hat{\bar{r}}$ of the emitter is estimated with the live position estimator $\hat{r}_{mLMS}^{(k=1)}$ defined in eq. (S50). This estimator is biased and works well in a region surrounding the beam pattern origin, as explained in section 3.2.3. This causes the MSD curves to deviate from the expected linear relationship of eq. (S63). Therefore, the raw trajectories are corrected in post processing using the numLMS estimator (see section 3.2.3).

### _MSD curves_

Figure S10A visualizes both, the MSD for a raw trajectory as well as the MSD for the corrected trace. It can be seen that the linear behavior of eq. (S63) is retrieved after the correction. At the same time, the offset of the corrected trace is higher, which suggests that the localization error $\sigma$ is increased after the numLMS correction. This is counterintuitive at first glance. The reason is related to the finite working region of the mLMSE. Outside of it, position estimations get considerably biased, especially radially, which reduces the spread of localizations.

### _Optimal fit length_

For each numLMS-corrected trajectory, the MSD curve (eq. (S64)) was fitted to eq. (S63) for different lengths $s \in [2,1200]$ (logarithmically spaced). Subsequently, the s values minimizing $\sigma_D/D$ were identified and found to follow the relationship:

$$s_{opt}(X) = 2 + 4.6X^{0.62} \qquad (S66)$$

The identified $s_{opt}$ values as well as eq. (S66) are visualized in fig. S10C.

Given that the $X$ value of a trajectory is not known a priori, the recursive algorithm proposed in (_44_) is used:

Start by setting $s_0 = S/10$ and extract $\hat{\sigma}_1$ and $\widehat{D}_1$ from eq. (S63). These estimates allow the calculation of a reduced square localization error estimate $\hat{X}_1$. Using the latter, a new length $s_1$ can be obtained from eq. (S66). Repeating this procedure results in a fast convergence to the optimal length value. Non-convergence is an exceptional case, and can be spotted by reoccurrences of $s$-values (i.e. optimization is trapped within the same two values). $\qquad$ (S67)

This algorithm was applied to trajectories with lengths $S$ of 100, 1200 and 5000 localizations, respectively. Note that in case that $s_i > S$, the number of points used in the MSD fit is set to $s_i = S$. The parameter choices (i.e. $D, \Gamma, ...$) were equal to the ones stated before in "Simulation". After initialization of the recursive algorithm (using $s_0 = S/10$), a total of 10 iterations were conducted. Figure S10B visualizes the median and the standard deviation of the values $s(\hat{X}_{11})$ for the three respective trace lengths. They are plotted against the estimation $\hat{X}_{12}$ to visualize the convergence (i.e., all the points are next to the optimal curve of eq. (S66)).

### Diffusion coefficients

The apparent diffusion coefficients were extracted using the last fit of the recursion. Subsequently, the variance $\sigma_D^2$ was calculated from the 100 repetitions of each parameter combination. The resulting values of $\sigma_D/D$ are visualized in fig. S10D. Though close to the CRB, the estimator is not optimal and could be improved. It deviates from its theoretical limit by a factor of about 1.4 (eq. (S62)). It should be noted that this CRB corresponds to the non-blinking case and should only be taken as a rough guide. The derivation of the likelihood function (and CRB) for the parameters of a trajectory of a blinking emitter surpasses the scope of this work.

### Estimation of the localization precision

Equation (S63) does not only permit to extract the diffusion coefficient, but also the localization precision $\sigma$. Figure S10E shows that the estimated localization precision $\hat{\sigma}$, obtained from eq. (S63), coincides very well with the ground truth tracking error $\epsilon$ (see section 6.4.5). Note that in (32) it is shown that the optimal trajectory length $s_{opt}$ for diffusion coefficient $D$ estimation differs from the one for localization precision $\sigma$ estimation. Nevertheless, the $[S, N]$-parameter region spanned in this work is well suited for precise $\sigma$ estimation.

In conclusion, the OLSF is well suited to extract the diffusion coefficient $D$ of trajectories with (i) blinking, (ii) non-uniform localization precisions and (iii) large values of the *reduced square localization error X* for the trajectory lengths that are present in this work.

# 6   Materials and methods

## 6.1   PSF and parameter vector $\bar{p}$

The PSF of the employed optical system was measured by scanning the excitation beam over a small fluorescent microsphere and detecting the emitted fluorescent photons. The sample consisted of immobilized fluorescent microspheres and immobilized Au nanorods to allow for a live drift measurement and compensation. Live drift measurement and compensation, by repositioning the piezo stage, was done with a custom LabView program, as described in section 6.5.2.

The measurement procedure is as follows:

- A fluorescent microsphere is placed at the center of the beam scanners by moving the manual and piezo stage.
- The stage lock system is set to lock the position of a nearby Au nanorod in three dimensions.
- 100-500 frames are recorded, in order to allow for additional drift correction in post-processing, and also compensation of fluorescent brightness changes of the microsphere during the measurement (i.e. blinking and photobleaching).

Scanning is performed with the non-descanned electro-optical beam scanners. Typically, a pixel size of 1 nm and a dwell time of 40 µs are used. The laser power is set to about 1 µW in the back focal plane of the objective.

### 6.1.1   Analysis

A functional expression for the relevant area of the experimental PSF is obtained from a fit of a model function to the centered frames.

*Centering*. Centering (and correction for residual sample drift) is performed by finding the center of the excitation beam in each frame, and shifting the frames with respect to each other, such that their centers overlap.

Second order 2D polynomial functions are fitted to the individual frames. The positions of the minimum of the fitted functions serve as estimates for the centers, as the doughnut beams in use have zeros of intensity at their center. The obtained center positions are smoothed in time (by using a moving average filter of typically 10 frames) to reduce the influence of noise. The frame coordinates are then shifted accordingly – rounded to integer pixel units, to avoid any interpolation of Poissonian counts. Finally, the centered frames are averaged, providing the (single) image of the PSF, which will be fit to a model function.

*Fitting*. For both imaging and tracking applications, the excitation beam pattern is placed on the fluorescent emitter, such that it is in proximity of the beam center. Therefore, a functional expression describing the beam shape in the central region is sufficient for any further analysis. The PSFs were fitted in a region of $200 \times 200\,\text{nm}^2$ (or $300 \times 300\,\text{nm}^2$ for EBP size $L > 100\,\text{nm}$) around the beam center. In this region, the PSF has mainly a parabolic profile. Deviations from a parabolic profile were accounted for by fitting a 2D polynomial up to order 4

$$I_{PSF}(\bar{r}) = I_{PSF}(x,y) = \sum_{i=0}^{4} \sum_{j=0}^{i} a_{ij}(x-x_0)^i (y-y_0)^{i-j} \qquad (S68)$$

with the coefficients $\{a_{ij}\}$, $x_0$ and $y_0$. Figure S7A-C shows a typical measured PSF and the fitted model function.

The parameter functions $p_i$, corresponding to a set of 4 doughnut shaped exposures, are obtained from the fitted PSF functions in a straight forward way. For displacements of the four doughnut beams as given by eq. (S23), the four beams can be written as

$$I_i(\bar{r}) = I_{PSF}^*(\bar{r} - \bar{r}_{b_i}) \qquad (S69)$$

where $I_{PSF}^*$ is an offset subtracted version of eq. (S68). This intensity is used to calculate the parameter vector $\bar{p}^{(0)}$, as defined in eq. (S4). For the localization of each molecule, its empirical SBR (eq. (S28)) has to be used in combination with $\bar{p}^{(0)}$ to calculate the parameter vector $\bar{p}$, as defined in eq. (S29). The four components of the parameter vector $\bar{p}^{(0)}$ for a beam separation of $L = 50\,\text{nm}$ are shown in fig. S7D-G.

## 6.2 Localization performance map

The experimental localization performance was characterized by repeatedly estimating the positions of single ATTO 647N molecules in ROXS buffer. The measurements were performed at different (known) positions with respect to the excitation beam pattern (EBP), in order to obtain the spatial dependence of the standard deviation and bias of the position estimation.

For characterizing the localization performance in a region of size $L$ (excitation beam separation), the multiplexed EBP was scanned in a $35 \times 35$ pixel grid (referred to as "frame" in the following) with a pixel spacing of $3\,\text{nm}$ (fig. 3A-D). For characterization of the localization error at the center of the EBP (fig. 3E), we scanned the EBP in a $9 \times 9$ grid with a spacing of $2\,\text{nm}$ with a change of $L$ after a full $9 \times 9$ grid scan.

In both cases, the scan region was centered on a single ATTO 647N molecule and scanning was performed with the electro-optical beam scanning system using the $125\,\text{kHz}$ bandwidth amplifiers. To ensure a stable beam position, multiplexing of the beam position was performed at $5\,\text{kHz}$ with a waiting time of $20\,\mu\text{s}$ after each beam movement. The average power of the modulated excitation laser was set to about $20\,\mu\text{W}$ resulting in a mean of about 0.3 to 2 collected photons per multiplexing cycle (dependent on $L$ and on the distance from EBP center).

### 6.2.1 Analysis

Obtaining the localization performance from the measured data requires a few processing steps. First, the recorded frames have to be centered. Due to the low number of photons recorded in each frame, noise is reduced by applying a sliding average filter on the data. The sliding window was 31 frames for the $35 \times 35$ pixel frames and 61 frames for the $9 \times 9$ pixel frames. A 2D polynomial function is fitted to the filtered frames to retrieve the beam center. In a next step, the individual unfiltered frames are shifted laterally to have their centers aligned. Shifting was performed in integer pixel units, to avoid combining or interpolating photon counts. The sum of all centered frames was use to obtain the experimental PSF. The PSF was modeled by a 2D polynomial function up to order four given by eq. (S68), whose coefficients were obtained from a fit to the data.

For a quantification of the localization performance for a specific number of detected photons, the data in each pixel is binned such that the photon number of interest is achieved. In this step, it is assured that every data point is only used once. Hence, many independent measurements with the same number of photons are obtained. This allows to localize the emitter many times using the same number of photons. Localization was performed with both, the MLE (see section 3.1.2) and the numLMS1 (see section 3.2.3). Figure S8 shows the experimental localization performance like in fig. 3 but for an extended range of detected photons and beam separations $L$.

The experimental localization performance is characterized by the mean 1D error defined as:

$$\sigma_{exp}^{(1D)} = \sqrt{\frac{1}{2\,M} \sum_{i=1}^{M} \left( \left( \hat{x}_m^{(i)} - x_0 \right)^2 + \left( \hat{y}_m^{(i)} - y_0 \right)^2 \right)} \qquad (S70)$$

with the true emitter position $(x_0, y_0)$, the number of independent localizations $M$ and the estimated emitter position $(\hat{x}_m^{(i)}, \hat{y}_m^{(i)})$.

## 6.3 MINFLUX nanoscopy

### 6.3.1 Data acquisition

In order to acquire an image of small objects, or to track short range movements, the excitation beam pattern (EBP) has to be centered on the object of interest. This is done by first taking a faint widefield image in which bright diffraction limited spots are selected manually. The centers of the bright spots are obtained by Gaussian fitting. Next, the EBP and detection volume are placed onto a spot center. Fine adjustment of the center of the four doughnut shaped excitation profiles is performed by a PI feedback loop that estimates the emitter location with the mLMS estimator (section 3.2.2). This adjustment is performed with the piezoelectric tip-tilt mirror. Typically, a beam separation of $L = 100$ nm was used. After a short time (~500 ms), the feedback loop is turned off and the parameters are changed to the parameters for the actual imaging measurement (e.g. laser power, beam separation $L$). The count traces for the four multiplexed excitation beams are then recorded. After the molecules in the illuminated region are bleached or enough data is recorded, the next region of interest is measured. The EBP is placed onto the next selected bright spot in the widefield image, and the whole procedure is repeated.

Separation of fluorescent molecules within the area illuminated by the four doughnut EBP is performed by stochastic switching of the molecules between fluorescent and non-fluorescent state. The off-switching of Alexa Fluor 647 is mainly driven by the excitation light of 642 nm. Transitions from the non-fluorescent state to the fluorescent state can be induced by illuminating with activation light of 405 nm wavelength. For minimizing both, the time with no molecule emission and the probability of having more than one molecule emitting at the same time, a conditional activation scheme was implemented. The 405 nm laser is switched on or off depending on the lowpass-filtered fluorescent count rate: it is switched on when the total count rate is below a threshold (~20 kHz) for a certain time (~50 ms), and it is turned off as soon as the count rate is above the threshold.

For the imaging, multiplexing of the EBP was performed with a period of 125 µs with a gate length of 13.6 µs and a gate delay of 13.6 µs (see fig. S5). The excitation laser power was 300 µW for measurements with $L = 70$ nm and 400 µW for measurements with $L = 50$ nm. The power of the 405 nm activation laser was set to about 1-2 µW. All powers were measured close to the back focal plane of the objective.

The process of obtaining an image from recorded count traces requires a few analysis steps which are described in the following.

### 6.3.2   Trace segmentation

Fluorescence dye molecule emission events have to be extracted from the recorded count traces. For this purpose, a simple hidden Markov model (HMM) for the sum of the fluorescence signal of the four exposures is employed. The model assumes three emission states (state 1: background emission; state 2: emission from a single emitter; state 3: emission from more than one emitter). State trajectory estimation is performed by applying the Viterbi algorithm.

The emission probability distribution of the three states are estimated in a two-step process. First, a mixture of Poisson distributions is fitted to the measured distribution of total count trace binned to 1 ms bins. The emission probability distribution is approximated by a Poisson distribution for the background with mean of $\lambda_1$, a Poisson distribution for one emitter with a mean of $\lambda_2$ and a Poisson distribution for two emitters with a mean of $\lambda_3 = 2\lambda_2$. To refine the emission distributions, an estimation of the most probable state path of the unbinned total count trace is performed using the Viterbi algorithm. We used the Matlab implementation "hmmviterbi" which is part of the Statistics Toolbox .The transition probability matrix ($T_{ij}$) for a transition from state $i$ to state $j$ is estimated to:

$$T = \begin{pmatrix} 0.9995 & 0.0005 & 0 \\ 0.00025 & 0.9995 & 0.00025 \\ 0 & 0.00025 & 0.99975 \end{pmatrix} \tag{S71}$$

From the most probable state path, an improved estimate for the emission distributions of the three states is obtained. This emission distribution is used in combination with the transition probability matrix (eq. (S71)) to re-run the Viterbi algorithm. The resulting estimation for the most probable state path is used to segment the count trace and extract single molecule emission events and the background level. Data points next to a state transition are discarded (as the emitter may have turned on during the four exposure cycle). Emission events with a total number of photons above a threshold, as indicated in fig. 4, are considered for further analysis. The trace segmentation for the measurement of the larger DNA origami (fig. 4A-G,M) is show in fig. S11F-G. The histogram of detected photons per emission event in the measurement of the larger origami is shown in fig. S11K.

### 6.3.3 Localization of emission events

For each single emitter emission event above the total count threshold, the position of the emitter is estimated. The counts for the four beam exposures are accumulated and the emitter position is estimated from the resulting count quartet $\bar{n}$. The $SBR$ is estimated from the data directly. From the HMM segmentation, the average background level $\lambda_{bg}$ can be estimated very well. For each emission event, the $SBR$ is estimated by

$$SBR = \frac{N}{\lambda_{bg} Q} \tag{S72}$$

where $N = \sum_{j=1}^{Q} \sum_{i=0}^{K-1} n_{ij}$ is the number of detected photons during the emission event of $Q$ multiplex cycles duration ($n_{ij}$ is the number of photons collected for exposure $i$ in multiplex cycle $j$) and $\lambda_{bg}$ is the average background counts per multiplex cycle ($K = 4$ exposures).

Position estimation is performed with the MLE, which is numerically obtained through a grid search, as described in section 3.1.2. Four successive 2D grids with spacings of 5 nm, 1 nm, 0.1 nm and 0.01 nm are used. The first grid spans a region of diameter 240 nm around the EBP center. An example of this grid search is shown in fig. S11E.

### 6.3.4   Clustering into nano-domains

In order to quantify the experimental localization precision, the estimated positions of the emission events are clustered into nano-domains. Clustering is performed with the k-means clustering algorithm with the squared Euclidean distance as distance metric. 100 sets of initial cluster centroid positions are used to avoid local extrema. Clustering was performed with a number of clusters between 1 and 12. The most likely number of clusters was determined by the highest mean cluster silhouette value.

### 6.3.5   Splitting events into equal number of photon events

A direct comparison between the experimental localization precision and its CRB requires the use of a fixed number of photons. For this purpose, each emission event is split into shorter emission events with a fixed number of photons. Splitting of the events is performed such that each beam multiplex cycle is used only once and that the total number of photons is exactly the number of interest.

### 6.3.6   Calculation of experimental covariance for fixed number of photons

The emission events splitted into equal numbers of photons are localized in the same manner as the original emission events. All localization events within a nano-domain are then used to calculate the *sample* covariance of the estimated emitter positon of the nano-domain. This covariance is a measure for the experimental localization precision for a fixed number of photons at the position of the nano-domain and can be compared with the CRB or the localization performance of single molecule localization using a camera. Details on the camera case can be found in section 4.

### 6.3.7   Drift correction in post processing

In addition to the active drift compensation with the system lock (see section 6.5.2) during the measurement process, a post processing drift compensation of the DNA origami imaging data can be performed. For post processing drift correction, a common drift function $\bar{x}_d(t)$ and the mean nano-domain centers $\bar{x}_{0i}$ are fitted simultaneously to all localizations:

$$\bar{x}(i,t) = \sum_{j=1}^{M} \delta_{ij}\,\bar{x}_{0j} + \bar{x}_d(t - t_0) \qquad (S73)$$

with the Kronecker delta $\delta_{ij}$, the number of nano-domains $M$, the mean time of the measurement $t_0$, the time $t$ and nano-domain identification $i$ of an emission event. The drift function $\bar{x}_d(t)$ was approximated by a polynomial up to order 4. Figure S11H-J shows the post processing drift correction of the imaging of the smaller origami (fig. 4H-L,N). No post processing drift correction was applied to data of the larger origami (fig. 4C-G,M).

## 6.4 MINFLUX tracking

### 6.4.1 Tracking initiation: intensity thresholding and Gaussian localizations

In order to track molecules with the excitation beam pattern (EBP) introduced in section 2.2 (and fig. 2B), an initial localization is needed to position the EBP on top of the molecule. This is related to the mLMSE (see section 3.2.2 and 3.2.3) working only in a finite region around the EBP origin.

In a first step, a Gaussian 560 nm excitation beam with $fwhm \approx 320$ nm is scanned over a field of view of $1.9 \times 1.9$ μm. In order to activate mEOS2 molecules, a Gaussian 405 nm activation beam is applied at the center of the scanned region with a period of 3 Hz and a pulse length of 1 ms. The pixel size of the scan was set to 127 nm and the pixel dwell time to 50 μs. The scan is repeated until a photon count threshold of $N_{th} = 15$ is surpassed. In that case, the scanning routine as well as the activation are stopped, and a Gaussian localization is launched around the center of the corresponding pixel.

This localization is implemented by four Gaussian beam exposures. Two are placed at $x_{b_0} = -L/2$ and $x_{b_1} = L/2$, in order to estimate the $x$ position of the molecule. The analogue is done with two further beams for the y position estimation. The beam separation $L$ was set to $L = 300$ nm. The multiplex cycle timings (see fig. S5) were: 10 μs of gate delay, 38 μs of excitation and detection gate and 2 μs for the localization window. Live position estimation employs the MLE introduced in section 3.1.3. The minimum number of photons used was $N = 60$. This ensures a localization precision of at least $\sigma \lesssim 30$ nm in a field of view of about 300 nm surrounding the Gaussian EBP origin.

The fastest diffusion coefficients expected are on the order of $D = 1\ \mu m^2/s$. In order not to be dominated by motion blurring for these $D$ values, the beam powers were chosen such that the count rate would stay above $\Gamma \gtrsim 130$ kHz. Once an emitter position $\hat{\vec{r}}_G$ is estimated, the EBP origin is repositioned to $\hat{\vec{r}}_G$.

Subsequently, the MINFLUX tracking routine is launched (see section 6.4.2). When terminated, the tracking initiation routine is restarted. Note that an additional 405 nm activation timeout of 10 s was introduced, in order to reduce the probability of having multiple activated emitter present.

### 6.4.2 The MINFLUX tracking routine

This tracking routine uses 560 nm excitation and the EBP introduced in section 2.2. The multiplex cycle rate (see fig. S5) was set to 8 kHz, where a gate delay of 7.4 μs, an excitation and detection gate of 23.2 μs and a localization window of 2 μs were employed. L was set to $L = 130$ nm. We used between 50-100 μW excitation powers. Live position estimation is calculated using the mLMS estimator $\hat{\vec{r}}_{mLMS}^{(k=1)}$ introduced in section 3.2.2. The $\bar{\beta}$ parameters are set to $\beta_0 = 1.27$ and $\beta_1 = 38$. Given that the mEOS2 proteins shows prominent blinking (see fig. 5C and S12), a reaction threshold of 6 counts is set for the total number of photons collected during each multiplex cycle. This is to prevent the EBP from being background-driven when the emitter is in a fluorescent off-state, as it would drift away. The threshold was set such that the probability of reacting to background counts is in the order of 2 %.

In order to enable single emitter tracking on a large field of view, a set of scanners and high voltage amplifiers are utilized as explained in section 6.5.1

### 6.4.3   Trace segmentation and blinking statistics

*Trace segmentation*

Measured trajectories consist of two segments. A time interval in which an emitter is tracked and, after bleaching or losing, a segment that is driven by background counts only. The respective segments were isolated by manual inspection. To be accepted as a successful trajectory, the average central doughnut ratio $\langle n_0/\sum n_i \rangle$ needs to fulfill: $\langle n_0/\sum n_i \rangle < 0.23$, with $n_i$ the counts obtained in the $i^{\text{th}}$ doughnut exposure of a multiplex cycle. Note that in the case of no emitter being present, a value of $\langle n_0/\sum n_i \rangle = 0.25$ is expected. Trajectories with unusually long off-times (the emitter in an off-state on the order of 100 ms) were separated into two independent trajectories. Likewise, long events of unusually high count rates ($\gtrsim 150$ kHz) with central donut ratio values $\gtrsim 0.25$ were cut out, to exclude the possibility of multiple emitters being present.

*Blinking statistics*

In order to identify molecule blinking events, a two stage hidden Markov model (HMM) is employed on $\bar{N}$ (see section 6.3.2). Two emission states (emitter on, off) are assumed. Estimation of the most probable state path employs the Matlab 'hmmviterbi' algorithm. In the first stage the transition matrix is assumed to be:

$$T = \begin{pmatrix} \dfrac{\Delta t}{\hat{t}_{off}} & 1 - \dfrac{\Delta t}{\hat{t}_{off}} \\ \dfrac{\Delta t}{\hat{t}_{on}} & 1 - \dfrac{\Delta t}{\hat{t}_{on}} \end{pmatrix} \tag{S74}$$

with $\hat{t}_{off} = 1$ ms and $\hat{t}_{on} = 3$ ms and $\Delta t$ the time sampling of the trajectory. The emission probability distribution is assumed to be Poissonian for both, the on and the off state, with parameters $\hat{\lambda}_{on}$ and $\hat{\lambda}_{off}$:

$$\hat{\lambda}_{on} = \langle \bar{N} \rangle \frac{\hat{t}_{on} + \hat{t}_{off}}{\hat{t}_{on} + \dfrac{\hat{t}_{off}}{\widehat{SBR}}}, \quad \hat{\lambda}_{off} = \frac{\hat{\lambda}_{on}}{\widehat{SBR}} \tag{S75}$$

and $\widehat{SBR} = 4$. The resulting state path is then used to improve the *state emission* estimates. The on and off times $\bar{t}_{on}$ and $\bar{t}_{off}$ are extracted (as arithmetic means) and assumed to follow truncated exponential distributions (as time intervals shorter than the sampling time are not collected). The updated parameter estimates $\hat{t}_{on}$ and $\hat{t}_{off}$ are obtained from the MLE:

$$\hat{t}_{on} = \langle \bar{t}_{on} \rangle - \Delta t, \ \ \hat{t}_{off} = \langle \bar{t}_{off} \rangle - \Delta t \tag{S76}$$

Furthermore, using $\bar{N}$ and the state path, the parameter estimate $\widehat{SBR}$ is recalculated. The new estimates are employed to update eq. (S74) and (S75), and the Viterbi algorithm is applied a second time. From which the final estimates $\hat{t}_{on}$, $\hat{t}_{off}$ and $\widehat{SBR}$ are extracted. The state path allows the determination of the total number of localizations $S$ in each trajectory. Note that all localizations with less than 3 total counts were discarded.

### 6.4.4 Trajectory reconstruction

As explained in section 3.2 and in section 6.4.2, the recorded trajectories are obtained employing the live position estimator $\hat{r}_{mLMS}^{(k=1)}(\hat{\bar{p}}, \hat{\beta})$, which is biased. In order to obtain an improved trajectory estimation, the trajectories are corrected using the numLMSE $r_{numLMS}^{(k=2)}(\hat{\bar{p}})$, introduced in section 3.2.3.

The numLMSE requires knowing the $\bar{p}(\bar{r})$ parameters and the SBR. The spatial dependence of the $\bar{p}$ parameter is measured following the procedure in section 6.1. The background $\lambda_b$ is estimated from the isolated trajectory (second) segment that is driven by background counts only (see section 6.4.3). The $\bar{p}$ parameter estimates $\hat{\bar{p}}$ are obtained from the measured count quartet using the MLE of the underlying multinomial distribution (see eq. (S44)).

### 6.4.5 Error definition

In the tracking application, the single emitter is free to move during the multiplexing cycle. This motion blurring leads to a reduction of the localization precision compared to a static emitter localization. The tracking error $\epsilon$ employed in this paper is defined as the distance between the *average emitter position during the multiplexing cycle* and its respective localization.

$$\epsilon = \sqrt{\frac{1}{d} \sum_{i=1}^{d} (\hat{r}_i - \langle r_i \rangle)^2}, \ \ \text{with } \bar{r} \in \mathbb{R}^d \tag{77}$$

where $\hat{\bar{r}}$ is the position estimate and $\langle \bar{r} \rangle$ the average emitter position during the multiplex cycle of duration $T$:

$$\langle \bar{r} \rangle = \frac{1}{T} \int_0^T \bar{r}(t) dt \tag{78}$$

### 6.4.6 Diffusion coefficient estimation

In order to attribute a diffusion constant $D$ to the numLMSE corrected trajectory (see section 6.4.4), the adapted OLSF algorithm described in section 5.2 is employed. The MSD is calculated using eq. (S64), where the emitter on and off states are obtained from the HMM state path (see section 6.4.3). Note that all localizations with less than 3 total counts are discarded.

The apparent diffusion constant plot in fig. 5D was calculated by using a sliding time window of 35 ms with a spacing of 3.75 ms. For every time window, the error bars ($\pm\sigma$) were approximated as the CRB times 1.4. This is because that is the precision obtained from simulations (see section 5.2).

The apparent diffusion histogram in fig. 5F was calculated by estimating $D$ values from a sliding window of 35 ms length every 17.5 ms on the respective tracks. $D$ estimations utilizing less than 100 valid localizations were discarded.

### 6.4.7 Localization precision

Figure 5H shows the localization precision $\sigma$ against the mean number of photons per localization $N_l$. The localization precision $\sigma$ was estimated for each track using the adapted OLSF algorithm described in section 5.2 (see also 6.4.6). Furthermore, $N_l$ was estimated using the total counts $\bar{N}$ and the state path (see 6.4.3). Subsequently, the precisions were binned according to their $N_l$ values, and the respective mean and standard deviations were calculated and plotted.

## 6.5 Experimental setup

The setup consists of a custom-built scanning microscope with fast beam scanning and modulation capabilities. A schematic illustration is shown in fig. S13.

### 6.5.1 Optical setup

Excitation of the sample can be performed by a Gaussian or doughnut-shaped beam or a wide field illumination. Two main lasers are available (Laser 1 and 2 in fig. S13), providing light with a wavelength of 642 nm or 560 nm. The beams are focused to a Gaussian or doughnut-shaped excitation spot with circular polarization. Amplitude modulation and switching between both beams is performed with electro-optical modulators (EOM).

Lateral scanning of the beam position in the sample is performed by two orthogonal electro-optical deflectors (EOD) and a piezoelectric tip/tilt mirror. Each EOD is driven by two high voltage amplifiers (Falco Systems WMA-300 and Trek PZD700A) in a differential arrangement. The first amplifier provides scanning with a bandwidth of 5 MHz, whereas the second provides a bandwidth of 125 kHz. In combination with the piezoelectric tip-tilt mirror, this provides a high-dynamic beam scanning system with a range of about $20 \times 20\,\mu\text{m}^2$. The sample is mounted on a piezoelectric stage which allows fine positioning in three dimensions.

Multiplexing of the excitation beam pattern (EBP) is done with the EODs and the fast amplifiers. The feedback controller is implemented with a combination EODs driven by the slow controller and the piezoelectric tip-tilt mirror. All scanners are controlled by the field programmable gate array (FPGA) board. The control signal is split into two (for the EODs and the tip-tilt mirror) with a highpass/lowpass filter bank, with a time constant of 10 ms.

The laser beams are focused onto the sample by an oil immersion microscope objective. Fluorescence photons are separated from the excitation light by a dichroic mirror (DM1 in fig. S13), spectrally filtered (F1 or F2 in fig. S13) and detected by an EMCCD camera or an APD. The APDs are coupled to multimode fibers which act as detection pinhole for spatial filtering. The effective pinhole diameter for APD 1 and APD 2 are 420 nm and 2.5 µm, respectively.

APD 1 is used for imaging whereas APD 2 is used for tracking applications. Switching between the camera and the APDs is done with a mirror on a motorized flip mount.

In addition to the main lasers 1 and 2, there are three other lasers (laser 3 to 5 in fig. S13) for wide field or focused illumination of the sample. Switching between the two illumination modes is performed by a lens (FL in fig. S13) on a motorized flip mount. The lasers in the setup are spectrally filtered (if necessary) by dichroic clean-up filters. Polarizers and wave plates are used to ensure a proper polarization.

### 6.5.2 System lock

The system is equipped with a system lock that measures and corrects drifts of the sample in all three dimensions.

Axial measurement of the sample position is based on measuring the displacements of a total internal reflected (TIR) beam on the coverslip-media interface. For this purpose, an infrared laser beam (laser 6 in fig. S13) is focused off-center into the back focal plane of the objective lens. The TIR signal is detected by a CMOS camera (Camera 2 in fig. S13). The center of mass in the camera image is used as measure for the axial sample position.

Lateral measurement of the sample position is performed by darkfield imaging scattering nanorods onto another CMOS camera (Camera 3 in fig. S13). A 2D Gaussian function is fitted to the image of a nanorod and the center position serves as measure for the lateral sample position.

The axial and lateral positions of the sample are kept constant by commanding the xyz piezo stage with a PI feedback loop written in LabView. Camera images are acquired at a framerate of about 90 fps with camera 2 and with about 160 fps with camera 3. The images are exponentially averaged and the position of the stage is updated every 100 ms.

### 6.5.3   Measurement control software

Measurements are performed with custom programs written in LabView (National Instruments, Austin, TX, USA). These programs control the devices directly, via the data acquisition (DAQ) boards or the FPGA board. The LabView programs perform the measurements as described in the previous sections. The main components of the programs are: control the laser beam position and modulation, multiplex the EBP, acquire the counts from the APD, perform a live position estimation using the mLMSE, reposition the beams and save the recorded data. In addition, the sample position is kept stable by a custom LabView program as described in section 6.5.2.

## 6.6   Sample preparation

### 6.6.1   Buffers

*Reducing and Oxidizing System buffer (ROXS):* ROXS buffer is prepared according to (*18*).

*Imaging buffer (IB):* Imaging buffer for Alexa Fluor 647 is prepared according to (*29*) with slightly different final concentrations of the ingredients: 0.4 mg/ml glucose oxidase, 64 µg/ml catalase, 50 mM TRIS/HCl pH 8.0, 10 mM NaCl, 10 mM MEA (Cysteamine) and 10 % (w/v) glucose. Additionally, 10 mM $MgCl_2$ is added.

*Folding buffer (FB):* DNA origami folding buffer is prepared from 1xTAE by addition of 10 mM $MgCl_2$.

### 6.6.2   Cleaning of coverslips and objective slides

In order to reduce background from unwanted fluorescent particles on the coverslips, all coverslips are cleaned prior to sample preparation. Cleaning is performed with a 2 % dilution of Hellmanex® III (Hellma GmbH & Co. KG, Müllheim, Germany) in Milli-Q water. The coverslips are sonicated inside the cleaning solution two times for 15 min. After the first 15 min, they are rinsed with Milli-Q water and the cleaning solution is replaced. After the second 15 min sonication step, the coverslips are rinsed with Milli-Q water and dried with compressed $N_2$. Objective slides are cleaned in the same manner but with only a single 15 min sonication step instead of two.

For measurements with drift compensation via the system lock, Au nanorods with their scattering peak at 980 nm (A12-25-980; Nanopartz Inc., Loveland, CO, USA) are immobilized on the cleaned coverslips. Prior to use, the nanorod stock solution is sonicated for 5 min, vortexed for 10 s and diluted in Milli-Q water (1: 3 to 1: 5). 10 µl of this solution are placed on a cleaned coverslip. After 1 min of incubation, the coverslip is rinsed with Milli-Q water and dried with compressed $N_2$.

### 6.6.3  Flow channel

Several samples are prepared with self-assembled flow channels. Flow channels are formed by gluing a coverslip with double sided tape (Scotch®, 3M France) to an objective slide.

### 6.6.4  Fluorescent microspheres

Samples for PSF measurements are prepared on cleaned coverslips with immobilized Au nanorods. The coverslips are coated with Poly-L-Lysine by placing 50 μl 0.01 % Poly-L-Lysine solution (Sigma Aldrich) on the coverslips. After 5 min, the coverslip is rinsed with Milli-Q water and dried with compressed $N_2$. 20 nm fluorescent microspheres (FluoSheres®, 0.02 μm, dark red fluorescent; Thermo Fischer Scientific Inc., Waltham, MA, USA) are sonicated for 10 min and diluted in PBS $10^6$ times. 20 μl bead solution is placed on the Poly-L-Lysine coated coverslip and rinsed with Milli-Q water after 1 min. The coverslip is dried with compressed $N_2$. The coverslip is mounted on an objective slide forming a flow channel. The channel is filled with PBS and sealed with epoxy glue (Hysol®, Locktite).

### 6.6.5  Immobilized single ATTO 647N molecule

Samples with individual immobilized ATTO 647N molecules are prepared from labeled DNA oligonucleotides. The sample for static localization measurements (fig. 3) is formed by annealing two 31 base long oligonucleotides of single stranded DNA (ssDNA) labeled with a single ATTO 647N molecule. For immobilization via biotin-streptavidin interaction, one end of the DNA is labeled with a biotin molecule. The dye is bound to the base at position 4. Single stranded DNA oligonucleotides were purchased from IBA GmbH (Göttingen, Germany). The sequences (5' to 3') are: ATA A(ATTO647N)TT TCA TTG CCA TAT ACT ACA GGA ATA A and TTA TTC CTG TAT ATG GCA ATG AAA TTA T(Biotin). The parentheses mark the bases labeled with ATTO 647N or Biotin, respectively. The oligonucleotides are mixed at equal concentrations and diluted to 100 nM concentration in 10 mM TRIS (pH 8), 10 mM NaCl and 1 mM EDTA. Annealing is performed by heating to ~95°C and gradually cooling down to room temperature in 45 min.

Samples are prepared according to the following protocol. From a cleaned objective slide and a cleaned coverslips with immobilized Au nanorods a flow channel is formed. The channel is filled with 15 μl biotinylated BSA (Albumin, biotin labeled bovine, A8549-10MG, Sigma Aldrich) dissolved in PBS, 2 mg/ml. After an incubation time of 15 min the channel is flushed with 400 μl PBS. The PBS in the flow channel is replaced by 15 μl Streptavidin (Streptavidin, recombinant, 11721666001, Sigma Aldrich) dissolved in PBS, 0.5 mg/ml. After an incubation time of 15 min the channel is flushed with 400 μl PBS. The PBS in the channel is replaced by 15 μl annealed DNA solution (50 pM in PBS). After an incubation time of 15 min the channel is flushed with 200 μl

PBS and afterwards with 200 μl ROXS. The sample is immediately sealed with epoxy glue (Hysol®, Locktite).

### 6.6.6   Immobilized labeled DNA origami

Custom DNA origami internally labeled with fluorescent dye molecules were purchased from GATTAquant GmbH (Braunschweig, Germany). The DNA origami have a size of about 60 nm × 80 nm and are labeled with up to nine Alexa Fluor 647 molecules with arrangements as indicated in fig. 4C,H. Each origami has six biotin attached which allow immobilization on a surface via a biotin-streptavidin interaction. Fluorescence on/off switching of Alexa Fluor 647 is facilitated by embedding the sample in an imaging buffer (IB) containing MEA and an oxygen scavenger system.

Samples are prepared according to the following protocol. From a cleaned objective slide and a cleaned coverslips with immobilized Au nanorods a flow channel is formed. The channel is filled with 15 μl biotinylated BSA (Albumin, biotin labeled bovine, A8549-10MG, Sigma Aldrich) dissolved in PBS, 2 mg/ml. After an incubation time of 15 min the channel is flushed with 400 μl PBS. The PBS in the flow channel is replaced by 15 μl Streptavidin (Streptavidin, recombinant, 11721666001, Sigma Aldrich) dissolved in PBS, 0.5 mg/ml. After an incubation time of 15 min the channel is flushed with 400 μl PBS. Prior to addition of the DNA origami, the channel is flushed with 200 μl FB. The FB in the channel is replaced by 15 μl DNA origami solution (5 pM in FB). After an incubation time of 15 min the channel is flushed with 200 μl FB and afterwards with 200 μl IB. The sample is immediately sealed with picodent twinsil® speed 22 (picodent® Dental-Produktions- und Vertriebs-GmbH, Wipperfürth, Germany).

### 6.6.7   E. coli sample preparation

Strains with 30S ribosomal protein S2 fused to mEos2 were obtained following the protocol described in (*24*).

Overnight cultures were grown from freezer cultures by 37°C incubation in a growing media. The latter consists of M9 media supplemented with 0.4 % glucose and 1640 Amino Acids (RPMI 1640 Amino Acids Solution (50x), Sigma-Aldrich). Cells were grown from overnight culture by the analog process, until an optical density at 600 nm of $OD_{600} = 0.15$ is obtained. The cells were separated from the used growing media by centrifuging at 4000 rpm for 4 min, and embedded in fresh growing media, to attain again $OD_{600} = 0.15$.

For imaging, these cells are placed on 2.5 % agarose pads (SeaPlaque GTG Agarose, Lonza) obtained using fresh growing media. Immobilization is attained by squeezing the agarose pad between a clean microscope slide and a clean cover slip (see section 6.6.2). The latter is attached using a frame seal (Frame-Seal Slide Chambers SLF0601, Bio-Rad Laboratories GmbH).

# Supplementary Figures

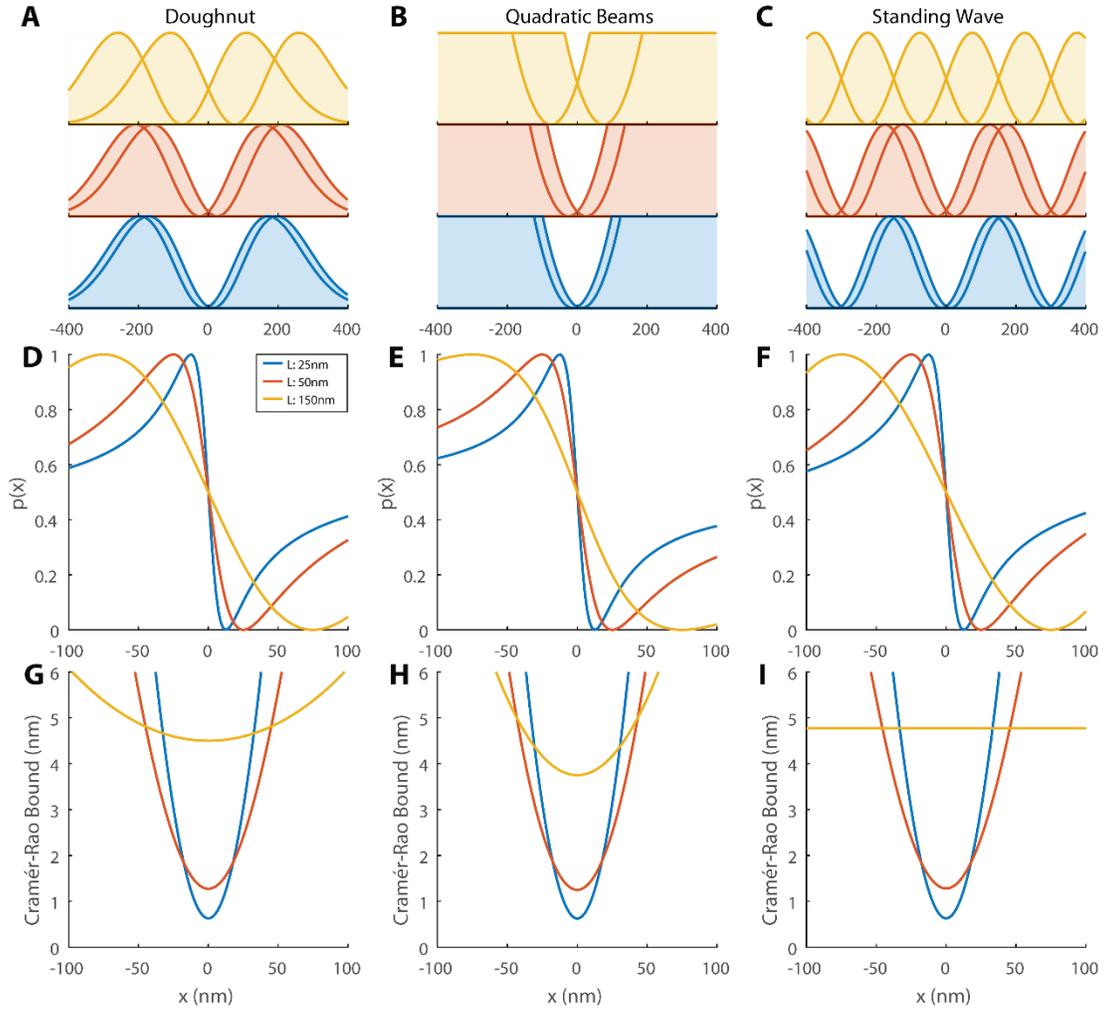

**Fig. S1**

**One dimensional p functions and CRB**. One-dimensional localization of an emitter with two (**A,D,G**) doughnut-, (**B,E,H**) quadratic beams and (**C,E,I**) two standing waves. For each case, the beam intensities $I_i(x)$, the binomial parameters $p(x)$ and the CRBs are shown for a beam separation $L$ of 25 nm (blue), 50 nm (red) and 150 nm (yellow), and the use of $N = 100$ photons. The size parameter $fwhm$ is 300nm. The parameter k is set to $k = 2\pi/\lambda$, with $\lambda = 600$ nm.

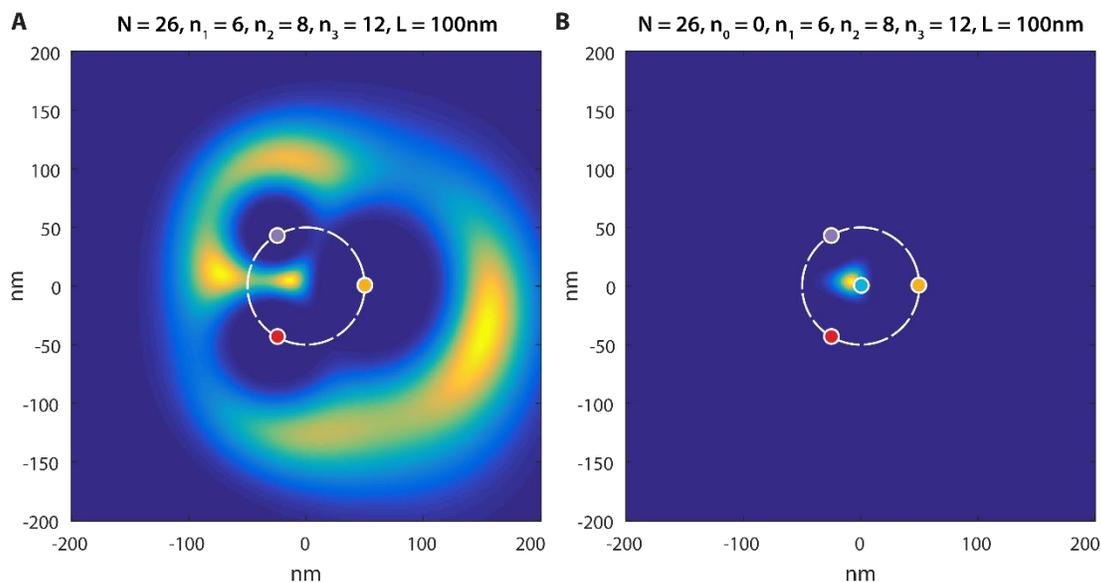

**Fig. S2**

**Visualization of localization indeterminations**. (**A**) Likelihood function $\mathcal{L}(\bar{r}_m|\bar{n})$ for the case of a localization with three exposures to doughnut beams (parameter $fwhm = 200$ nm) with their zeros at the marked colored points and collected photons of $\bar{n} = [6, 8, 12]^T$ with a total number $N = 26$. The likelihood function is badly behaved for creating a maximum likelihood estimator. (**B**) Same function for the case of a four doughnut localization and collected photons $\bar{n} = [0, 6, 8, 12]^T$. Though the same number of photons is collected, the radial information encoded in $n_0 = 0$ concentrates the relevant support of the likelihood function to a much smaller region with a clearly defined maximum; which has a better behavior in terms of defining a maximum likelihood estimator.

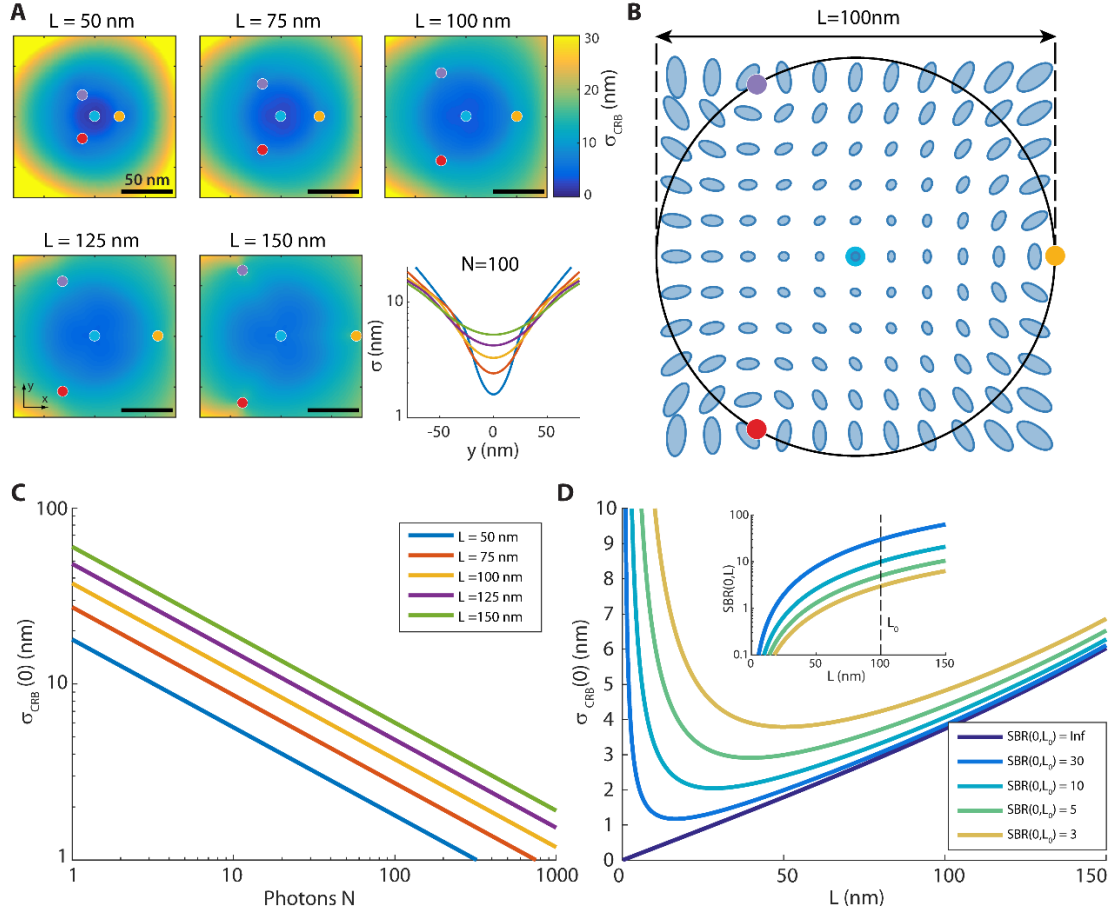

**Fig. S3**

**CRB for the two-dimensional case**. (**A**) Position dependent CRB $\tilde{\sigma}_{CRB}(\vec{r})$ for $N = 100$ photons. The employed EBP is depicted in section 2.2. The doughnut shaped excitation profiles are set to have $fwhm = 360$ nm. Colored dots indicate the respective excitation beam origins. The peripheral beams have triangular symmetry and are inscribed in a circle of diameter $L$. The CRB is minimal at the EBP origin and grows with increasing distance to it. Reduction of the parameter $L$ enables to increase the localization precision, especially in the vicinity of the origin. The last panel shows a $y$ cut through $\tilde{\sigma}_{CRB}(\vec{r})$, were the coloring indicates different $L$ values (see legend in (**C**)). (**B**) The covariance matrix $\Sigma(\vec{r}_m)$ plotted at different positions for a total number of $N = 1000$ photons. The covariance is visualized as ellipses (contour level $e^{-1/2}$). Note, that the localization precision is not isotropic. (**C**) CRB at the origin of the beam pattern as a function of the total photon number $N$ for different beam pattern diameters $L$ and an infinite SBR. (**D**) Influence of the $SBR(\vec{0}, L)$ (see eq. (S31)) on the CRB at the origin $\vec{r} = \vec{0}$, for $N = 100$ photons and for different $SBR(\vec{0}, L_0 = 100$ nm$)$ values. Depending on the value of the latter, a reduction of $L$ improves the localization precision only up to a minimal $L$. A further diminution results in an increase of $\tilde{\sigma}_{CRB}$. The effective signal-to-background ratio $SBR(\vec{0}, L)$ is visualized in the inset.

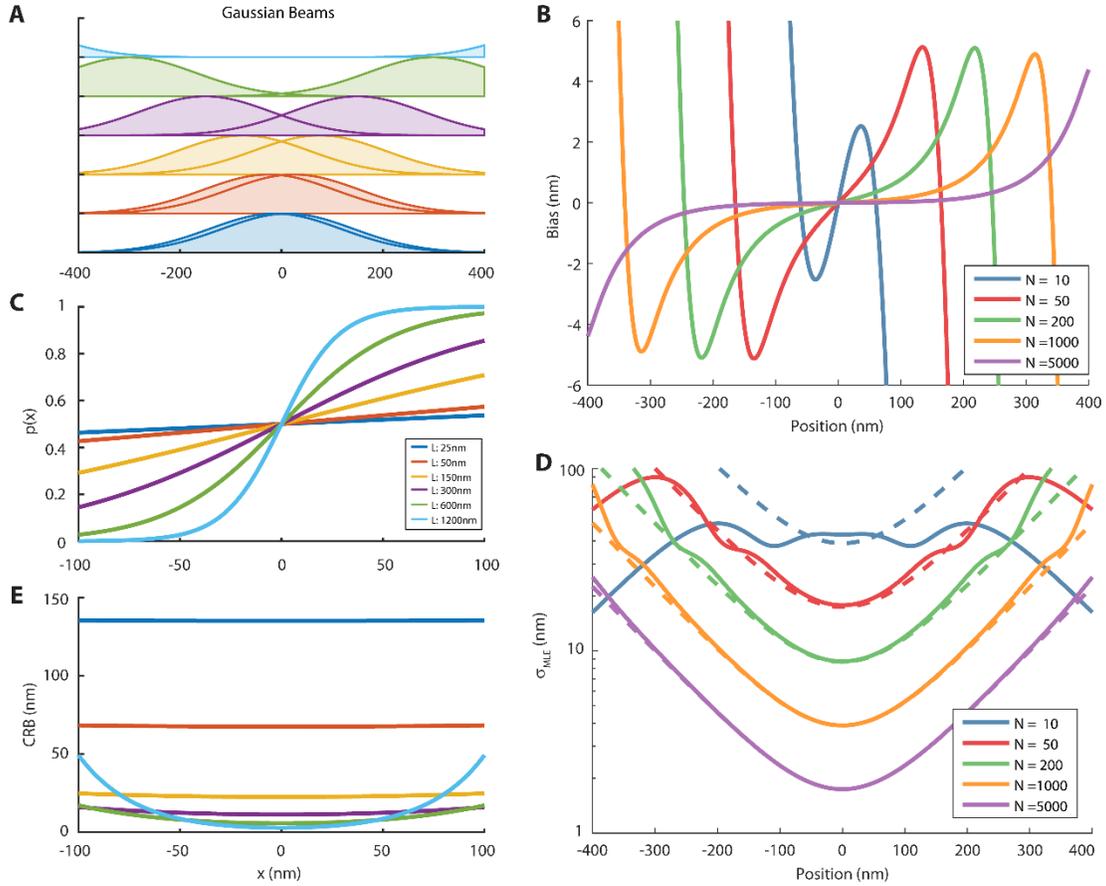

## Fig. S4

**One-dimensional localization performance with two Gaussian beams**. (**A,C,E**) One-dimensional localization of an emitter with two Gaussian beams. The beam intensities $I_i(x)$, the binomial parameters $p(x)$ and the CRBs are shown for different beam separation $L$ of 25 nm to 1200 nm and the use of $N = 100$ photons. The size parameter $fwhm$ is 300 nm. (**B**) Bias as a function of the emitter position for different total number of photons $N$. Excitation geometry: two 1D Gaussian beams with $fwhm = 320$ nm separated by a distance $L = 300$ nm. The region with acceptable bias grows with increasing number of photons, whereas the origin stays unbiased. (**D**) Comparison of the standard deviation $\tilde{\sigma}_{MLE}$ (solid lines) to the Cramér-Rao bound $\tilde{\sigma}_{CRB}$ (dashed lines) Higher values of $N$ make $\tilde{\sigma}_{MLE}$ approach the information theoretical limit. Especially at outer positions, $\tilde{\sigma}_{MLE}$ falls below the CRB, as the estimator $\hat{x}_m^{MLE}$ is biased.

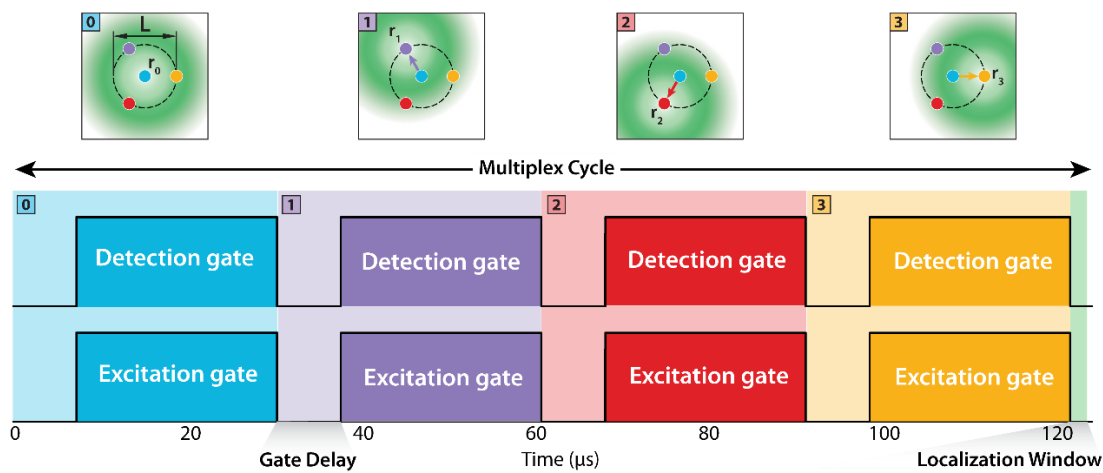

**Fig. S5**

**Visualization of the Multiplex Cycle**. An FPGA board controls the modulation and deflection of the laser beam, and the acquisition of the fluorescence photon-counts. The excitation is gated by EOMs in order to assure that the fluorophore is not excited during beam repositioning (see section 6.5). The excitation gate off-time is given by the *gate delay*. Photons are acquired only if the detection gate is enabled. The delay of the latter is equal to the excitation gate off-time. After 4 successive exposures, a live position estimation can be calculated employing the FPGA board. The calculation time slot is given by the *localization window* (green). Note that the *multiplex cycle* is visualized depicting the excitation pattern described in section 2.2, but is not restricted to it.

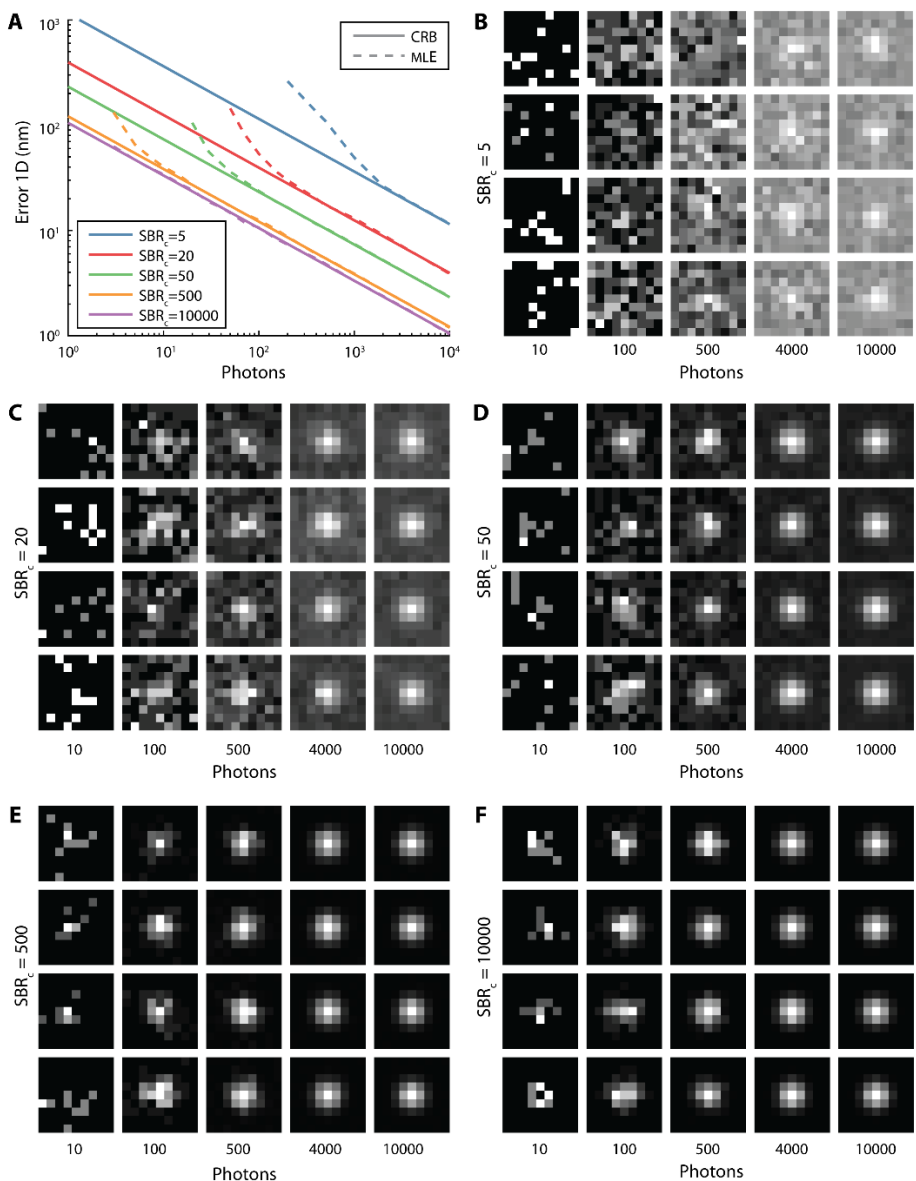

**Fig. S6**

**CRB and MLE performance for an ideal camera**. (**A**) CRB (solid line) and performance of the MLE (dashed line) for single emitter localization using a perfect camera. The $SBR_c$ is define as in eq. (S60). The PSF is modeled as a symmetrical Gaussian function with width $\sigma_{PSF} = 100$ nm. The pixel size of the camera is 100 nm. Camera size 9x9 pixels. (**B**)-(**F**) Simulated camera images for different $SBR_c$ and number of detected photons $N$. $9 \times 9$ pixels, pixel size 100 nm. The emitter is located in the center of the image. All images are normalized to the maximal pixel value.

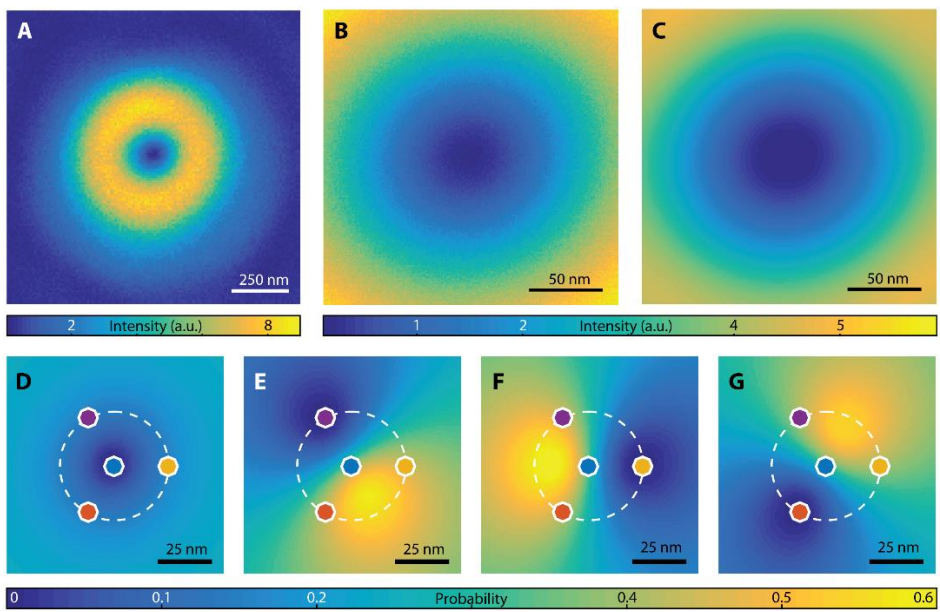

**Fig. S7**

**Experimental PSF**. (**A**) Typical measured PSF of the employed doughnut-shaped excitation beam. (**B**) Mean image of the drift corrected and centered frames of the central region of doughnut-shaped excitation beam. (**C**) Fit of the model function to (**B**). (**D**)-(**G**) $\{p_i^{(0)}\}$ functions of the four multiplexed doughnut-shaped excitation beams for a beam separation of $L = 50$ nm. The positions of the beam centers are indicated in the same way as in fig. 2.

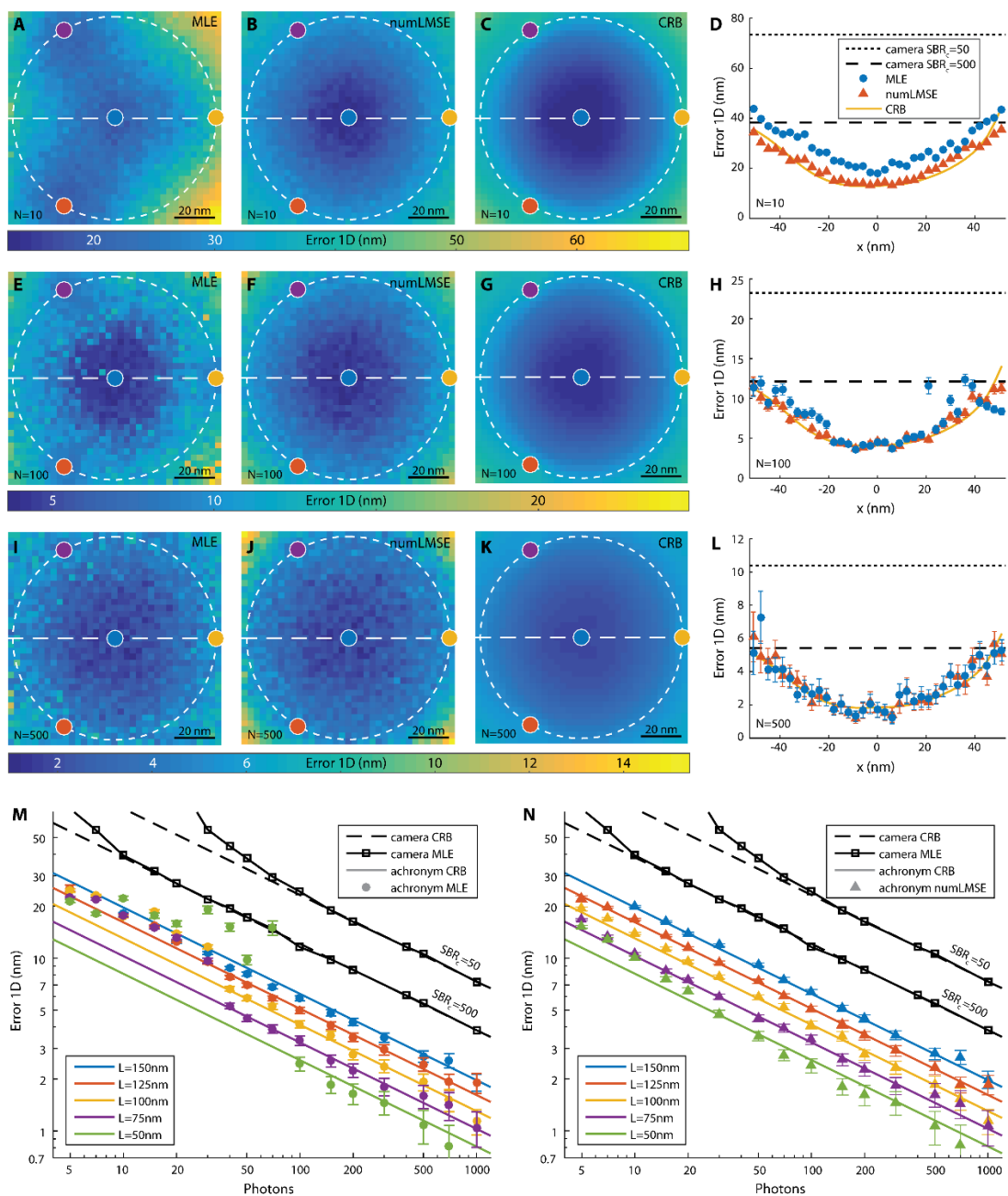

**Fig. S8**

**Details on the localization precision measurements**. (**A**)-(**L**) Static localization performance map for a single ATTO 647N molecule. Measurements were performed with $L = 100$ nm. (A), (E) and (I) show the measured performance of the MLE for three different numbers of detected photons. The corresponding performance of the numLMSE is shown in (B), (F) and (J). The CRB is shown in (C), (G) and (K). Profiles along the dashed line in the maps are shown in (D), (H) and (L). The camera localization performance was calculated using the same parameters as for fig. S6 (see also section 4). (**M**) and (**N**) show the localization performance of at the center of the beam pattern for different beam separations $L$. Position estimation was done with the MLE (see section 3.1.2) and the numLMSE (see section 3.2.3), respectively.

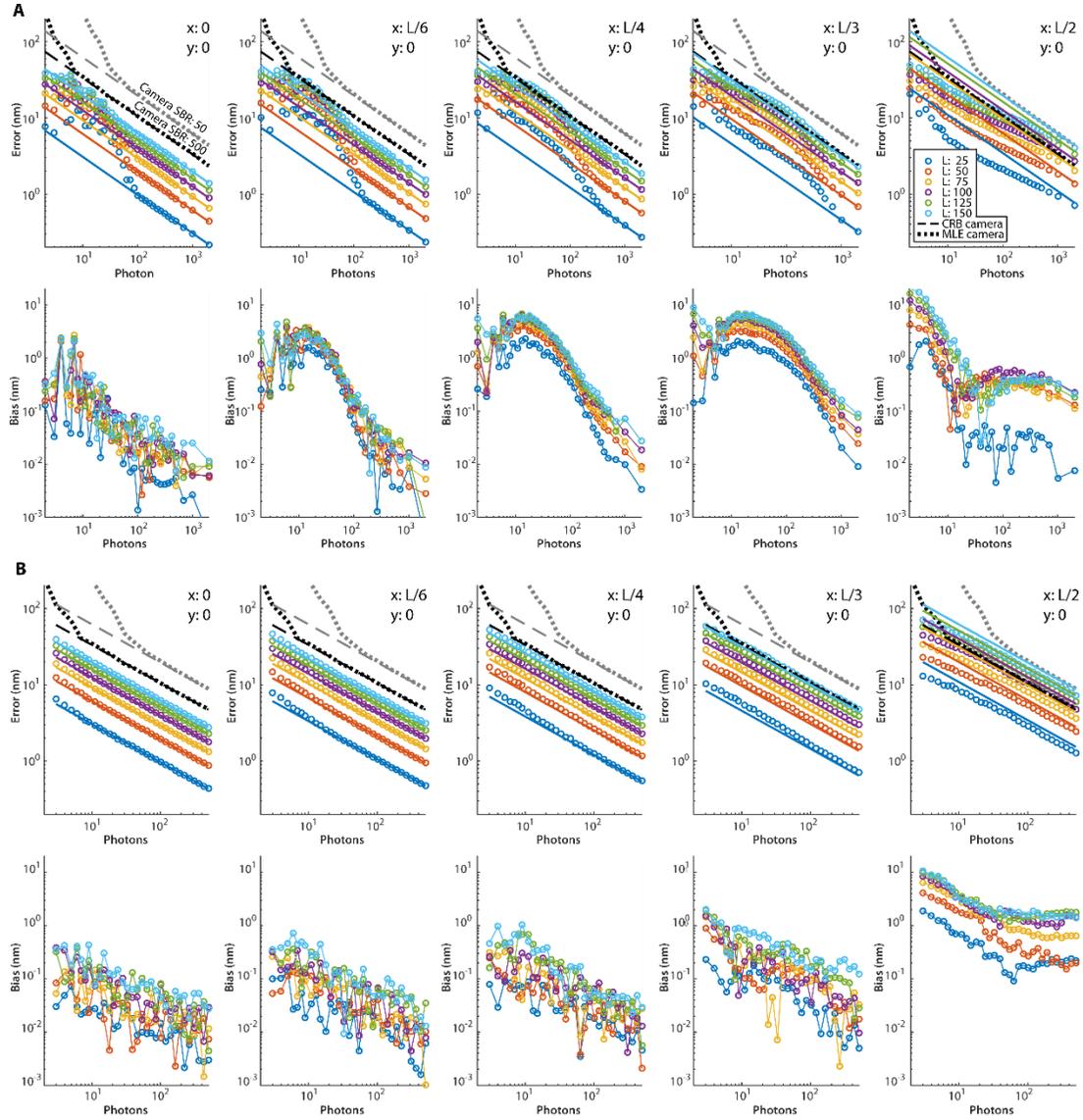

**Fig. S9**

**Convergence of position estimators.** Localization performance of the MLE and the numLMSE as a function of the number of photons $N$ evaluated at different positions $\bar{r}$. The excitation PSF as well as the EBP are the ones introduced in section 2.2. The SBR is set to 10 in (**A**) and (**B**). For every pictured data point, $10^4$ $\bar{p}(\bar{r})$ parameter estimates $\{\hat{\bar{p}}\}_{\bar{r}}$ were generated. The camera CRB and MLE were calculated as depicted in section 4.1. Parameters were set to: $\sigma_{PSF} = 87$ nm, pixel size $a = 100$ nm, number of pixels $K = 9x9$, and two values of $SBR_c$ (50 in gray and 500 in black), respectively. Note that no read noise is incorporated. (**A**) The MLE of section 3.1.2 was used to localize the respective $\hat{\bar{p}}$ values. It can be seen that the MLE converges to the CRB for $N \gtrsim 100 - 500$ photons only, depending on $\bar{r}$ as well as on $L$ (units of $L$ in legend: nm). For smaller $N$ the MLE deviates considerably from its information theoretical limit. (**B**) In this case the numLMSE (see section 3.2.3) was used to localize the $\hat{\hat{p}}$ values of the respective sets $\{\hat{\hat{p}}\}_{\bar{r}}$. The divergence to the CRB is strongly reduced, compared to the MLE. Especially for low photon numbers $N$ the numLMSE performs better. Note that this estimator is designed to work in a region $|\bar{r}| \lesssim L/2$.

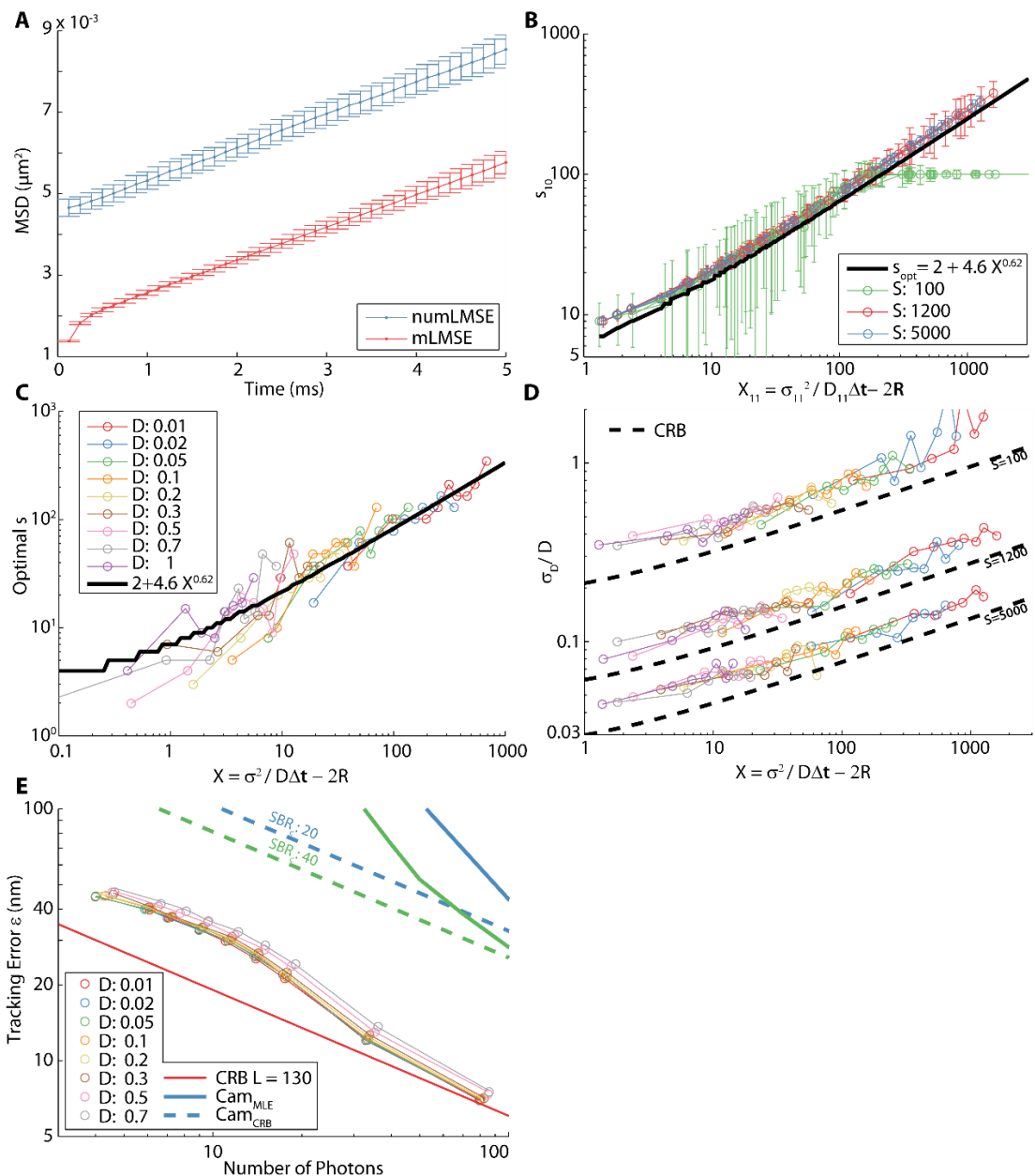

**Fig. S10**

**Estimation of the apparent diffusion coefficient and tracking error**. Data in this figure is based on a tracking simulation using MINFLUX (see section 5.2). The single emitter movement follows free isotropic Brownian motion in two dimensions. (**A**) Comparison of the mean of 100 MSD curves for trajectories obtained using the mLMS estimator (red) and the numLMS estimator (blue). The total width of the error bars represents twice the standard deviation. (**B**) Convergence of $s$ to $s_{opt}$ using the recursive algorithm in eq. (S67) for different number of trace length $S$. Each data point represents the median of 100 trajectories. The total width of the error bars is twice the standard deviation. The pictured $s$ values $s_{10}$ were obtained after 10 iterations, and are plotted against the estimated reduced square localization error $\hat{X}_{11}$. The latter was obtained from the results of the $10^{\text{th}}$ iteration. (**C**) Optimization of the number of MSD points $s$ used in the fit to eq. (S62). Optimization was conducted by minimizing the deviation of the estimated $\hat{D}$ value to its ground truth $D$ value. The optimal $s$ values can be described by the relationship $s_{opt}(X) = 2 + 4.6X^{0.62}$. (**D**) Comparison of the relative standard deviation $\sigma_D/D$ of the estimated $D$ values to the CRB. Each data point is calculated from 100 trajectories. The results of 3 different trajectory length $S$ are visualzed. 10 iterations of eq. (S67) were conducted. (**E**) Comparison of the real tracking error $\epsilon$ (thin colored lines), to the value extracted by the OLSF method (colored circles). Additionally, the CRB (eq. (S26)) for $L = 130$ nm and $fwhm = 450$ nm is shown as thick red line. The information theoretical limit for camera localization for two $SBR_c$ are shown as thick dotted lines. The MLE camera performance (for the same two $SBR_c$) is pictured as thick colored lines.

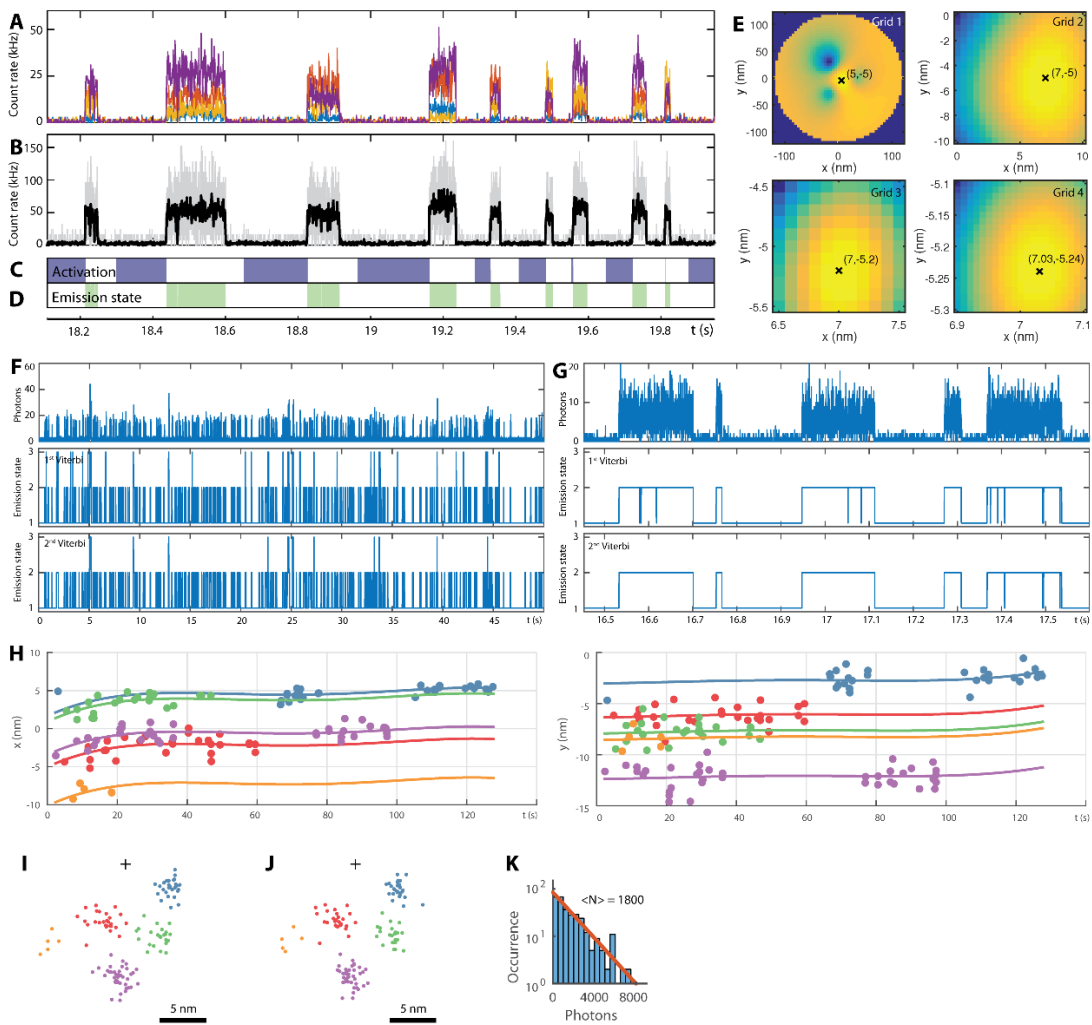

**Fig. S11**

**Details of MINFLUX nanoscopy processing**. (**A**) Typical trace of recorded counts during imaging of a labeled DNA origami. Time bins: 1 ms. (**B**) Total count rate of recorded photons (gray: raw; black: binned to 1 ms) of the same data as in (**A**). (**C**) Conditional activation of molecules with illumination of 405 nm light. The laser is switched on and off according to the count rate of detected photons as described in section 6.3.1. (**D**) Single molecule emission states detected by an HMM trace segmentation as shown in (F) and (G). (**E**) Position estimation by maximization of the likelihood function in a successive grid search algorithm. It shows a typical localization in the imaging of the origami. The quartet of detected photons is $\bar{n} = (124, 609, 695, 1382)$ and the *SBR* is 17.75. (**F**) HMM trace segmentation on the total count trace of detected photons as described in section 6.3.2. (**G**) Zoom into (F). (**H**) Post processing drift correction of the 6 nm origami imaging as described in section 6.3.7. The colors are consistent with (I). (**I**) Localizations of 6 nm origami imaging clustered into nano-domains before drift correction. The black cross indicates the position of the central excitation beam. (**J**) Data of (I) after drift correction. (**K**) Histogram of detected photons per emission event in the measurement of the larger origami.

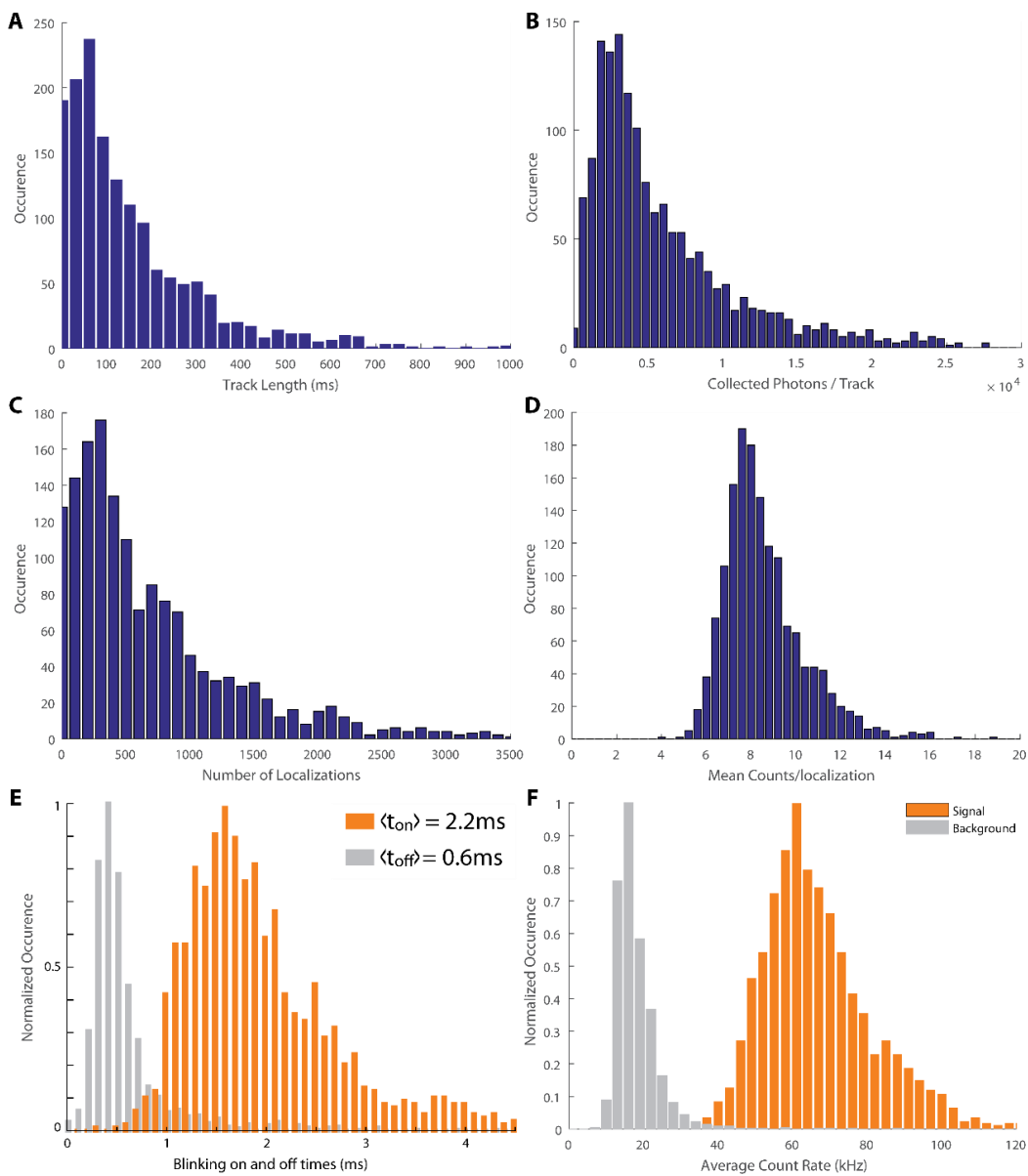

**Fig. S12**

**MINFLUX tracking in living E. coli.** (**A**) Histogram of the track length with mean of 157 ms. (**B**) Collected photons (background corrected) per trace with mean of 5803. (**C**) Valid localizations per trace with mean of 742 after application of a HMM and discarding of all localizations with less than 3 total counts (see 6.4.3). Note that the distributions in (A-C) extend further from the plot limits. All stated mean values of the distributions in (A-C) are the parameter value (i.e. average) of the respective non-truncated exponential distributions (truncated distributions have higher averages than reported). The longest measured track was 1444 ms with 86009 photons and 7503 valid localizations. (**D**) Mean counts per localization of the respective tracks, with an ensemble average of 9. (**E**) Blinking on and off times of mEos2 extracted from a two stage HMM on the counts $\overline{N}$ of each track (see 6.4.3). (**F**) Normalized occurrences of the signal and background count rates for the respective traces extracted from $\overline{N}$ and the state path (see 6.4.3). The average $SBR$ is 3.7.

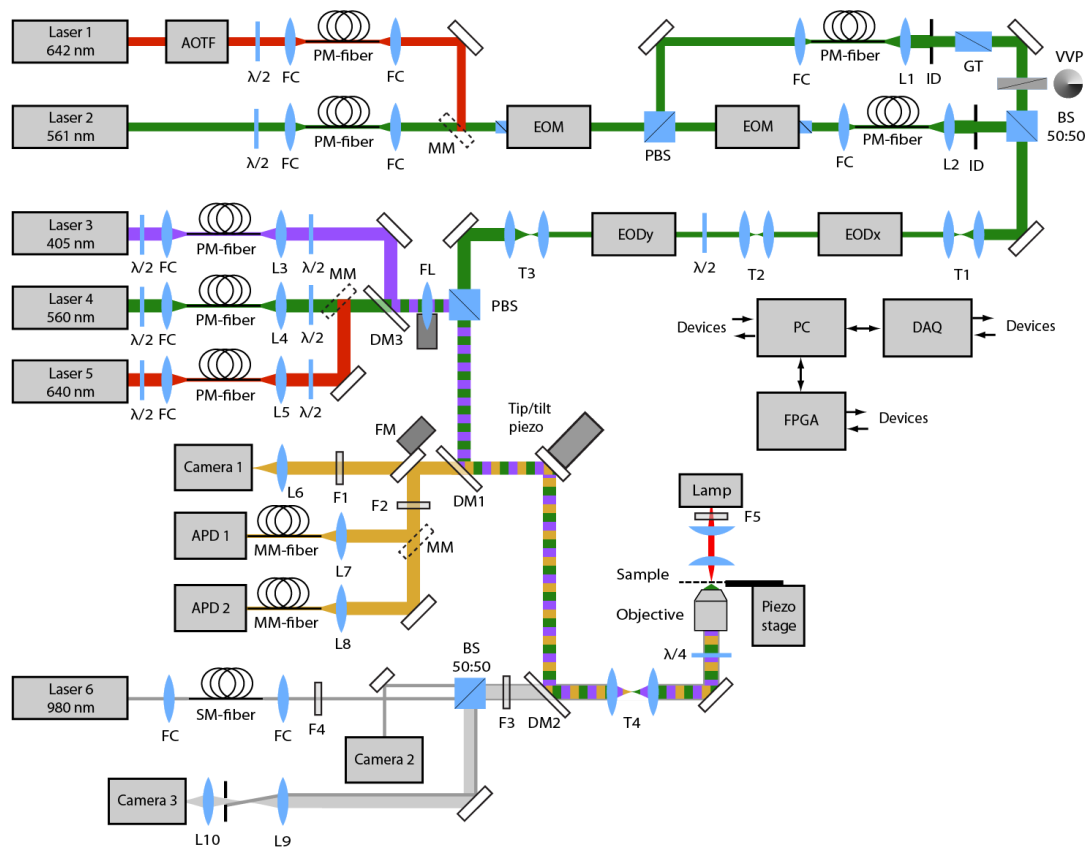

**Fig. S13**

**Schematic illustration of the experimental setup**. The following components are used: *Laser 1:* VFL-P-1500-642 (MPB Communications Inc., Pointe-Claire, Quebec, Canada) or Koheras SuperK Extreme (NKT Photonics, Birkerød, Denmark), *Laser 2:* Cobolt Jive™ (Cobolt AB, Solna, Sweden). *Laser 3:* 405-50-COL-004 (Oxxius, Lannion, France), *Laser 4:* LDH-D-C-485 (PicoQuant, Berlin, Germany), *Laser 5:* VFL-P-1000-560 (MPB Communications Inc., Pointe-Claire, Quebec, Canada), *Laser 6:* LDH-D-C-640 (PicoQuant, Berlin, Germany), *Laser 7:* LP980-SF15 (Thorlabs Inc., Newton, NJ, USA), *AOTF:* AOTFnC VIS (AA Sa, Orsay, France), *EOM:* LM 0202 P 5W + LIV 20 (Qioptiq Photonics GmbH & Co. KG, Göttingen, Germany), *EODx and EODy:* M-311-A (Conoptics Inc., Danbury, CT, USA) + PZD700A (Trek Inc., Lockport, NY, USA) + WMA-300 (Falco Systems BV, Amsterdam, The Netherlands), *Tip/tilt piezo:* PSH-10/2 + EVD300 (both piezosystem jena GmbH, Jena, Germany), *Piezo stage:* P-733.3-DD + E725 (both Physik Instrumente (PI) GmbH & Co. KG, Karlsruhe, Germany), *GT:* Glan-Thompson prism (B. Halle Nachfl. GmbH, Berlin, Germany) *PBS:* polarizing beam splitter cube (B. Halle Nachfl. GmbH, Berlin, Germany), *BS:* beam splitter cube 50:50, *FC:* fiber collimator 60FC-* (Schäfter+Kirchhoff, Hamburg, Germany), *λ/2:* half wave plate (B. Halle Nachfl. GmbH, Berlin, Germany), *λ/4:* quarter wave plate (B. Halle Nachfl. GmbH, Berlin, Germany), *VVP:* VVP 1a (RPC Photonics, Rochester, NY, USA), *SM-fiber:* single mode fiber (Thorlabs Inc., Newton, NJ, USA), PM-fiber: polarization maintaining fiber (Thorlabs Inc., Newton, NJ, USA or Schäfter+Kirchhoff, Hamburg, Germany), *MM-fiber*: multimode fiber M31L01 (Thorlabs Inc., Newton, NJ, USA), *L:* achromatic lens with VIS or NIR AR coating (Thorlabs Inc., Newton, NJ, USA or Qioptiq Photonics GmbH & Co. KG, Göttingen, Germany), *T*: telescope, *FL*: lens of flip mount, ID*: iris diaphragm, *MM:* mirror on magnetic mount, *FM:* mirror on motorized flip mount, *DM1:* Z488/633RDC (Chroma Technology Corp., Bellows Falls, VT, USA) or ZT405/488/561 (Chroma Technology Corp., Bellows Falls, VT, USA), *DM2:* BB1-E02P (Thorlabs Inc., Newton, NJ, USA), *DM3:* Z500RDC-XT (Chroma Technology Corp., Bellows Falls, VT, USA), *F1:* FF01-635/LP-25 (Semrock Inc., Rochester, NY, USA) or BLP02-561R-25 (Semrock Inc., Rochester, NY, USA) + FF01-842/SP-25 (Semrock Inc., Rochester, NY, USA), *F2:* ET700/75m (Chroma Technology Corp., Bellows Falls, VT, USA) + ZET642NF (Chroma Technology Corp., Bellows Falls, VT, USA) or BLP02-561R-25 (Semrock Inc., Rochester, NY, USA) + FF01-775/SP-25 (Semrock Inc., Rochester, NY, USA), *F3:* FEL850 (Thorlabs Inc., Newton, NJ, USA), *F4:* LL01-980-12.5 (Semrock Inc., Rochester, NY, USA) + FB980-10 (Thorlabs Inc., Newton, NJ, USA), *F5:* Z635/10, *APD 1,2:* SPCM-AQRH-13-FC (Excelitas Technologies, Waltham, MA, USA), *Camera 1:* Luca S, (Andor Technology Ltd., Belfast, UK), *Camera 2,3:* DMK 22BUC02 (The Imaging Source Europe GmbH, Bremen, Germany), *Objective:* HCX PL APO 100x/1.40-0.70 Oil CS (Leica Microsystems GmbH, Wetzlar, Germany), *Lamp:* LQ 1100 (Fiberoptic − Helm AG, Bühler, Switzerland), *PC:* personal computer running Windows 7 (Microsoft Corp., Redmond, WA, USA) and LabView 2013 (National Instruments, Austin, TX, USA), *DAQ:* NI PCIe-6353 + NI PCI-6259 (both National Instruments, Austin, TX, USA) + USB-3103 (Measurement Computing Corporation, Norton, MA, USA), *FPGA:* NI PCIe-7852R (National Instruments, Austin, TX, USA)

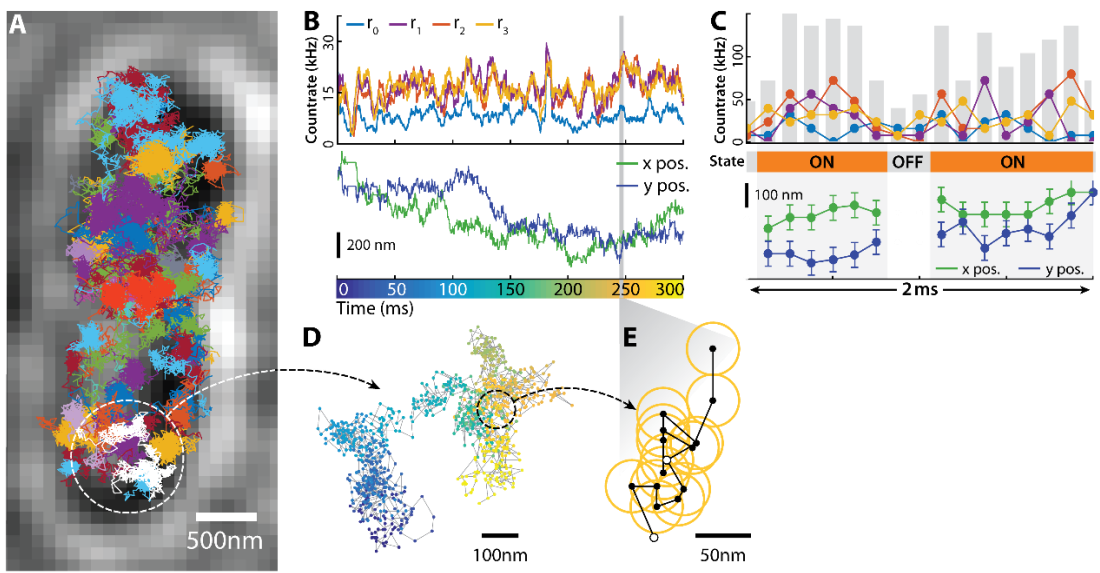

**Fig. S14**

**MINFLUX tracking.** (**A**) Transmission image of a bacterium overlaid with 77 independent tracks. (**B**) Details of the track highlighted in white in (A). Upper panel: low pass filtered count rate of the four exposures (blue: $\bar{r}_0$, violet: $\bar{r}_1$, red: $\bar{r}_2$, yellow: $\bar{r}_3$), average total count rate 58 kHz. The lower counts of the blue trace indicate that the proximity of the single molecule to the central zero of the doughnut quadruple. <u>Lower panel</u>: Extracted $x$ and $y$ coordinates of the trajectory. (**C**) 2 ms excerpt of the trace in (B) (marked in gray at time point 248 ms). <u>Upper panel</u>: counts per exposure are shown together with their sum (gray bars) used for on/off classification. <u>Lower panel</u>: Extracted $x$ and $y$ coordinates. (**D-E**) Trajectories shown in (B) and (C), respectively. Diameter of the open circles in (E) visualize the average tracking error.

# Supplementary Tables

| $d$ | $K$ | $p_i$ | $F_{\bar{p}}/N$ | $\mathcal{J}^*$ | $\widetilde{\sigma}_{CRB}$ |
|---|---|---|---|---|---|
| 1 | 2 | $\dfrac{I_i}{I_0 + I_1}$ | $p_0^{-1} + p_1^{-1}$ | $\dfrac{\partial p_0}{\partial r_0}$ | $\sqrt{\operatorname{tr}\left(F_{\bar{r}_m}^{-1}\right)}$ |
| 1 | 3 | $\dfrac{I_i}{I_0 + I_1 + I_2}$ | $\begin{bmatrix} p_0^{-1} & 0 \\ 0 & p_1^{-1} \end{bmatrix} + p_2^{-1}\mathbf{1}$ | $\begin{bmatrix} \frac{\partial p_0}{\partial r_0} \\ \frac{\partial p_1}{\partial r_0} \end{bmatrix}$ | $\sqrt{\operatorname{tr}\left(F_{\bar{r}_m}^{-1}\right)}$ |
| 2 | 3 | $\dfrac{I_i}{I_0 + I_1 + I_2}$ | $\begin{bmatrix} p_0^{-1} & 0 \\ 0 & p_1^{-1} \end{bmatrix} + p_2^{-1}\mathbf{1}$ | $\begin{bmatrix} \frac{\partial p_0}{\partial r_0} & \frac{\partial p_0}{\partial r_1} \\ \frac{\partial p_1}{\partial r_0} & \frac{\partial p_1}{\partial r_1} \end{bmatrix}$ | $\sqrt{\dfrac{1}{2}\operatorname{tr}\left(F_{\bar{r}_m}^{-1}\right)}$ |
| 2 | 4 | $\dfrac{I_i}{I_0 + I_1 + I_2 + I_3}$ | $\begin{bmatrix} p_0^{-1} & 0 & 0 \\ 0 & p_1^{-1} & 0 \\ 0 & 0 & p_2^{-1} \end{bmatrix} + p_3^{-1}\mathbf{1}$ | $\begin{bmatrix} \frac{\partial p_0}{\partial r_0} & \frac{\partial p_0}{\partial r_1} \\ \frac{\partial p_1}{\partial r_0} & \frac{\partial p_1}{\partial r_1} \\ \frac{\partial p_2}{\partial r_0} & \frac{\partial p_2}{\partial r_1} \end{bmatrix}$ | $\sqrt{\dfrac{1}{2}\operatorname{tr}\left(F_{\bar{r}_m}^{-1}\right)}$ |
| 3 | 4 | $\dfrac{I_i}{I_0 + I_1 + I_2 + I_3}$ | $\begin{bmatrix} p_0^{-1} & 0 & 0 \\ 0 & p_1^{-1} & 0 \\ 0 & 0 & p_2^{-1} \end{bmatrix} + p_3^{-1}\mathbf{1}$ | $\begin{bmatrix} \frac{\partial p_0}{\partial r_0} & \frac{\partial p_0}{\partial r_1} & \frac{\partial p_0}{\partial r_2} \\ \frac{\partial p_1}{\partial r_0} & \frac{\partial p_1}{\partial r_1} & \frac{\partial p_1}{\partial r_2} \\ \frac{\partial p_2}{\partial r_0} & \frac{\partial p_2}{\partial r_1} & \frac{\partial p_2}{\partial r_2} \end{bmatrix}$ | $\sqrt{\dfrac{1}{3}\operatorname{tr}\left(F_{\bar{r}_m}^{-1}\right)}$ |
| 3 | 5 | $\dfrac{I_i}{I_0 + I_1 + I_2 + I_3 + I_4}$ | $\begin{bmatrix} p_0^{-1} & 0 & 0 & 0 \\ 0 & p_1^{-1} & 0 & 0 \\ 0 & 0 & p_2^{-1} & 0 \\ 0 & 0 & 0 & p_3^{-1} \end{bmatrix}$ $+ p_4^{-1}\mathbf{1}$ | $\begin{bmatrix} \frac{\partial p_0}{\partial r_0} & \frac{\partial p_0}{\partial r_1} & \frac{\partial p_0}{\partial r_2} \\ \frac{\partial p_1}{\partial r_0} & \frac{\partial p_1}{\partial r_1} & \frac{\partial p_1}{\partial r_2} \\ \frac{\partial p_2}{\partial r_0} & \frac{\partial p_2}{\partial r_1} & \frac{\partial p_2}{\partial r_2} \\ \frac{\partial p_3}{\partial r_0} & \frac{\partial p_3}{\partial r_1} & \frac{\partial p_3}{\partial r_2} \end{bmatrix}$ | $\sqrt{\dfrac{1}{3}\operatorname{tr}\left(F_{\bar{r}_m}^{-1}\right)}$ |

**Table S1.**
Relevant quantities for characterizing the localization scheme performance. $d$ is the dimensionality of the localization, $K$ is the number of exposures of the emitter, $p_i$ are the components of the multinomial vector parameter $\bar{p}$, $F_{\bar{p}}$ is the Fisher information matrix on $\bar{p}$, $\mathcal{J}^*$ is the reduced Jacobian matrix for the change of variables from the reduced $\bar{p}$ space to the molecules position $\bar{r}_m$ space and $\widetilde{\sigma}$ is the arithmetic mean of the eigenvalues of the CRB for the covariance matrix of the molecules position estimation. $\mathbf{1}$ is an all-ones matrix. The shaded cases are the most relevant for this work.

| Ref. | Probe(s) | $D$ ($\mu m^2/s$) | | $\sigma$ (nm) | $\Delta t$ (ms) | Length (ms) | Average Localizations | Cutoff |
|---|---|---|---|---|---|---|---|---|
| a (45) | RNAP-PAmCherry | 7-8 | | 40 | 15 | 85 | 5.6 | 4 |
| b (24) | L1-mEos2 | 0.055 | bound | 20 | 20 | 180 | 9 | 5 |
| | S1-mEos2 | 0.4 | free | 60 | | | | |
| c (46) | RelA-YFP | 1.52 | | 45 | 10 | 90-150 | >9-15 | 6-10 |
| | RelA-mEos2 | 0.64 | | | | | | |
| | RelA-Dendra2 | 0.32 | | | | | | |
| | S2-mEos2 | 0.05 | | | | | | |
| d (47) | S2-YFP | 0.04 | | 10-30 | 30 | 150-180 | 5-6 | 8-13 |
| e (48) | VSVG-EosFP | 0.14 | | 25 | 50 | 250 | 4-5 | 15 |
| | Gag-EosFP | 0.11 | | | | | | |

**Table S2.**
Results and parameter values of typical camera tracking experiments in living cells using fluorescent protein labels. The cutoff states the minimum number of localizations used for $D$ estimation.